\documentclass{aa}  

\usepackage{graphicx}
\usepackage{txfonts}
\usepackage{color}
\usepackage{hyperref}

\usepackage{ulem}
\usepackage[below]{placeins}

\begin{document}

   \title{Ten new, ultracompact triply eclipsing triple star systems}

   \author{T. Borkovits\inst{1,2,3}  %\fnmsep\thanks{Just to show the usage of the elements in the author field}
	  \and
	  S. A. Rappaport\inst{1,4}
          \and
	  T. Mitnyan\inst{1,5}
	  \and
	  R. Gagliano\inst{6}
	  \and
	  T. Jacobs\inst{7}
	  \and
	  B. Powell\inst{8}
	  \and
	  V. Kostov\inst{8,9}
	  \and
	  M. Omohundro\inst{10}
	  \and
	  M. H. Kristiansen\inst{11,12}
	  \and
	  I. Terentev\inst{13}
	  \and
	  H.M. Schwengeler\inst{10}
	  \and
	  D. LaCourse\inst{13}
	  \and
	  Z. Garai\inst{14,15}
	  \and
	  T. Pribulla\inst{14}
	  \and
	  I. B. B\'\i r\'o\inst{1,2}
          \and
          I. Cs\'anyi\inst{2}
	  \and
	  Z. Dencs\inst{3,15}
	  \and
	  A. P\'al\inst{3}
          }

   \institute{HUN--REN--SZTE Stellar Astrophysics Research Group, H-6500 Baja, Szegedi \'ut, Kt. 766, Hungary 
         \and
   	   Baja Astronomical Observatory of University of Szeged, H-6500 Baja, Szegedi \'ut, Kt. 766, Hungary.\\
             \email{borko@bajaobs.hu} 
	\and
	   Konkoly Observatory, Research Centre for Astronomy and Earth Sciences,  H-1121 Budapest, Konkoly Thege Mikl\'os \'ut 15-17, Hungary	   
	\and
	  Department of Physics, Kavli Institute for Astrophysics and Space Research, M.I.T., Cambridge, MA 02139, USA \\
	   \email{sar@mit.edu}	   
	\and
	   Department of Experimental Physics, University of Szeged, H-6720, Szeged, D\'om t\'er 9, Hungary
	\and
	   Amateur Astronomer, Glendale, AZ 85308 
	\and
	   Amateur Astronomer, Missouri City, Texas 77459 USA
	\and
	   NASA Goddard Space Flight Center, 8800 Greenbelt Road, Greenbelt, MD 20771, USA 
	\and
	   SETI Institute, 189 Bernardo Avenue, Suite 200, Mountain View, CA 94043, USA
	\and
	   Citizen Scientist, c/o Zooniverse, Dept,~of Physics, University of Oxford, Denys Wilkinson Building, Keble Road, Oxford, OX1 3RH, UK 
	\and
	   Brorfelde Observatory, Observator Gyldenkernes Vej 7, DK-4340 T\o ll\o se, Denmark 
	\and 
	   National Space Institute, Technical University of Denmark, Elektrovej 327, DK-2800 Lyngby, Denmark 
	\and
	   Amateur Astronomer, 7507 52nd Place NE Marysville, WA 98270, USA 
	\and
	   Astronomical Institute, Slovak Academy of Sciences, 05960 Tatransk\'a Lomnica, Slovakia
        \and
	   ELTE Gothard Astrophysical Observatory, Szent Imre herceg u. 112, H-9700 Szombathely, Hungary
       }

   \date{Received  .., 2025; accepted ... , ...}

  \abstract
  {}
   {We have identified more than a hundred close triply eclipsing hierarchical triple star systems from data taken with the space telescope TESS. Many of them have outer periods less than or, close to 100\,days, hence, we call them `ultracompact hierarchical triples'. These systems are noteworthy in that we can potentially determine their dynamical and astrophysical parameters with a high precision, in many cases even without radial velocity data. In the present paper we report the comprehensive study of ten new ultracompact triply eclipsing triple star systems, located in the northern ecliptic hemisphere, taken from this larger sample: TICs 198581208, 265274458, 283846096, 337993842, 351404069, 378270875, 403792414, 403916758, 405789362, 461500036.}
   {Most of the data for this study come from TESS observations, but we obtained supplemental ground-based photometric measurements for two of the systems.   The eclipse timing variation curves extracted from the TESS and the ground-based follow up data, the photometric light curves, and the spectral energy distribution are combined in a complex photodynamical analysis to yield the stellar and orbital parameters of all ten systems.}
   {The outer periods are in the range of 46.8-101.4 days. We found third-body forced, rapid apsidal motion in four systems. Moreover, TIC~403916758 was found to be a double twin triple (i.e. both the inner and the outer mass ratios are close to unity). All of the systems are substantially flat, with mutual inclination angles of $\lesssim 5\degr$. Finally, we have taken the results for the ten systems in the present paper, and combined them with the system parameters for more than 30 other compact triples that we have reported on in previous work, in order to examine some of the global properties of these systems on a statistical basis. }
   {}


   \maketitle
   
\nolinenumbers

\section{Introduction}  
\label{sec:intro}

Triple and multiple stellar systems are quite frequent. Their fractional abundance grows quickly with the mass of the primary component. For example, according to the recent review of \citet{offneretal23}, the bias corrected triple/high-order faction of brown dwarfs and main-sequence (MS) stars exceeds 10\% for solar-type stars, and may  reach even 60-70\% for multiple systems that consist of at least one O-type component.  Restricting ourselves only to triple stars (or, at least the innermost triple subsystems of higher-order hierarchies), the characteristic sizes of such systems may span several magnitudes from the regime of some tens of millions of kilometers (that is, scalesizes smaller than the orbit of Mercury or, at least, Venus) to some parsecs.  These correspond to outer periods, $P_{\rm out}$, of just a few weeks to billions of years.  

The astrophysical, dynamical, as well as evolutionary significances of binary and multiple systems which belong to different size scales are discussed in several works, \citep[e.g.][]{kisseleva-eggletoneggleton10,grishinperets22,offneretal23,sagliaetal25}. The various observing techniques which may be best for one or the other kinds of multiple stellar systems (or, even, different subsystems within the same multiple star system) are also discussed in a number of papers \citep[see, e.g. the chart of][]{tokovinin14}.

In this paper we concentrate on triple systems that have the smallest physical sizes (or, what is practically the same, which have the shortest outer periods). These are referred to as `compact' or even `ultracompact' triples, depending on the outer period.  Before the advent of the recent planet-hunter space-telescopes, e.g., Kepler \citep[see][]{borucki10} and TESS \citep[see][]{ricker15} which produce(d) nearly uninterrupted high-precision photometric observations (from months to years) for millions of stars, it was exceedingly rare to serendipitously identify such triple star systems where the period of the outer, third star did not exceed a number of years.

In contrast to this, the {\textit Kepler} spacecraft identified many compact triple systems, including more than a dozen such systems where the distant third star periodically eclipses, or is eclipsed by, the inner binary members \citep[see][for the discovery of the first such object, KOI-126]{carter11}. Naturally, the presence and discovery of such `triply eclipsing triple stars' are only a matter of purely geometric effects (that is, the outer orbit should be seen almost exactly edge-on by an observer), and the chances of observing third-body eclipses are roughly proportional to $P_{\rm out}^{-2/3}$.  The relatively easy detection of such systems opens up a new window for the discovery of compact triple star systems.

Besides, or more precisely after, these {\textit Kepler} discoveries, a number of other triply eclipsing triples were also reported \citep[see the compilation of][]{borkovits22}. The real breakthrough, however, came with the regular operation of the TESS spacecraft in 2018. Since the beginning of the survey observations with this instrument, the number of known triply eclipsing triple stars started to grow quickly and, by now, our group has identified more than a hundred such new systems. (Of course, one should keep in mind, that this number of the identified or, at least suspected, triply eclipsing triple systems continues to be very low compared to the total number of known or, at least hypothesized, multiple stars; but, this is now quite sufficient to carry out detailed studies of a substantial number of individual triples, as well as to carry out statistical investigations of the group.) Formerly, in a series of earlier papers, we have carried out homogeneous photodynamical analysis of 32 TESS-discovered triply eclipsing triple systems \citep{borkovitsetal20b,borkovitsetal22a,mitnyanetal20,rappaportetal22,rappaportetal23,rappaportetal24,czavalingaetal23,kostovetal24}. In the current paper we introduce a similar investigation of ten additional newly discovered triply eclipsing triple stellar systems.

In Section \ref{sec:10triples} we describe the collection of 10 ultra-compact triply eclipsing triple systems that we have selected for this detailed study.  We explain briefly how these sources were discovered.  In Section \ref{sec:lightcurves}, the light curves exhibiting third body eclipses are introduced along with model fits, and briefly discussed.  The eclipse timing variation curves (ETVs) are introduced and discussed in general in Section \ref{sec:etvs}.  The photodynamical code, with which the system parameters are extracted, is briefly reviewed in Section \ref{sec:photodynamics}.  In Section \ref{sec:results} we summarize the system parameters in a set of comprehensive tables, and the results for each individual system are presented. 
   
\section{The ten triply eclipsing triples}  
\label{sec:10triples}

In this compilation we select ten potentially interesting, formerly unanalysed, triply eclipsing triples from the northern ecliptic hemispheres.  We focused our attention on those systems with outer periods of $\lesssim 100$ days.\footnote{Originally we had intended to use the exact value of $P_\mathrm{out} \le 100$\,days as the upper limit on the outer periods of the triple systems considered in this paper. Finally, however, we decided not to exclude TIC\,337993842 for which the outer period is longer than this by only $\sim1.4\%$.} The main catalogue parameters for these ten triples can be found in Tables~\ref{tbl:mags10-1} and \ref{tbl:mags10-2}. We refer to these systems as `ultra compact hierarchical triple' (UCHT) star systems.  Choosing northern ecliptic systems has two purely practical aspects. First, according to the currently available observing schedule of the TESS mission, the spacecraft will not observe any northern ecliptic sectors until at least the end of Cycle 8 (September 7th, 2026). Therefore, we cannot expect any newer TESS observations in the forthcoming year. Second, objects in the northern ecliptic hemisphere are easily available for follow up ground-based observations (at least in some parts of the year) with telescopes in Central Europe.

The discoveries of the triply eclipsing nature of the currently investigated ten triples were made in three different ways. Six of the ten systems were found by our `Visual Survey Group' \citep[VSG;][]{kristiansen22} in the manner described in detail, e.g., in \citet{rappaportetal24}. Moreover, two triples, TICs\,265274458 and 351404069, were first identified as doubly eclipsing 2+2 quadruples, and the former was even catalogued as such by \citet{kostovetal21}. Finally, the presence of likely third-body eclipses in the early TESS light curves of the previously known eclipsing binaries (EBs) (TIC\,198581208 = CSS\,J170425.5+463533 and TIC\,461500036 = ASASSN-V\,J221919.64+850413.4) were first reported by \citet{zascheetal22}.

The analysis of UCHT objects has both theoretical and practical aspects. Regarding the theoretical aspects, (i) these systems have small characteristic sizes, e.g., they would typically fit within Venus' orbit around the Sun. Therefore, we may expect, that the components of such systems formed in a different way than is the case for wider triples \citep[see, e.g.,][]{tokovinin21}, and perhaps the orbital and dynamical configurations of such systems retained some relics of these formation mechanism(s).  Moreover (ii), in these systems we may expect some rare and extreme stellar evolutionary end states (such as multiple common envelope stages, solo or multiple stellar mergers, etc.), which might help to explain the origin of some extreme stellar systems or phenomena.  Finally, (iii) the vast majority of such systems are not only compact or, ultra compact, but also `tight' enough for the continuous occurrence and detection of gravitational perturbations (even higher order ones, see, e.g. \citealp{borkovitsmitnyan23}), which are not only interesting in themselves, but may also lead to more accurate dynamical determinations of the stellar and orbital parameters.  In this regard, note that formerly, tight triples were defined as outer-to-inner period ratios less than 100 (i.e., $P_\mathrm{out}/P_\mathrm{in}\lesssim100)$ \citep[see, e.g.][]{borkovitsetal22b}, but in the newer works a value of $P_\mathrm{out}/P_\mathrm{in}\lesssim50)$ is considered `tight'. In such triples, the third-body perturbations substantially affect the orbit of the inner binary, and, at least in the case of compact systems, such dynamical effects are observable within months or years \citep{borkovitsetal25}.  Regarding these definitions, one can see that the former definition for tightness is naturally satisfied in all ten of our selected UCHT systems, as the inner EB period in all these triples is longer than 1\,day. Considering the newer, and more strict definition, the ratio of the outer and the inner periods is less than 50 for all but one system amongst our ten UCHTs. And, as will be shown below in Sect.~\ref{sec:etvs}, the ETVs in these nine systems are clearly dominated by the short-timescale third-body perturbations.

The practical aspects of the very short outer periods also manifest themselves in at least two ways. First, (i) the shorter the outer orbital period (and, hence, the separation of the third star from the members of the inner binary), the larger the chance for third-body eclipses \citep[see, quantitatively, e.g. in][]{borkovitsetal22b}. Second, (ii) TESS has revisited the northern ecliptic hemisphere several times since 2019. Despite the fact that the nominal duration of each sector is only $\sim$27.5\,days (and, practically, even less), there is a good chance, at least for the shortest outer period systems, that the total outer orbital phase has been covered with observations (even several times).  Therefore, numerous third-body eclipse events have been observed.  In this context we refer to Table~\ref{tbl:sectors} where one can see that all of our targets were measured during 4--15 sectors and, moreover, at least four third-body eclipsing events were observed for all but one of our targets. (The sole exception is, our longest period triple, TIC\,337993842, for which only two third-body eclipse events were detected with TESS and, moreover, both of them belong to the superior conjunction of the third star.) In the case of the shortest outer period system, TIC~405789362, 15 such outer eclipsing events were detected (see examples in Figs.\,\ref{fig:283846096lc} and,\,\ref{fig:lcs1}--\ref{fig:lcs2}).  

Adding to the important third-body eclipses are the multiply covered ETV curves (Figs.\,\ref{fig:351404069ETV} and \ref{fig:etvs}). It is not surprising that, for all systems, we were able to determine the outer orbital period (which is a key parameter to start the complex, photodynamical analysis) with the TESS ETV observations alone. This is valid even in the case of TIC\,337993842 where, despite the lack of any observed third-body events at the inferior conjunction of the outer orbit, due to the well-covered, dynamically dominated ETV curve, we were able to determine the orbital parameters exclusively from the available TESS data.) Therefore, for the current analysis, the importance of the ground-based archival data was less important than in our previous studies of triples.  On the other hand, however, these statements are valid only a posteriori, i.e., after collecting several cycles of TESS data. When we first discovered this newest set of triples, however, we followed the same steps of the preliminary period determinations as before (that is, we used the archival data to find reliable input periods for the first, analytic ETV studies, as it was described e.g., in \citealp{rappaportetal22}).

\section{Light curve and model fits}  
\label{sec:lightcurves}

\begin{figure*}[ht]
   \centering
\includegraphics[width=0.32\textwidth]{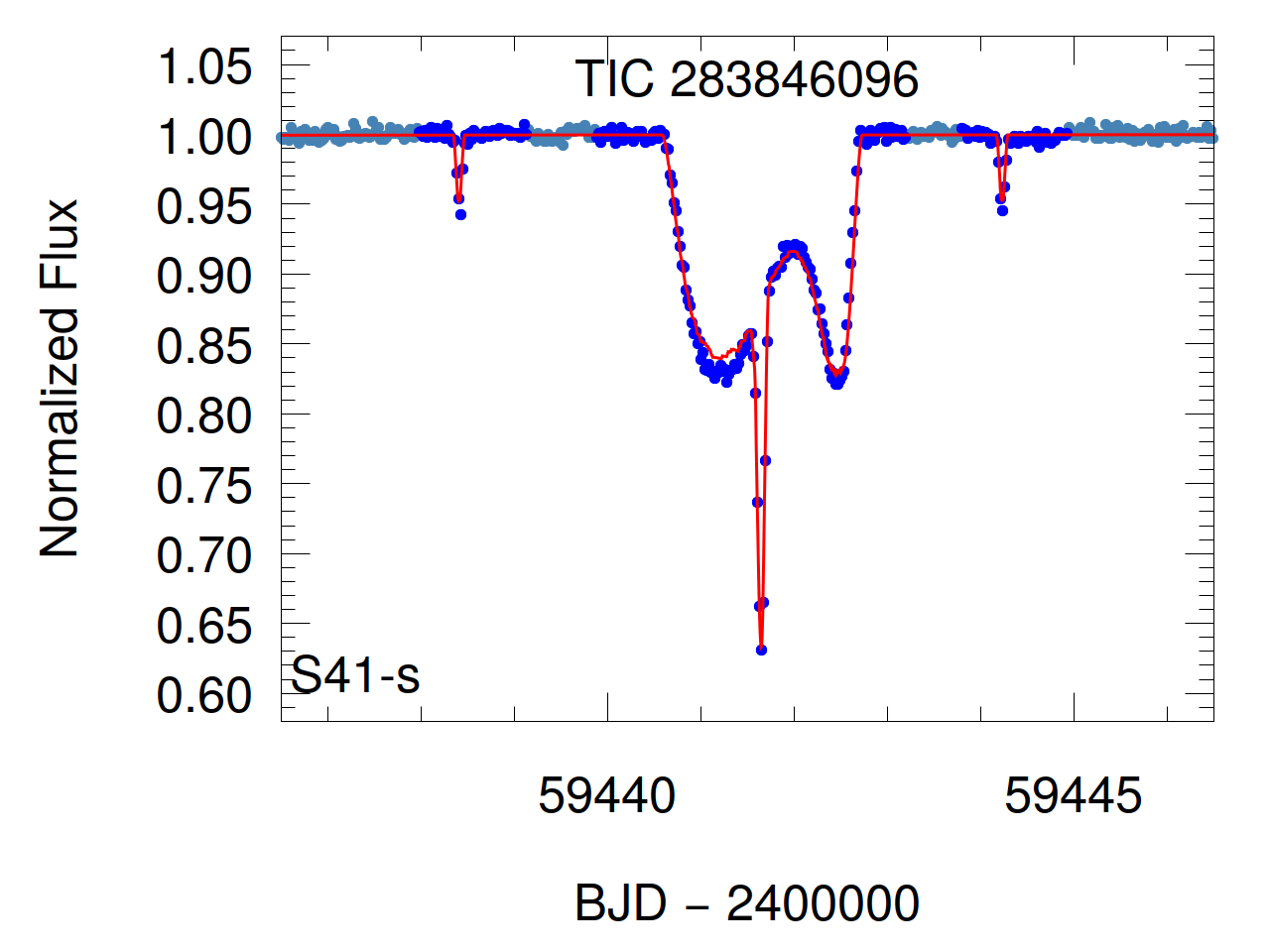}\includegraphics[width=0.32\textwidth]{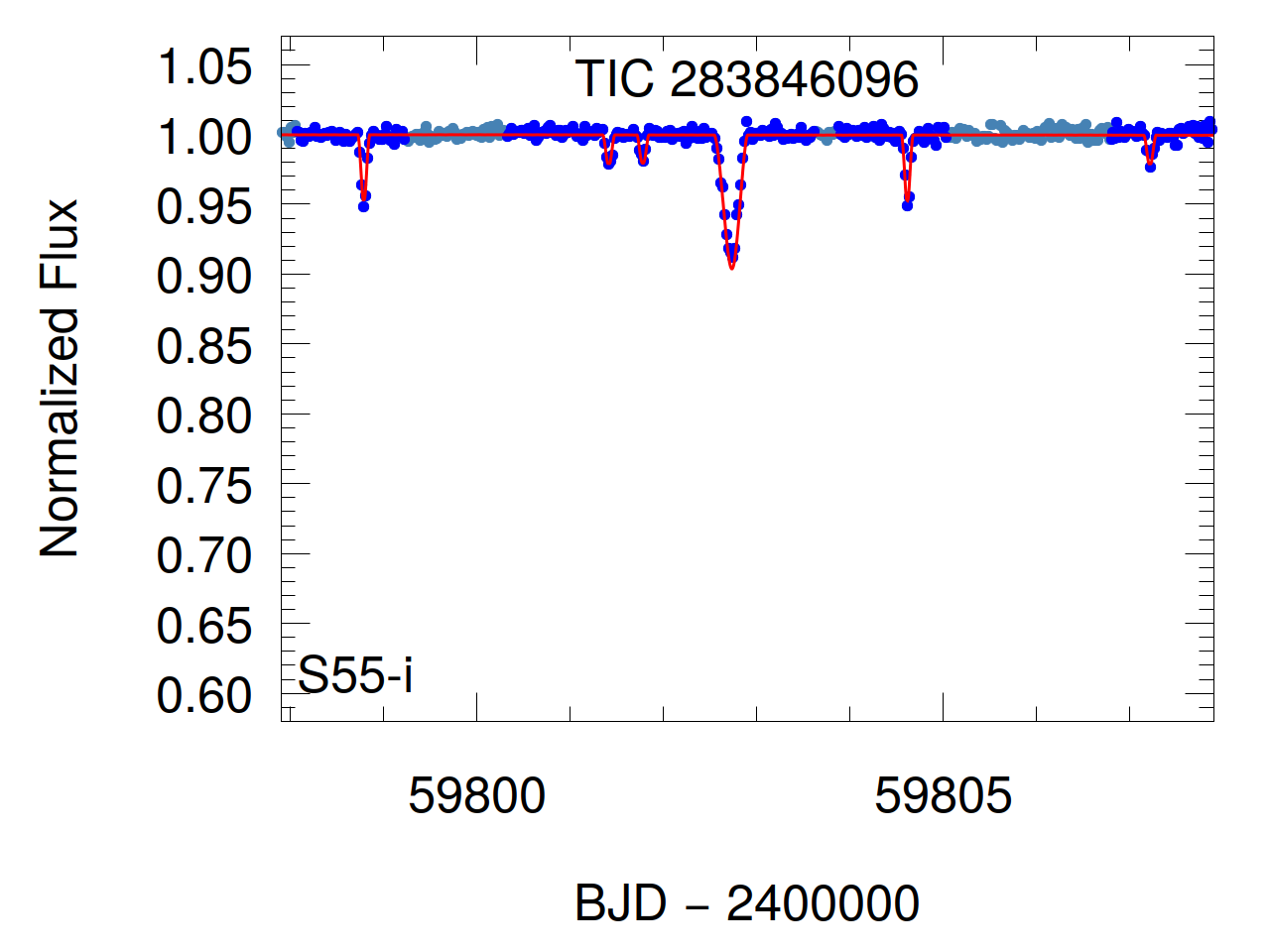}\includegraphics[width=0.32\textwidth]{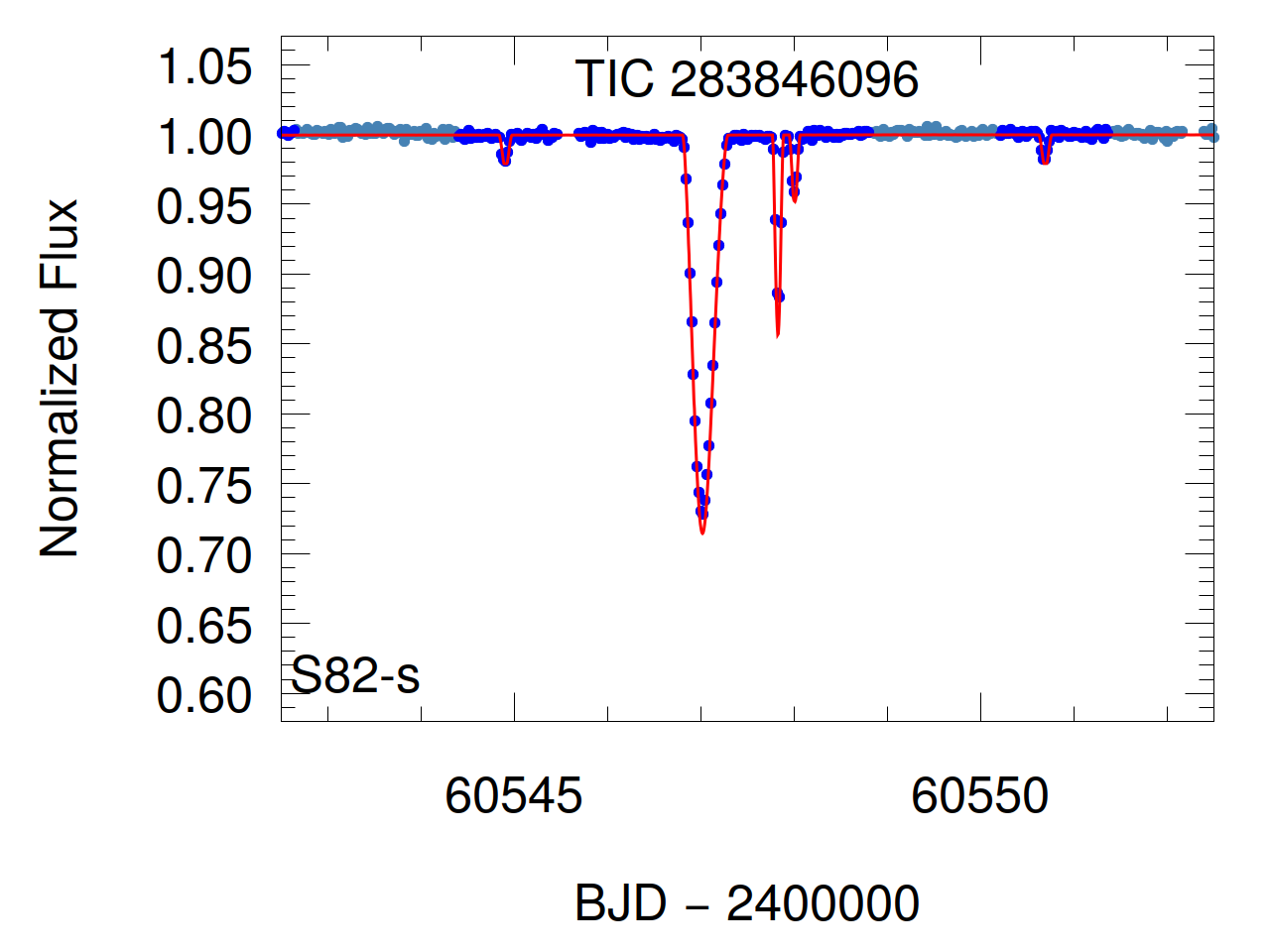}
   \caption{Light curves (blue points) and model fits (smooth red curves) near three illustrative third-body eclipses of TIC\,283846096. Dark/pale blue points are for those light curve sections which were used/not-used for the photodynamical solution. The sector numbers are indicated in the lower left corner of each panel. Letters `i' or `s' after the sector numbers refer to the inferior or superior conjunction of the third star, respectively.}
   \label{fig:283846096lc}
\end{figure*}

TESS observed 70 third-body eclipsing events from the currently investigated group of ten triples, of which we have selected three such events for TIC\,2838846096, in Fig.\,\ref{fig:283846096lc}, to illustrate the main properties of the different third-body eclipses.  Moreover, we present 26 additional third-body events in Figs.\,\ref{fig:lcs1}--\ref{fig:lcs2} for the remaining nine systems. The blue points represent the TESS measurements. For the light curve analysis and modelling we used 30-min cadence data. For those sectors where  shorter cadence data were available, we binned them to 30-min cadence. Therefore, all the blue points are at the same 30-min cadence. In the vicinity of the third-body, or `extra'', eclipses, naturally, several regular binary eclipses are also shown. These latter eclipses are generally self-evident, while the `extra' eclipses are, for the most part, all the dips in flux that cannot be ascribed to the regular EB eclipses. In some cases, especially, when the third-body eclipses (or eclipsed by) the members of the inner binary system during a regular two-body eclipse (that is, when the three stars, from the point of view of the observer are aligned), the third-body eclipses are quite irregular and anomalous looking (see the left panel of Fig.\,\ref{fig:283846096lc}), while in other cases, in general, around the quadratures of the inner binary, the eclipses look almost `normal' but occur in rapid succession and/or are clearly out of place. This latter situation is illustrated in the middle and right panels of Fig.\,\ref{fig:283846096lc}. (The difference in the depths of the extra dips between the two panels, is due to the fact that in the middle panel the cooler inner binary members occulted the third star, separately, while in the right panel, this latter, hottest star eclipsed the two cooler EB members, one by one.)

We also plot the light curve solution taken from the joint photodynamical analysis, as a smooth red curve. These solutions will be discussed below, in Sect.~\ref{sec:photodynamics}.

These third-body eclipses, especially their shapes, durations, as well as their occurrence times, contain crucial information about both the orbits and the properties of the stars themselves (e.g., the relative sizes and effective temperatures).

\section{ETV curves}
  \label{sec:etvs}
  
\begin{figure}
\centering
     \includegraphics[width=\hsize]{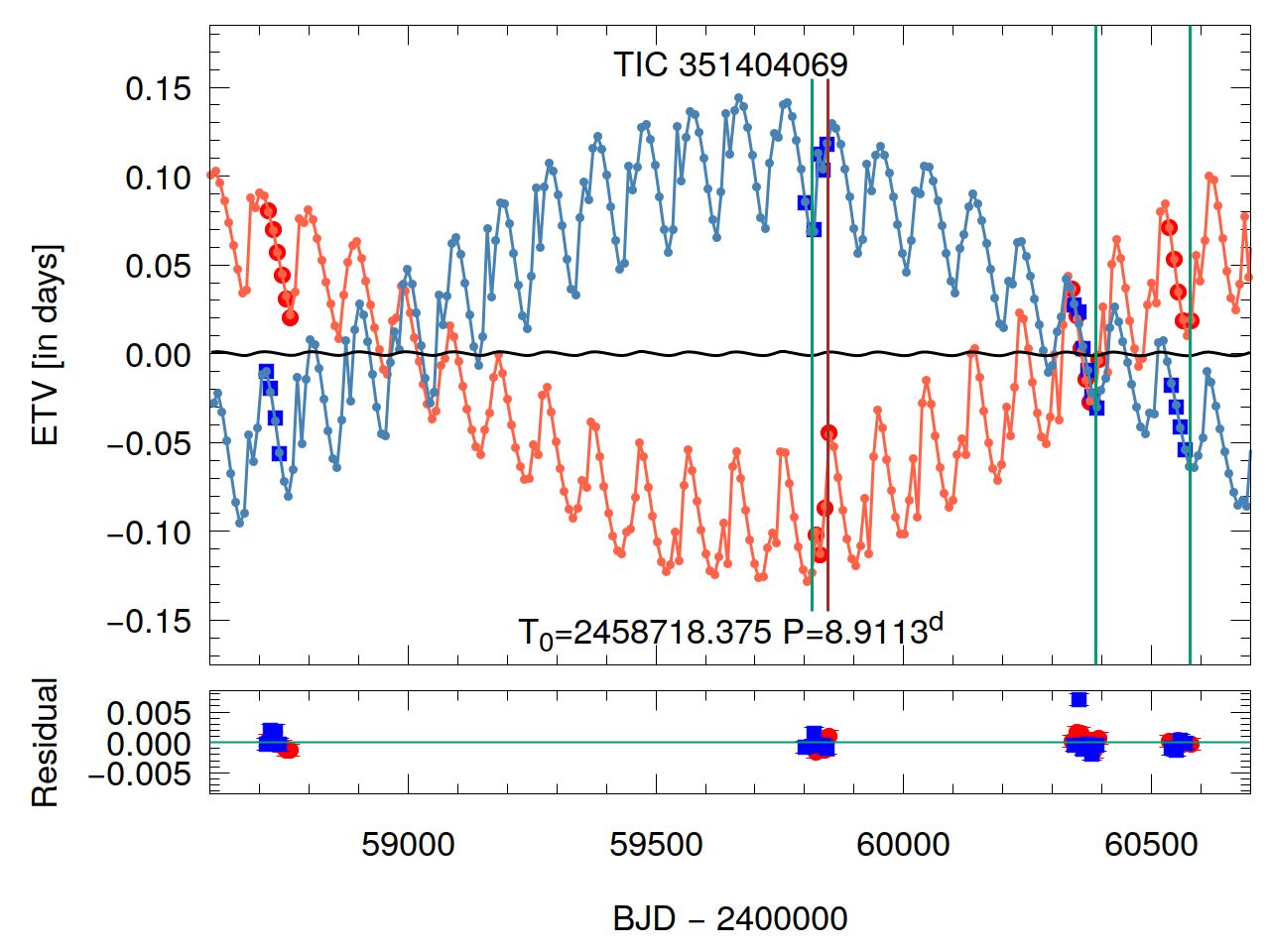}
     \caption{Primary and secondary ETV curves (red and blue circles, respectively) formed from the TESS observations with the best-fit photodynamical solution for TIC 351404069 (see Sect.~\ref{sec:photodynamics}). The dynamically forced, rapid apsidal motion of the inner, eccentric EB is nicely visible. (Be aware of the huge amplitude of the ETV!) The horizontally centred black curve represents the pure LTTE contribution. Vertical lines mark the times of the observed outer eclipses (green -- the binary occulting the tertiary star and, brown -- vica versa).}
\label{fig:351404069ETV}
\end{figure}

In addition to the light curves of the third-body eclipses discussed above, as well as the regular eclipses, another very important input for the comprehensive photodynamical analysis (to be discussed in Sect.~\ref{sec:photodynamics}) comes from the ETV curves. These are based on the mid-times of both kinds of eclipses (primary and secondary) of the inner EB in each triple.  These mid-eclipse times are extracted in the manner discussed previously in \citet{borkovitsetal15,borkovitsetal16}. As one can see in Fig.\,\ref{fig:351404069ETV} for TIC\,351404069 and, in Fig.\,\ref{fig:etvs} for the other nine triples, well characterised non-linear behavior can be seen all but one of these ETV curves. These features come predominantly from three basic effects, as follows.  First is the classical light-travel-time effect \citep[LTTE;][]{roemer1677} due to the changing distance to the EB as it is pulled around in its outer orbit by the tertiary star. The amplitude of this effect is proportional to $P_\mathrm{out}^{2/3}$ \citep[see, e.g.][]{borkovitsetal16} and, therefore, due to the short outer periods of the current systems, this effect is generally the least significant in these particular systems as can nicely be seen in Fig.\,\ref{fig:351404069ETV}, where the cyclic, sinusoidal nature of the black horizontal LTTE curve is almost unnoticeable).  

Second are the `dynamical' delays which, in nearly coplanar orbits, are caused largely by the lengthening of the EB orbit due to the presence of the tertiary star \citep[see, e.g.,][]{rappaportetal13,borkovitsetal15}. This effect manifests itself in Fig.\,\ref{fig:351404069ETV} as the $P_\mathrm{out}$-period wobbles. The magnitude of this effect depends on the instantaneous separation between the EB and the more distant third star, and hence, it varies with the phase of the outer orbit (at least when it is eccentric). (3) Finally, there is the so-called apsidal motion (AM), which may occur in eccentric binaries. This is a longer timescale effect which has three main types: (i) the classical, tidal AM, caused by the non-spherical mass distributions of the tidally distorted binary components; (ii) the general relativistic AM and; (iii) the dynamically driven one, forced by the perturbations of the tertiary star. The timescale of the dynamically driven AM is on the order of $P_{\rm out}^2/P_{\rm in}$ \citep[see, e.g.][]{soderhjelm75}. The dominant driver of AM in the currently investigated systems is the third-body forced AM; however, in a minority of these ten systems, the mutual tidal deformations of the two EB stars are also significant. The largest amplitude, longer period anti-correlated nature of the primary and secondary ETV curves in Fig.\,\ref{fig:351404069ETV} is due to this, dynamically driven AM.

As was mentioned above, the ETV curves themselves are shown in Figs.\,\ref{fig:351404069ETV} and \ref{fig:etvs}, while the mid-eclipse times used for the derivation of these ETV curves are tabulated in Appendix~\ref{app:ToMs}.

\section{Photodynamical models}
\label{sec:photodynamics}

Similar to our former works on triply eclipsing triples, the ten multiple systems considered in this work have been subjected to a detailed photodynamical analysis with the use of our own developed software package {\sc Lightcurvefactory}.  The details as well as the input data sets and the input/output parameters have been explained in several of our earlier papers \citep[see, e.g.][]{borkovitsetal18,borkovitsetal19a,borkovitsetal19b,borkovitsetal20a,borkovitsetal20b,borkovitsetal21,mitnyanetal20}. Therefore, here we note only that the code contains four basic features. First, there are emulators for multi-passband light curve(s), the ETVs, and radial velocity (RV) curve(s) (the latter feature was unused in the current work due to the absence of any RV data).  Second, the main astrophysical parameters of the stars are calculated with the use of built-in, tabulated \texttt{PARSEC} isochrones \citep{PARSEC} and, therefore, we are able to produce theoretical combined spectral energy distributions (SED) for the given system under investigation (this feature is optional). Third, there is a built-in numerical integrator (a seventh-order Runge-Kutta-Nystr\"om algorithm) to calculate the instantaneous (Jacobian) coordinates and velocities of the stars along their perturbed three-, or multiple-body orbits. Finally, there a Markov Chain Monte Carlo (MCMC)-based search routine for determining the best-fit system parameters, as well as the statistical uncertainties.  The latter feature uses our own implementation of the generic Metropolis-Hastings algorithm \citep[see, e.g., ][]{ford05}.

Note also that all the essential details of how this code was used to analyse compact triply eclipsing triple systems, especially those which were found with the TESS spacecraft, were described in \citet{rappaportetal22}. Here we provide only a very concise overview of the inputs to the code and the parameters that are either fitted or constrained by the MCMC fit. Altogether, for a hierarchical triple configuration, there are 25 -- 27 system parameters that result directly from the analysis. These are nine stellar parameters (masses, radii, and effective temperatures of all the three stars), all 12 of the elements of the inner and outer orbits (or, some equivalents of the classic orbital elements, e.g. orbital periods instead of semi-major axes), as well as the 4 system parameters: distance to the source and the interstellar extinction, as well as the system metallicity and age.  

\begin{figure*}
\centering
     \includegraphics[width=0.7\textwidth]{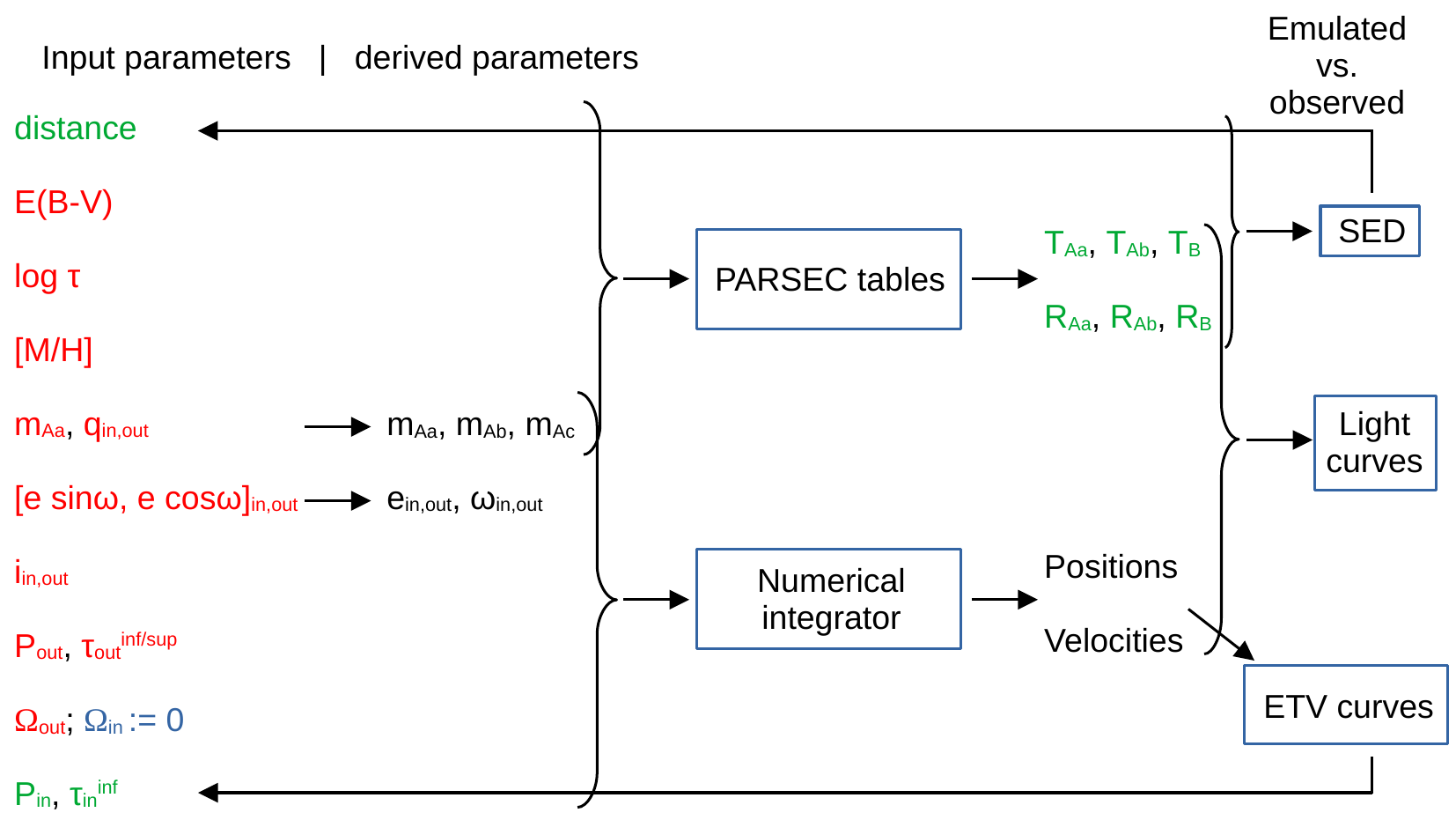}
     \caption{Schematic flow chart of the entire photodynamical fitting process. Parameters marked in red font give those input parameters that are allowed to adjust during each MCMC trial step. Green symbols stand for the constrained quantities, while the black symbols denote quantities derived directly from the (adjustable, red) input parameters just at the beginning of each trial step and used during the subsequent part of the given step. The other quantities, not shown in the chart, but listed in the result tables below are computed a posteriori, at the end of the entire photodynamical analysis process. Note also, the only parameter with fixed input value ($\Omega_\mathrm{in}=0$) is given in blue font. (For the meaning of each symbol, see Table\,\ref{tbl:definitions}.)}
\label{fig:schematic}
\end{figure*}

In Fig.\,\ref{fig:schematic} we illustrate schematically the process of the entire photodynamical analysis, denoting all the initially adjusted, constrained and fixed input parameters. Another, more detailed flow chart can be found in Fig.\,5 of \citet{borkovitsetal20a}. Finally, one may optionally adjust the amount of any passband-dependent contaminated (extra) light $\ell_\mathrm{x}$ if it is necessary. We note that, in the case of TESS observations, due to the large pixel sizes, the contamination might come from other nearby stars, which may affect the eclipse depths of the investigated source. And, even in the absence of other nearby stars, due to the unavoidable stray light, some extra flux can be expected in the light curve and, therefore, it is useful to allow the extra flux parameter to vary. (And, moreover, note that another source of such contaminated flux might be an unknown, unresolved, more distant bound stellar component.) Hence, in all ten of our sources, we set and adjust the extra light ratio in the TESS band, as the 26th input parameter. Moreover, note that for two of our ten targets (TICs 405789362 and 461500036) we used a second light curve, compiled from ground-based observations in Sloan $r'$-band, too. In these two systems, though there was no a priori information about any contaminating sources in the aperture of the CCD photometry, for homogeneity in the analysis we also allowed (and adjusted) a second extra light parameter and, therefore, for these two triples we used 27 input parameters. (Note, these latter one or two optionally used and adjustable parameters are not shown in Fig.\,\ref{fig:schematic}.)

The TESS light curves which primarily determine the EB and third-body eclipse profiles, as well as the ETV curves via the timing data used in our analysis, were taken from the TESS full-frame images (`FFI').  For eight of the ten systems, the photometry of the FFIs up to Sector 77 was done with Andr\'as P\'al's {\sc FITSH} package \citep{pal12}. The two exceptions are TICs 337993842 and 351404069 for which all-sector photometry was processed with the publicly available software {\sc Lightkurve} \citep{2018ascl.soft12013L}.  For technical reasons, the most recent sectors (from 78 to 86) in the case of all ten triples were processed with this latter software package. Note that changing to another photometry pipeline for the last set of sectors might introduce smaller inhomogeneities into the TESS light curves. Therefore, we made some steps to `validate', or synchronize, the {\sc FITSH} and {\sc Lightkurve} processed light curves to each other. For this reason we processed with the {\sc Lightkurve} pipeline some of those earlier sectors as well, which had been made with the {\sc FITSH} pipeline. In the process, we tested the parameters to be used for the {\sc Lightkurve}-processed light curves to obtain outputs which are close to the {\sc FITSH} light curves (especially in eclipse depths) of the same sectors. 

Again, similar to our previous works, in order to save on computational time, we binned the 200-sec and 10-min cadence data to 30-min cadence\footnote{We note, however, that this binning was applied only to the light curve analysis. The mid-eclipse times for the ETV curves were calculated from the original shorter cadence data sets.}, and dropped the out-of-eclipse regions of these light curves, keeping only the $\pm0\fp15$ phase-domain regions around the EB eclipses themselves.  This latter process of dropping the out-of-eclipse regions was not applied in the vicinities of any third-body eclipses, where we keep the data for an entire binary period before and after the first and last contacts of that particular third-body eclipse. Keeping the out-of-eclipse light curve points for at least two binary periods around each expected third-body eclipse was done for two reasons. First, due to the continuously varying configurations of the three stars, the occurrence times of the outer eclipses are not strictly periodic. Some small shifts in the outer eclipse features may occur which would be taken into account in such a manner. Second, complete omission of the out-of-eclipse light variations would result in the suppression of such lower amplitude effects, as ellipsoidal variations, reflection effects, and Doppler-boosting, which would arise from the binarity, and would introduce some smaller bias in our results. With such a decision, we retained the signals of these possible effects in our investigated light curves, but rendered smaller weights to them with respect to the most important eclipsing patterns.

We also note, that in the case of two targets, TICs\,405789362 and 461500036, ground-based photometric follow-up observations were carried out with the two identical 80-cm RC telescopes of Gothard Astrophysical Observatory (GAO80) Szombathely, Hungary and Baja Observatory of Szeged University (BAO80), Baja, Hungary. The details of these instruments, the methods of observation, and the data processing were described in detail in \citet{borkovitsetal22a}. The sloan $r'$-band light curves that were obtained were also included in our complex analysis, though with half the weight as that of the TESS photometry.

Regarding the mass determination of the investigated stars, we should make some additional comments. In the absence of RV data, the question naturally arises: how are we able to derive absolute stellar masses, temperatures, and radii? This was done with the use of the above mentioned \texttt{PARSEC} isochrones as proxies. The details of this process, together with its limits were described in \citet{borkovitsetal22a}. Moreover, we also discussed this question in \citet{rappaportetal24}. Therefore, we suggest that the interested reader might consult these two earlier papers.

\section{System parameters}
\label{sec:results}

\subsection{Tables of fitted parameters}

In what follows, we discuss the astrophysically and/or dynamically most interesting findings regarding the currently investigated ten UCHT systems. In addition to these discussions, similar to our former works, we also give our results in tabulated form.  These tables do not contain all the directly fitted (i.e. adjusted, constrained) parameters but, in several cases, they instead give parameters that are calculated from those directly derived parameters.  Naturally, we include all the basic stellar parameters in our results but, regarding the orbital elements, for example, instead of the adjusted parameters $e\sin\omega$ and $e\cos\omega$, we give directly the eccentricities ($e$) and arguments of periastron ($\omega$) for each orbit. We also put into our tables such additional calculated or derived geometrical parameters as the relative orientations of the orbits.   For a better comparison with the traditional EB light-curve fitting codes and, also for better accuracy in several non-dimensional relative quantities (which are not adjusted directly in our method but, indirectly, strongly constrain our solution) we calculate and give, e.g., fractional radii and relative temperatures (e.g., $R/a$, $T_B/T_{Aa}$), and some others.  Our tables contain dozens of different parameters, which were defined in \citet{rappaportetal23}, while the methods of the calculations of the indirectly derived quantities were described or referenced in \citet{borkovitsetal15} and \citet{kostovetal21}. Here, however, similar to \citet{rappaportetal24}, for the sake of completeness, we tabulate the meaning of each parameter (generally noted only with symbols in the results tables) in Table~\ref{tbl:definitions}. Finally, note that the system parameters that are derived from the photodynamical analyses are listed in Appendix\,\ref{app:resulttables}, in Tables \ref{tab:syntheticfit_TIC198581208265274458} through \ref{tab:syntheticfit_TIC405789362461500036}.

\subsection{Results for the individual systems}

\subsubsection{TIC 198581208}
\label{Sect:discussion_TIC198581208}

The triply eclipsing nature of this formerly known EB (CSS J170425.5+463533) was first reported in \citet{zascheetal22}; however, the correct outer orbital period, as well as the first complete analysis of this triple, are given only in this paper.  The out-of-eclipse light curve of this EB shows variations whose amplitude is larger than the secondary eclipses themselves. We assumed that this signal had come from chromospheric activity in any of the late-type stars in this triple.  Nevertheless, independent of its origin, it was clear that this signal, as it regards the eclipsing light curve analysis, simply caused additional noise. Therefore, we made efforts to remove these distortions with the use of medium-order local smoothing polynomials, separately for each TESS sector, and naturally excluding the eclipsing sections of the TESS light curves. 
For this process we used 4th to 18th-order polynomials at different sections of the TESS light curve. Such polynomials effectively smooth out the high frequency modulations from the light curve, while better retaining the lower frequency variations which come from the binarity for which the frequencies are nearly equal to or double of the orbital frequency. Despite this, it is clear to us that such polynomials would erase not only those variations which arise from the strongly irregular chromospheric activity, but also they may affect other out-of eclipse variations which may arise from the binarity, such as, e.g., ellipsoidal light variations, reflection effects, Doppler-boosting, etc. Therefore, we selected the order of the fitting polynomials in such a manner that we checked continuously the phase-folded light curve, and retained the shape and structure of the average eclipses, especially close to the first and last contact points. 

According to our results, the most massive component of the system is the primary of the inner EB, with $M_\mathrm{Aa}=1.04\pm0.06\,\mathrm{M}_\sun$, while the third component has only a slightly lower mass $M_\mathrm{B}=0.94\pm0.05\,\mathrm{M}_\sun$. Besides these two sun-like stars, the secondary of the inner EB is a quite low mass with $M_\mathrm{Aa}=0.58\pm0.03\,\mathrm{M}_\sun$, which nicely explains the very shallow secondary eclipses.  As usual, the mass ratios can be determined with higher accuracies than the individual physical masses, and these are $q_\mathrm{in}=0.56\pm0.01$ and $q_\mathrm{out}=0.576\pm0.004$.

The spatial configuration of this old ($\tau=6.2\pm1.7$\,Gyr) system was found to be quite flat with $i_\mathrm{mut}=0\fdg2\pm0\fdg1$. This substantial flatness, however, is likely a residual of the formation processes of this triple. Despite the relatively small characteristic size of the inner EB ($a_\mathrm{in}=10.0\pm0.2\,\mathrm{R}_\sun$), due to the smaller radii of the EB members, the tidal forces should have remained small during the entire MS lifetime of the system \citep[see, e.g.][]{correiaetal11}. In this regard, however, we note that currently the primary star is close to the end of its MS evolution and, therefore, as it evolves, the currently moderate fractional radius of $r_\mathrm{Aa}=R_\mathrm{Aa}/a_\mathrm{in}=0.127\pm0.002$ is expected to grow rapidly, causing more and more significant tidal effects.

The tightness ratio is $P_\mathrm{out}/P_\mathrm{in}\simeq25.3$, and though quite small, is still the third largest in our sample. As the inner orbit is almost circular ($e_\mathrm{in}=0.0141\pm0.0006$) and the outer one has only moderate eccentricity ($e_\mathrm{out}=0.289\pm0.001$), this triple looks to be dynamically quite stable.

Finally, we note that for all ten ETV curves (Figs.\,\ref{fig:351404069ETV} and \ref{fig:etvs}), we also plot the best photodynamically modelled ETV curves, as well as the pure geometrical, LTTE, part of this best-fit solution (black curve). Comparing the amplitudes of the entire primary (red) and secondary (blue) ETV curves with the LTTE contribution (black) one can see that the current ETV curve (top left panel of Fig.\,\ref{fig:etvs}), as was expected theoretically  \citep[see, e.~g.,][]{borkovitsetal15} for such a tightness ratio (see in the paragraph above), is dominated by the dynamical effects. Note also that the maxima and minima of the black LTTE contribution (which are not necessarily coincident with the extrema of the entire ETV curves) represent the largest and the smallest distances of the EB from the observer, and hence one may expect that the third-body should transit in front of the EB stars (vertical brown lines), and eclipsed by either of the EB stars (vertical green lines) around these LTTE-ETV extrema, respectively.

\subsubsection{TIC 265274458}
\label{Sect:discussion_TIC265274458}

This system was listed as a 2+2 type quadruple system candidate in \citet{kostovetal22} which was based upon only the Year 2 TESS data. Newer observations, however, made it clear that this is an UCHT, exhibiting both kinds of third-body eclipses.

TIC\,265274458 has the most extreme inner mass ratio ($q_\mathrm{in}=0.229\pm0.003$) in our sample.  As a consequence, the secondary eclipses are hardly visible and, therefore, this was the only EB in the current sample where we were unable to measure a useful secondary ETV curve. This triple system was found to be quite young ($\tau\approx270\pm30$\,Myr), and dominated clearly by the hot primary of the inner EB ($M_\mathrm{Aa}=1.90\pm0.03\,\mathrm{M}_\sun$) which emits $\approx93\pm1$\% of the total flux of the triple, at least in the photometric band used by TESS. The other two late type stars have substantially lower masses:  $M_\mathrm{Ab}=0.434\pm0.007\,\mathrm{M}_\sun$ and $M_\mathrm{B}=0.98\pm0.02\,\mathrm{M}_\sun$. Here we also call attention to the fact that, while the mass of the third star was determined to only $\sim$5\,\% fractional error, the relative uncertainty in the much more accurate outer mass ratio is about 1\%, which follows from $q_\mathrm{out}=0.420\pm0.004$. This emphasizes again that the dimensionless relative quantities, such as the mass ratios and the fractional radii (determined themselves mainly by dynamical and light curve effects) have much better accuracies than those of the absolute masses and radii.  The latter determinations depend strongly on the \texttt{PARSEC} isochrones which served mainly as proxies for the mass determinations in the absence of any RV data.  Regarding the radii, the $1\sigma$ statistical uncertainties of the absolute values can exceed even $\sim$2\,\%, being: $R_\mathrm{Aa}=1.70\pm0.03\,\mathrm{R}_\sun$, $R_\mathrm{Ab}=1.70\pm0.03\,\mathrm{R}_\sun$ and, $R_\mathrm{B}=0.86\pm0.02\,\mathrm{R}_\sun$, while their relative, i.e., dimensionless scaled counterparts were obtained with considerably smaller relative uncertainties: $r_\mathrm{Aa}=0.146\pm0.001$, $r_\mathrm{Ab}=0.0365\pm0.0005$ and, $r_\mathrm{B}=0.0092\pm0.0001$, respectively. 

Interestingly, we found both the inner and the outer orbits to be extremely close to circular ($e_\mathrm{in}=0.0025\pm0.0004$ and, $e_\mathrm{out}=0.003\pm0.002$). Such a doubly circular, almost flat ($i_\mathrm{mut}=0\fdg7\pm0\fdg3$) configuration would be far from surprising in the case of an UCHT formed by old, and at least partially evolved, stars (see, e.g., the cases of HD~181068 \citealp{borkovitsetal13}; TIC~242132789 \citealp{rappaportetal22}; and TIC~332521671 \citealp{rappaportetal23}) where the tidal interactions were strong enough and had time to flatten and circularise the whole systems. In the current situation, however, with young stars having small fractional radii, this is not the case. Therefore, we might argue that the current flat and doubly circular configuration is most likely primordial for this triple.

\subsubsection{TIC 283846096}
\label{Sect:discussion_TIC283846096}

This triple can claim several superlatives amongst our ten sample systems. From a dynamical point of view, this (i) is the tightest triple, having $P_\mathrm{out}/P_\mathrm{in}\approx9.71$, (ii) has the most eccentric inner orbit ($e_\mathrm{in}=0.1057\pm0.0009$), (iii) is the only one where the outer eccentricity is found to be smaller than the inner one ($e_\mathrm{out}=0.068\pm0.002$), and (iv) has the second highest outer mass ratio ($q_\mathrm{out}=0.880\pm0.002$). And, therefore, naturally this triple exhibits the most rapid dynamical AM with $P_\mathrm{apse}^\mathrm{obs}\approx1.8$\,yr\footnote{This value, which can be seen directly in the corresponding top right panel of Fig.~\ref{fig:etvs}, differs substantially from the theoretical AM period given in Table~\ref{tab:syntheticfit_TIC283846096337993842}, where the tabulated value is $P_\mathrm{apse}=4.228\pm0.005$\,yr. The reason for this discrepancy is due to the fact that the tabulated theoretical values of $P_\mathrm{apse}$ are calculated from the lowest quadrupole-order perturbation theory, which clearly fails for such a very tight triple system, as was shown in \citet{borkovitsmitnyan23}. Corrected formulae for the theoretical AM periods, which take into account higher-order terms (in the perturbation function) and, moreover, non-linear approximations will be presented soon in a separate paper (Deme et al., in prep.).}, which means that more than two complete rotations of the orbital ellipse have been completed since the first TESS observations in the summer of 2019. Moreover, from an astrophysical point of view, a further superlative is that this UCHT has both the lowest total mass out of the current ten systems and, also, it contains the smallest mass star in our sample.

The individual masses of the current triple are: $M_\mathrm{Aa}=0.59\pm0.03\,\mathrm{M}_\sun$, $M_\mathrm{Ab}=0.38\pm0.02\,\mathrm{M}_\sun$ and $M_\mathrm{B}=0.86\pm0.05\,\mathrm{M}_\sun$, that is, all three stars are low-mass cool red M and K dwarfs.  Here we stress again what we have discussed about the much higher accuracy of the relative, or dimensionless, quantities (as opposed to the absolute, or physical, quantities) such as the mass ratios, which are constrained mainly through the dynamics, i.e.,  the perturbations.  In this triple the mass ratios are orders of magnitude more accurate than the masses themselves, e.g., $q_\mathrm{in}=0.636\pm0.004$.  Turning to the other set of the fundamental parameters and their relative dimensionless counterparts, i.e., the physical and the fractional radii, we find that the relative accuracy difference is less significant, as $R_\mathrm{Aa}=0.59\pm0.03\,\mathrm{R}_\sun$, $R_\mathrm{Ab}=0.37\pm0.03\,\mathrm{R}_\sun$ and, $R_\mathrm{B}=0.83\pm0.03\,\mathrm{R}_\sun$, while the fractional radii are: $r_\mathrm{Aa}=0.044\pm0.001$, $r_\mathrm{Ab}=0.028\pm0.001$ and $r_\mathrm{B}=0.0113\pm0.0003$.  Note, in contrast to the two previously discussed triples, here even the uncertainties of the more accurate dimensionless quantities are also a bit higher, especially in the case of the two inner binary members. In our view, these are mainly due to the very shallow regular eclipses, which were insufficient to better constrain the fractional radii. Of course, this shallowness can be well explained by the large outer mass ratio. In this case, the more distant tertiary component is much more massive than the EB members, and therefore it emits more than 86\% of the total flux of the triple.

Finally we note another `superlative' of sorts due to the fact that this is the least observed system in our sample---TESS observed it only during four sectors. Despite this, the outer period ($P_\mathrm{out}=55\fd954\pm0\fd003$) is quite short, which led to third-body eclipses in all four sectors, and the rapid dynamical AM as well as the large-amplitude $P_\mathrm{out}$-period ETV-wobbles strongly constrain much of the dynamical parameters.  We were thereby able to find a robust and satisfying photo-dynamical solution purely from such a small set of observations.

\subsubsection{TIC 337993842}
\label{Sect:discussion_TIC337993842}

This is the longest outer period triple in our sample with $P_\mathrm{out}=101\fd4$ and, therefore, strictly speaking, this already exceeds the formal definition we set for UCHTs with a limit of $P_\mathrm{out}=100$\,days, but only by $\sim1.4$\%. Only two third-body events were detected with TESS (though the target was observed in six sectors) and, both of them belong to the superior conjunction of the third star. Despite this, similar to the previous target, due to the well-covered, characteristic ETV pattern, we were able to find a robust photodynamical solution simply from these six sectors of TESS observations.

We found that the distant third component of this triple (with $M_\mathrm{B}=2.3\pm0.1\,\mathrm{M}_\sun$) is the most massive object amongst all the thirty stars investigated in the current ten UCHTs. The other two stars of the inner EB are similar to each other, and are also more massive than our Sun ($M_\mathrm{Aa}=1.36\pm0.04\,\mathrm{M}_\sun$ and, $M_\mathrm{Ab}=1.34\pm0.04\,\mathrm{M}_\sun$). The fractionally more accurate mass ratios are $q_\mathrm{in}=0.98\pm0.01$ and $q_\mathrm{out}=0.85\pm0.03$. The physical dimensions as well as the temperatures of the three stars are also larger than that of our Sun, being $R_\mathrm{Aa}=1.36\pm0.04\,\mathrm{R}_\sun$, $R_\mathrm{Ab}=1.32\pm0.04\,\mathrm{R}_\sun$, $R_\mathrm{B}=2.8\pm0.2\,\mathrm{R}_\sun$; and $T_\mathrm{Aa}=6650\pm100\,\mathrm{K}$, $T_\mathrm{Ab}=6570\pm100\,\mathrm{K}$ and $T_\mathrm{B}=8800\pm400\,\mathrm{K}$.

The ETV curve shows evidence of AM and, therefore, some eccentricity of the inner orbit ($e_\mathrm{in}=0.0040\pm0.0002$).  In the absence of any observed third-body events at the inferior conjunction, the amplitude, shape and phase(s) of the $P_\mathrm{out}$-period ETV wobbles give the chief constraints on the outer eccentricity, which was found to be $e_\mathrm{out}=0.214\pm0.007$.

Finally, we note that a slightly problematic issue with our solution for this system is that the photodynamically obtained distance ($d_\mathrm{phot}=1770\pm70$\,pc) differs quite significantly from the distance of \citet{bailer-jonesetal21} that was obtained from the normally accurate Gaia DR3 parallaxes ($d_\mathrm{EDR3}=2186\pm64$\,pc). We will discuss the question of the sometimes discrepant parallactic and photometric distances in Sect.\,\ref{sec:discuss}.

\subsubsection{TIC 351404069}
\label{Sect:discussion_TIC351404069}

This is one of the tightest triples in our sample, with $P_\mathrm{out}/P_\mathrm{in}=10.79$. Moreover, it contains the second most eccentric EB, with $e_\mathrm{in}=0.0389\pm0.0002$. The rapid, dynamically forced AM is also readily visible. The numerical integrations give its period as $P_\mathrm{aps}^\mathrm{obs}=7.4$\,yr indicating, again, the insufficiency of the lowest-order, quadrupole-level approximation, which yields a theoretical value of $P_\mathrm{apse}=10.92\pm0.03$\,yrs.

According to our results, the system contains three quite similar stars. The primary of the EB is a slightly evolved F-star, while the secondary EB star, and also the more distant tertiary are two G-type stars.  Their masses are radii are $M_\mathrm{Aa}=1.16\pm0.06\,\mathrm{M}_\sun$, $M_\mathrm{Ab}=0.95\pm0.04\,\mathrm{M}_\sun$, and $M_\mathrm{B}=0.97\pm0.05\,\mathrm{M}_\sun$; and $R_\mathrm{Aa}=1.99\pm0.05\,\mathrm{R}_\sun$, $R_\mathrm{Ab}=0.98\pm0.05\,\mathrm{R}_\sun$, and $R_\mathrm{B}=1.01\pm0.05\,\mathrm{R}_\sun$.  The more accurate relative quantities are: $q_\mathrm{in}=0.826\pm0.005$, and $q_\mathrm{out}=0.460\pm0.002$; and $r_\mathrm{Aa}=0.086\pm0.001$, $r_\mathrm{Ab}=0.042\pm0.002$, and $r_\mathrm{B}=0.0079\pm0.0003$, respectively.  Interestingly, despite the fact that the most massive component has a higher mass by $\sim15-17$\% than the other two stars, due to its slightly evolved nature, the absolute temperatures of all three stars are similar (within their $1\sigma$ uncertainties), being $T_\mathrm{Aa}=5900\pm100\,\mathrm{K}$, $T_\mathrm{Ab}=5865\pm85\,\mathrm{K}$ and $T_\mathrm{B}=5930\pm80\,\mathrm{K}$, respectively.

Finally, we note that this triple was found to be the second most inclined in our sample with $i_\mathrm{mut}=3\fdg2\pm0\fdg2$.

\subsubsection{TIC 378270875}
\label{Sect:discussion_TIC378270875}

This faint triple system consists of three similarly cool and less massive K dwarfs ($M_\mathrm{Aa}=0.83\pm0.06\,\mathrm{M}_\sun$, $M_\mathrm{Ab}=0.77\pm0.06\,\mathrm{M}_\sun$, $M_\mathrm{B}=0.86\pm0.06\,\mathrm{M}_\sun$, $R_\mathrm{Aa}=0.83\pm0.02\,\mathrm{R}_\sun$, $R_\mathrm{Ab}=0.76\pm0.03\,\mathrm{R}_\sun$, $R_\mathrm{B}=0.88\pm0.03\,\mathrm{R}_\sun$ and $T_\mathrm{Aa}=5490\pm80\,\mathrm{K}$, $T_\mathrm{Ab}=5210\pm80\,\mathrm{K}$ and $T_\mathrm{B}=5620\pm60\,\mathrm{K}$). Note, however, that the fractional accuracy of the dimensionless quantities are, again, much better determined as: $q_\mathrm{in}=0.936\pm0.005$, $q_\mathrm{out}=0.533\pm0.009$ and $r_\mathrm{Aa}=0.090\pm0.001$, $r_\mathrm{Ab}=0.082\pm0.001$ and $r_\mathrm{B}=0.0104\pm0.0003$, respectively.

As one can see in the top row of Fig.~\ref{fig:lcs2}, the TESS-observed third-body eclipses are very shallow and, their depths are usually smaller than the out-of-eclipse light curve variations, the latter of which, irrespective of their origin, we removed with the local fitting of medium-order smoothing polynomials. (The original, that is, unsmoothed, but 1800-sec cadence light curve, is marked with grey in the corresponding panels of the top row of Fig.~\ref{fig:lcs2}.)

We note also that this triple was found to be the oldest in our sample with $\tau=9.4^{+2.0}_{-4.1}$ Gyr, though the uncertainty in its age was found to be quite large and asymmetric.

\subsubsection{TIC 403792414}
\label{Sect:discussion_TIC403792414}

This is another system in our sample where the ETV reveals clear, rapid, dynamically forced AM. However, when we carried out a short timescale numerical integration of the motion, we see that in this dynamically forced AM, substantial higher-order effects are also present.  This can be nicely seen on the numerically generated ETV plot, calculated for the current century and shown in Fig.~\ref{fig:403792414ETV-100yr}. In Fig.~\ref{fig:403792414numint100yr} we illustrate that these interesting extra variations might arise in the AM due to the following reason.  The argument of pericentre of the inner orbit ($\omega_\mathrm{in}$) librates around the argument of pericentre of the outer orbit ($\omega_\mathrm{out}$), the latter of which revolves in the same direction as that of the orbital motion with a period of $P_\mathrm{apse,out}^\mathrm{obs}=52.4$\,yr. (Note, this latter value is very close to the theoretically calculated AM period of the outer orbit, which is $P_\mathrm{apse,out}=50.1\pm0.2$\,yr.) This libration of the inner major axis forces a cyclic variation in the inner eccentricity ($e_\mathrm{in}$) with the same period (that is, $P_\mathrm{e-in}^\mathrm{obs}=P_\mathrm{lib}^\mathrm{obs}=8.6$\,yr -- see in the right panel of Fig.~\ref{fig:403792414numint100yr}) and, finally, this latter variation results directly in the extra periodicity in the ETV curve (Fig.~\ref{fig:403792414ETV-100yr}). Here we note also, that a very similar behaviour was reported and discussed in the case of another tight, compact triple, KIC~9714358, in \citet{borkovitsmitnyan23}.

Regarding the deduced stellar parameters of this triple, this is also formed by three low-mass, cool dwarf stars with masses: $M_\mathrm{Aa}=0.84\pm0.06\,\mathrm{M}_\sun$, $M_\mathrm{Ab}=0.58\pm0.04\,\mathrm{M}_\sun$, and $M_\mathrm{B}=0.80\pm0.06\,\mathrm{M}_\sun$; radii: $R_\mathrm{Aa}=0.83\pm0.06\,\mathrm{R}_\sun$, $R_\mathrm{Ab}=0.58\pm0.04\,\mathrm{R}_\sun$, and $R_\mathrm{B}=0.76\pm0.05\,\mathrm{R}_\sun$; and $T_{\rm eff}$: $T_\mathrm{Aa}=5430\pm180\,\mathrm{K}$, $T_\mathrm{Ab}=3945\pm90\,\mathrm{K}$, and $T_\mathrm{B}=5180\pm140\,\mathrm{K}$, respectively.  The more accurate dimensionless quantities are $q_\mathrm{in}=0.69\pm0.02$ and $q_\mathrm{out}=0.56\pm0.01$. This is also a flat ($i_\mathrm{mut}=0\fdg7\pm0\fdg4$) and old ($\tau=6.3^{+3.3}_{-1.5}$\,Gyr) triple system.

\subsubsection{TIC 403916758}
\label{Sect:discussion_TIC403916758}

This triple star is very close to the class of objects which are known as `double twins', as both the inner and outer mass ratios are close to unity.  In this system, $q_\mathrm{in}=0.98\pm0.03$ and $q_\mathrm{out}=0.98\pm0.02$. Though the enrichment of `double twins' amongst the population of the triple stars is predicted by some of the multiple star evolution theories \citep[see][for further references]{offneretal23}, from an observational point of view, it is hard to detect them, at least photometrically. This is so because of the fact that the tertiary's mass is about twice that of the individual masses of the inner components. Therefore, the emitted fluxes of the inner stars and, in the case of an eclipsing configuration, their mutual eclipses may easily disappear in the glare of the much brighter third star.

Direct photometric discovery of such a double twin becomes easier when it is a relatively flat CHT seen nearly edge-on, and the more massive third component leaves the MS and becomes a red giant (RG).  The probability of the system then producing extra (third body) eclipses increases as the radius of the tertiary grows. This is exactly what happens in this triple, which is the only one in the present sample which contains a RG tertiary.

This triple system contains the shortest period inner EB ($P_\mathrm{in}=1\fd134$). Therefore, despite the fact that $P_\mathrm{out}=71\fd060\pm0\fd002<100$\,d, which classifies this triple as a UCHT, it cannot be considered to be tight, as $P_\mathrm{out}/P_\mathrm{in}=62.66>50$, which is the largest in our sample. Therefore, not surprisingly, the ETV is dominated by the small amplitude LTTE and hence, it is completely unusable (see the third row, left panel of Fig.\,\ref{fig:etvs}).

In this regard it is important to note that the TESS light curve (third row of Fig.\,\ref{fig:lcs2}) is very reminiscent of that of HD~181068 \citep{derekasetal11,borkovitsetal13}, one of the first triply eclipsing triples (found with \textit{Kepler}).  In both systems the outer eclipses reveal directly that the third most distant component must be an RG star, which strongly dominates the total light of the triple. For such kinds of UCHTs, one may expect (nearly) coplanar and circular inner and outer orbits (due to substantial tidal damping), where the dynamical ETV contribution is nearly zero \citep[see][]{borkovitsetal03}. Therefore, the ETV reflects only the low amplitude LTTE signal whose amplitude is much lower than the scatter of the individual ETV points.

In this context, however, it is perhaps somewhat surprising, that this is the least flat triple in our sample with $i_\mathrm{mut}=4\fdg1\pm0\fdg8$. In this regard, however, one should note that the inner eclipses are not deeper than $\sim$0.6\,\%.  Therefore, the variations in the eclipse depths, and the corresponding inferred non-alignment of the orbital planes might be strongly affected by other incidental out-of-eclipse light variations.  These could include chromospheric activity, different `extra fluxes' and/or stray light in the different sectors.

As was already mentioned above, this is the only system in the current sample where the most massive star (component B, with $M_\mathrm{B}=1.74\pm0.06\,\mathrm{M}_\sun$) is clearly evolved from the MS, and is now an RG star with $R_\mathrm{B}=6.9\pm0.3\,\mathrm{R}_\sun$. The other two K dwarf-type inner stars, however, are still on the MS ($M_\mathrm{Aa}=0.90\pm0.03\,\mathrm{M}_\sun$, $M_\mathrm{Ab}=0.87\pm0.03\,\mathrm{M}_\sun$ and $R_\mathrm{Aa}=0.81\pm0.03\,\mathrm{R}_\sun$, $R_\mathrm{Ab}=0.79\pm0.03\,\mathrm{R}_\sun$, respectively):

\subsubsection{TIC 405789362}
\label{Sect:discussion_TIC405789362}

This is the most compact triple system in the current sample (that is, it has the shortest outer period with $P_\mathrm{out}=46\fd810\pm0\fd003$), and also one of the tightest ($P_\mathrm{out}/P_\mathrm{in}\approx10.25$). Due to the small mass and, hence, the low temperature and surface brightness of the less massive tertiary star ($M_\mathrm{Aa}=1.54\pm0.08\,\mathrm{M}_\sun$, $M_\mathrm{Ab}=1.29\pm0.07\,\mathrm{M}_\sun$ and, $M_\mathrm{B}=0.82\pm0.05\,\mathrm{M}_\sun$ or, regarding the mass ratios: $q_\mathrm{in}=0.837\pm0.006$ and $q_\mathrm{out}=0.293\pm0.007$) the TESS light curve displays deep regular eclipses, while the third-body events are very shallow and, some of them are almost hidden by the other kinds of light curve variations and distortions.

The inner orbit has a very low, but definitely non-zero, eccentricity $e_\mathrm{in}=0.0072\pm0.0003$. In this regard we note that, in the case of such a tight triple system, an exactly circular inner orbit is impossible, due to the perturbing force of the tertiary, of which the strength, as well as the orientation varies moment by moment. As a consequence of this small, but non-zero, inner eccentricity, a slight and varying displacement occurs continuously at the times of the secondary eclipses. According to the ETV plot (third row, middle panel of Fig.\,\ref{fig:etvs}), during the first observing sectors in Year 2019, they were advanced by $\sim$15\,min relative to the primary eclipses (or, more strictly speaking, they occurred about a quarter of an hour before the mid time between two primary eclipses), while for the Year 2024 observations, the primary and secondary ETV curves overlap each other.  Specifically, the secondary eclipses occurred practically at the mid-times between consecutive primary eclipses.

Considering other dynamical properties, the outer orbit is found to be moderately eccentric with $e_\mathrm{out}=0.163\pm0.003$, and the entire triple is flat to within $2\degr$, being $i_\mathrm{mut}=1\fdg4\pm0\fdg4$.

\subsubsection{TIC 461500036}
\label{Sect:discussion_TIC461500036}

The features of the light curve from this last system in our sample strongly resemble the previous one. That is, the regular inner primary and secondary eclipses are quite deep (with depths greater than 40\%) and look rather similar.  By contrast, the third-body eclipses are shallow, however they are well observable due to the precise TESS photometry. The similarity to TIC~405789362, naturally arises from the very similar surface brightness ratios for the two systems.  Regarding the fact that both the current and the previous system are formed by three MS-stars, the similarities of the surface brightness ratios imply that both the inner and outer mass ratios of the current system are close to the corresponding quantities of the previous triple. In the current system this is $q_\mathrm{in}=0.962\pm0.008$ and $q_\mathrm{out}=0.309\pm0.006$. Interestingly, not only the mass ratios, but the individual stellar masses are also quite similar to the previous triple as $M_\mathrm{Aa}=1.29\pm0.08\,\mathrm{M}_\sun$, $M_\mathrm{Ab}=1.25\pm0.08\,\mathrm{M}_\sun$ and, $M_\mathrm{B}=0.79\pm0.05\,\mathrm{M}_\sun$.

The similarities, however, end there when one considers the dynamical properties. The current triple has a shorter inner and a longer outer period as compared to the previous system and, hence, is much less tight ($P_\mathrm{out}/P_\mathrm{in}=22.08$). Due to the nearly two-times lesser tightness, the third-body perturbation forces are weaker, and the inner orbit is allowed to be much closer to circular. However, as was discussed above, the osculating eccentricity cannot reach exactly zero, hence $e_\mathrm{in}=0.0013\pm0.0001$. More interesting, however, is that the outer orbit also has a low eccentricity with $e_\mathrm{out}=0.032\pm0.001$. We will return to the question of the relatively small outer eccentricities in Sect.\,\ref{sec:discuss}.

Finally, note another similarity to the previous system, as we obtained a very similar mutual inclination between the inner and outer orbital planes, as $i_\mathrm{mut}=1\fdg4\pm0\fdg2$.

\section{Summary, discussion, and conclusions}
\label{sec:discuss}

This work presents 10 new, ultracompact triply eclipsing triples with a full set of stellar and orbital solutions.  These systems were discovered as UCHTs by searching through several million TESS photometric light curves for third-body eclipses in what are otherwise seemingly normal binary systems (see e.g., \citealt{kristiansen22,rappaportetal22}). 

The analyses utilized the photometric light curves from TESS, the ETV curves derived from the TESS light curves, SED data found in the archives and, in a few cases, ground-based follow up eclipse photometry. In contrast to our previous works, however, we did not use photometric data from the archives of ASAS-SN and ATLAS, since the frequent TESS observations by themselves were quite satisfactory for the determination of the orbital solutions. Moreover, due to the faintness of the currently investigated triples, no RV observations were available for our targets. All the above mentioned data were analysed jointly with a complex photodynamical code wherein we solve for all the stellar and orbital system parameters, as well as the distance to the source.  Typical uncertainties on the masses and radii are of order a couple per cent to about $\sim$5 per cent.  Uncertainties on the angles associated with the orbital planes (e.g., $i_{\rm out}$ and $i_{\rm mut}$) range from a fraction of a degree to about a degree.  See Tables   \ref{tab:syntheticfit_TIC198581208265274458} through  \ref{tab:syntheticfit_TIC405789362461500036}.

\begin{figure*}
\begin{center}
\includegraphics[width=0.305 \textwidth]{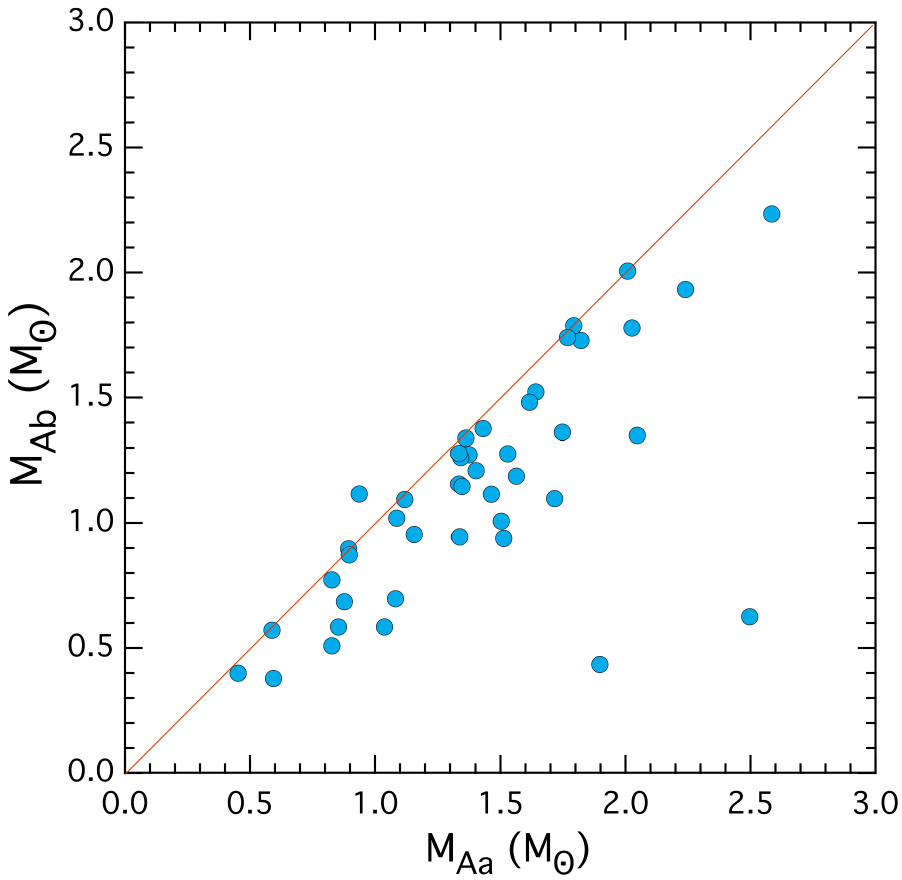}  \hglue0.1cm                        
\includegraphics[width=0.305 \textwidth]{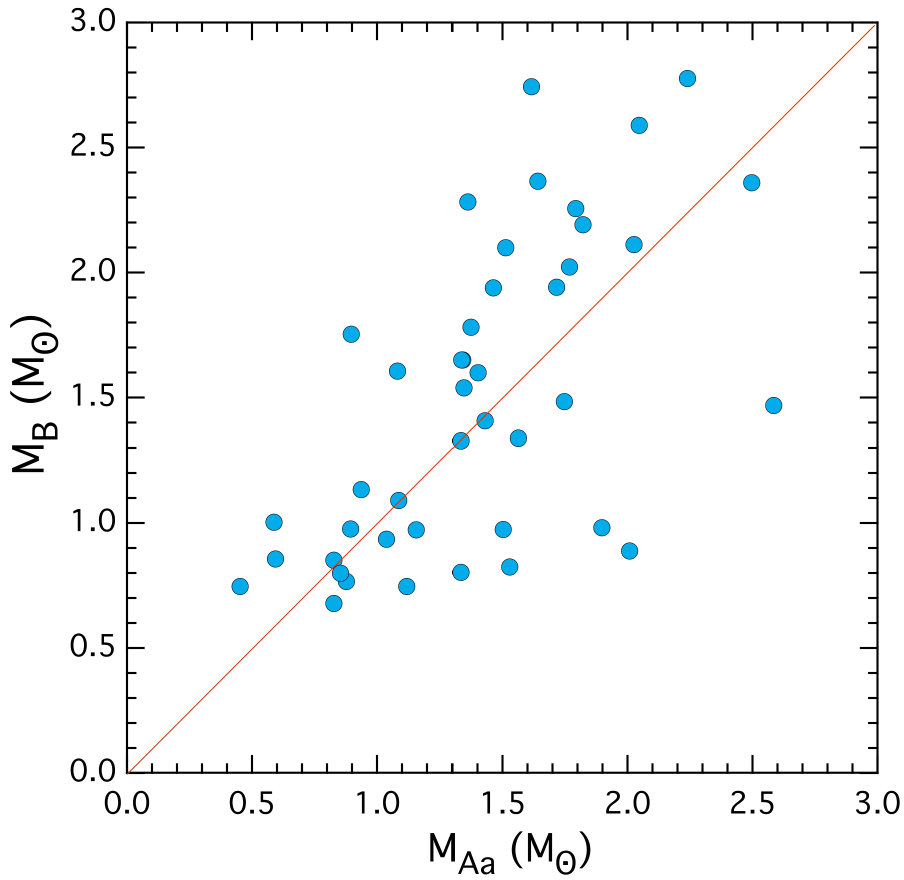}                                        
\includegraphics[width=0.311 \textwidth]{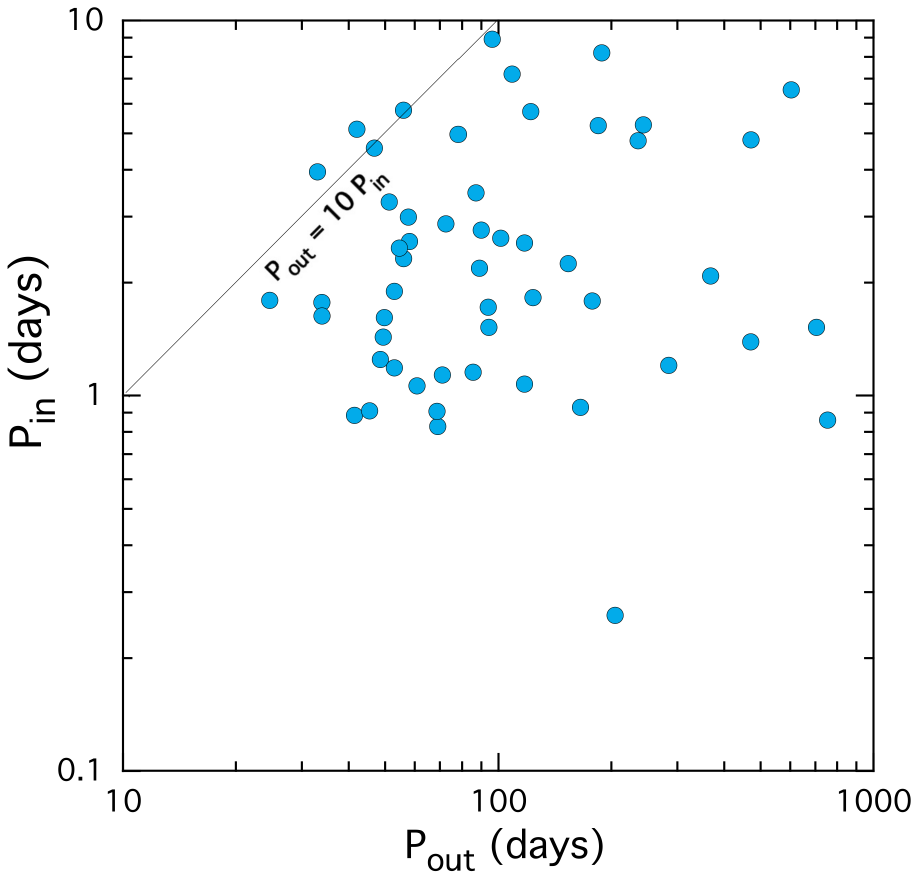}  \hglue0.00cm                  
\includegraphics[width=0.316 \textwidth]{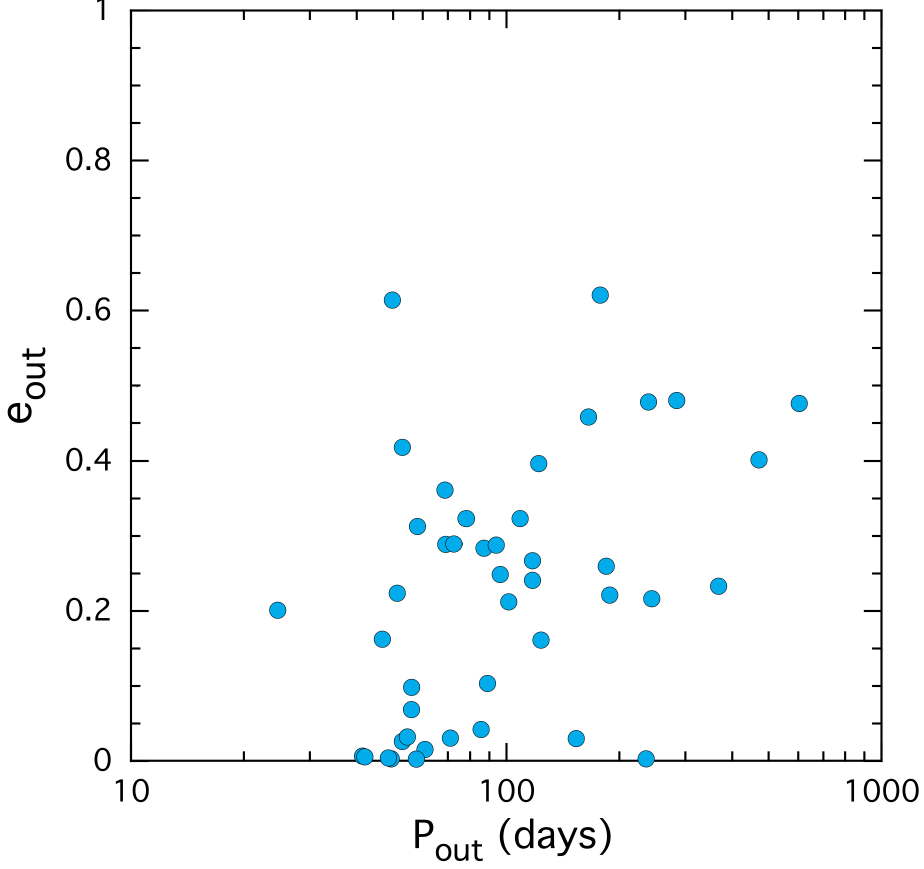} \hglue-0.20cm
\includegraphics[width=0.309 \textwidth]{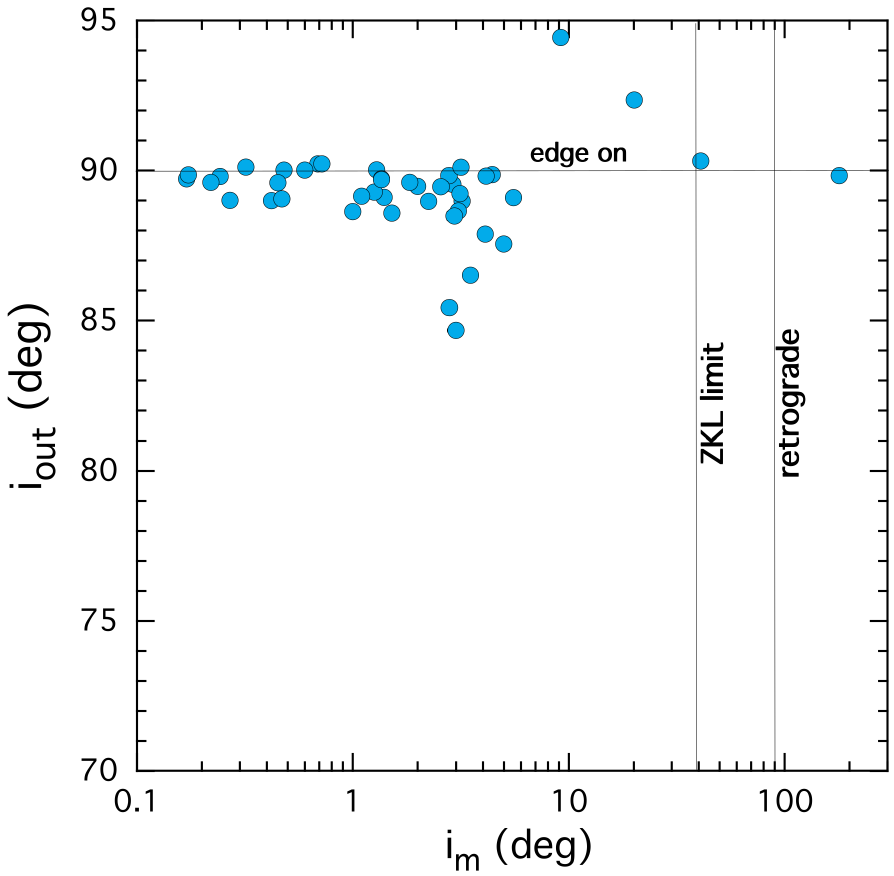} \hglue0.0cm   
\includegraphics[width=0.311 \textwidth]{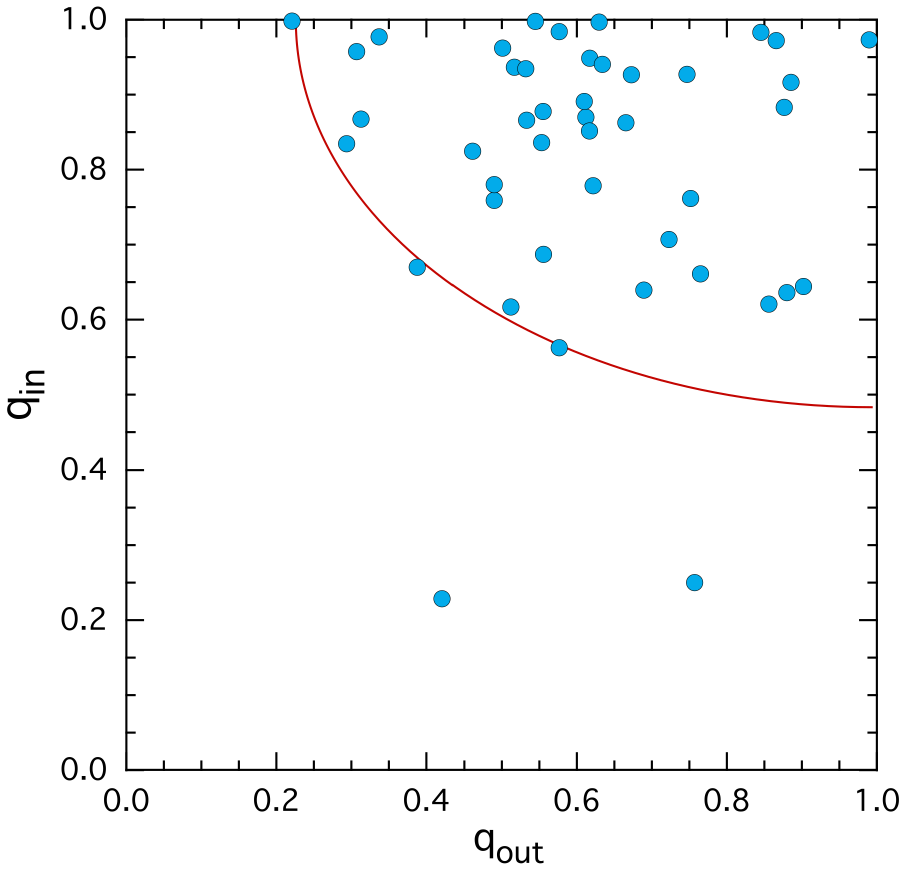} \hglue-0.05cm
\includegraphics[width=0.308 \textwidth]{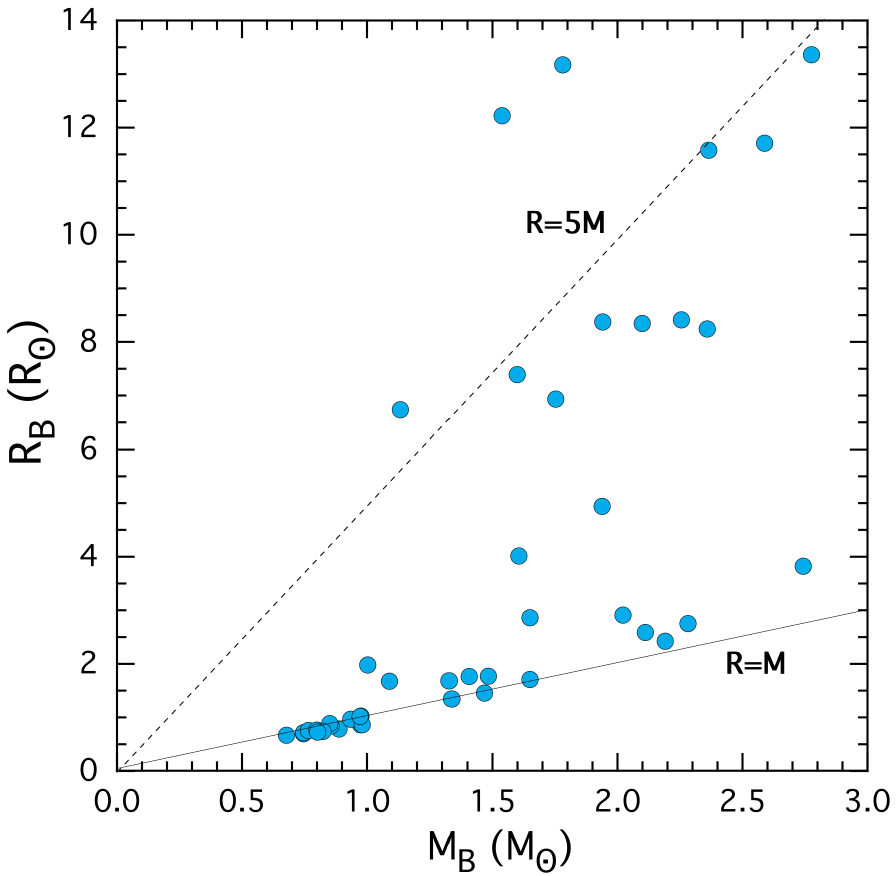} \hglue-0.10cm
\includegraphics[width=0.313 \textwidth]{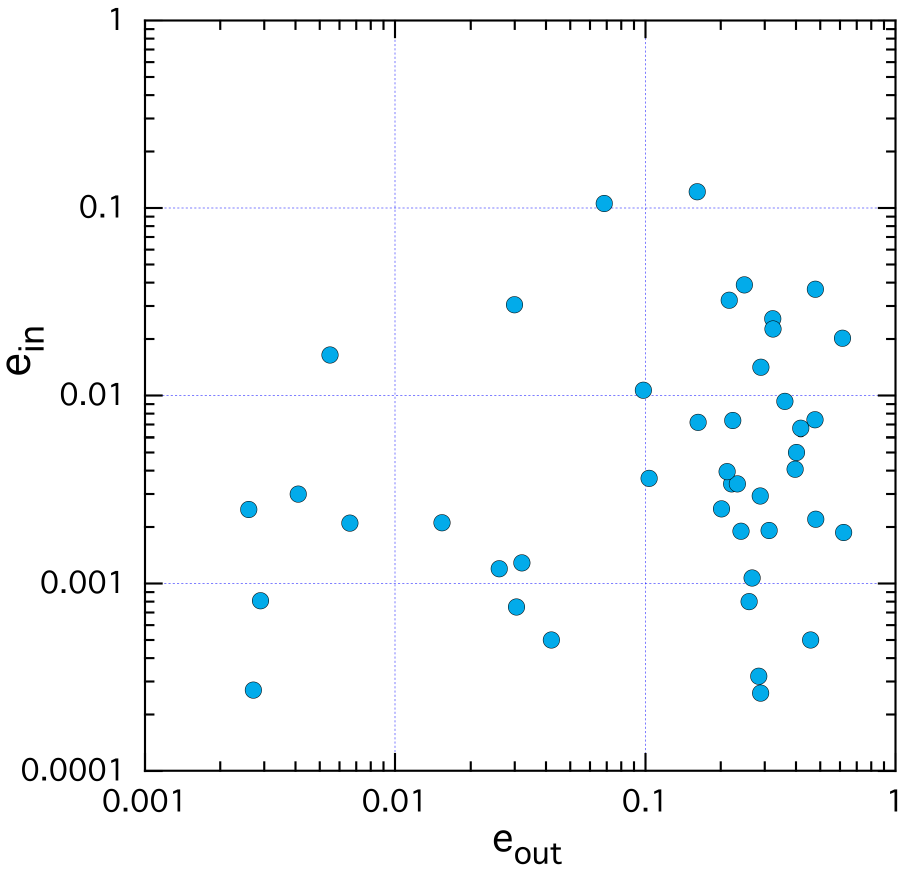} 
\includegraphics[width=0.307 \textwidth]{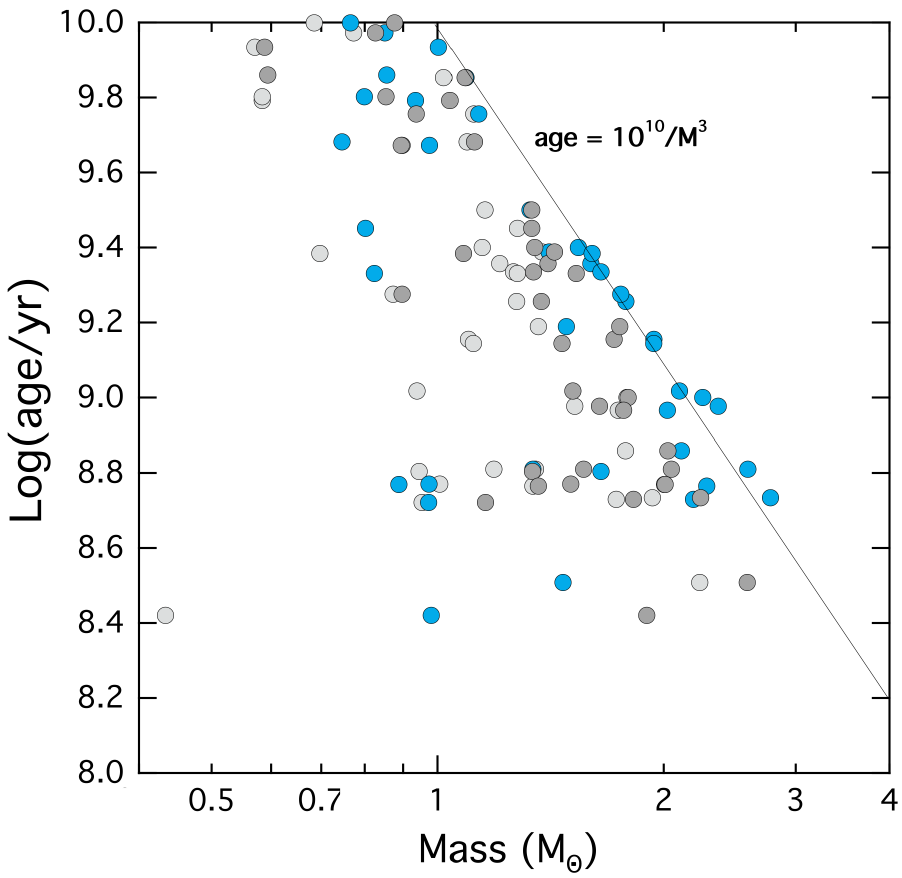}  
\caption{Statistical plots for properties of 44 triply eclipsing triples uniformly analysed (see text for references).  {\it Top-row panels:} $M_{\rm Ab}$ vs.~$M_{\rm Aa}$, $M_{\rm B}$ vs.~$M_{\rm Aa}$, and $P_{\rm in}$ vs.~$P_{\rm out}$.  {\it Middle-row panels:} $e_{\rm out}$ vs.~$P_{\rm out}$,  $i_{\rm out}$ vs.~ $i_{\rm mut}$, and $q_{\rm in}$ vs.~$q_{\rm out}$. {\it Bottom-row panels:} $R_{\rm B}$ vs.~$M_{\rm B}$, $e_{\rm in}$ vs.~$e_{\rm out}$, and the age of the systems vs.~the masses of the tertiary (blue), primary (dark grey) and secondary (light grey) EB stars. In this latter panel as well as the first two panels, the masses of TIC 290061484 are $\sim$7\,M$_\odot$ and are off the plots. The red curve in the middle right panel shows how nearly all the systems are confined to $0.2 <  q_{\rm out} < 1.0$ and $q_{\rm in} > 0.2$. In the central  panel the vertical lines denote the transition to the Von Zeipel-Lidov-Kozai (ZLK) cycles \citep[see][for a review]{naoz16}, and to retrograde orbits, respectively. The sloped dashed lines in the lower left panel are for $R_{\rm B}$ =1\,M$_{\rm B}$ and = 5\,M$_{\rm B}$ (both expressed in solar units), as rough guides of unevolved and quite evolved stars, respectively.}
\label{fig:statistics}
\end{center}
\end{figure*} 

In Fig.~\ref{fig:statistics} we present a set of nine correlation plots among some of the more physically interesting parameters associated with our collection of 44 triply eclipsing triples that we have analysed in a uniform way. To the ten sources studied in this paper, we have added the 22 triples from our previous closely related papers \citet{rappaportetal22,rappaportetal23,rappaportetal24}, as well as 12 triples studied in \citet{borkovitsetal19a,borkovitsetal20b,borkovitsetal22a,borkovitsetal22b,czavalingaetal23,kostovetal24}, and \citet{mitnyanetal20} using essentially the same selection criteria as well as methods of analysis.

In the top row of Fig.~\ref{fig:statistics}, the left two panels show correlations between the two masses of the inner binary and between the tertiary star and the primary of the inner binary. The former pair have a correlation coefficient of 0.71 with all the stars and 0.91 if we eliminate the two secondaries that are farthest from the red (equal mass) line\footnote{In both cases the correlation coefficient was calculated without including the massive ($\sim$7\,M$_\odot$) stars in TIC 290061485 which are located off the plot.}.  The correlation coefficient for the tertiary with primary of the EB is only 0.64. These also serve to show that most of the stars in our sample have masses between 0.5 and 2.5 M$_\odot$.

The top right and middle left panels plot the EB period and outer eccentricity vs.~ the outer orbital period, respectively.  Neither set of orbital parameters is particularly correlated.  The $P_{\rm in}$ vs.~$P_{\rm out}$ diagram does nicely show the rough empirical upper stability limit at $P_{\rm out} \gtrsim 7 P_{\rm in}$.

The middle panel shows the relation between the outer orbital inclination angle and the mutual inclination (i.e., the angle between the inner and outer orbital planes).  That most of the values of $i_{\rm out}$ are near $90^\circ$ is a selection effect since these triples were actually discovered from their third-body eclipses.  The same selection also holds, to some extent, for the low values of $i_{\rm mut}$, otherwise third body eclipses would be more difficult to detect.  Two of the systems have large enough $i_{\rm mut}$ (20$^\circ$ and 40$^\circ$) to undergo substantial precession of their orbital planes. Finally, there is one triple (TIC 276162169 = V493~Cygni) in a nearly flat system, but where the outer orbit is likely retrograde with respect to the inner EB; however, as was stated in \citet{rappaportetal23}, a verification of that finding will require further observations.  These are rare systems. 

The middle right panel shows the correlation between the inner mass ratio, $q_{\rm in} = M_{\rm Ab}/M_{\rm Aa} \equiv < 1.0$, and the outer mass ratio $q_{\rm out} \equiv M_{\rm B}/(M_{\rm Aa}+M_{\rm Ab}$).  With only two exceptions, $1.0 > q_{\rm in} \gtrsim 0.55$ while $q_{\rm out}$ is typically between 0.3 and 0.9, with no restrictions on it being larger than unity.

The lower left panel gives the relation between the radius of the tertiary star and its mass.  Stars lying approximately along the $R=M$ line (both in solar units), indicate largely unevolved stars.  Stars closer to the $R=5M$ line and above it are progressively more evolved.  The tertiary stars in UCHTs generally still have more room to evolve before filling their Roche lobes than do the primary stars of the inner binaries.  More evolved, and therefore more luminous, tertiaries make for easier detection; but, on the other hand, if they are too bright they can obscure the third-body eclipses by which these systems are found - at least in this work.

The bottom middle panel shows how the eccentricities of the inner and outer orbits are possibly related.  While they both range over about three orders of magnitude, the outer eccentricities span 0.001 to almost 1 while the inner eccentricities span a range that is shifted an order of magnitude lower: 0.0001 to 0.1.  They are otherwise uncorrelated.

Finally, in the bottom right corner we show how the ages of the systems that we infer for these compact triples are related to the masses of the tertiaries (blue points) and of the primary (dark grey) and secondary (light grey) stars in the EBs.  We omitted one very young (pre-MS) system, as well as TIC 290061484 with an age of 13 Myr and containing $\sim$7\,M$_\odot$ stars, and find that the remainder of the systems are modest-to-very old, spanning the range from 300 -1000 Myr.  The sloped line is a rough guide to the MS lifetime of stars, going as $10^{10}/M^3$ yr, where the mass is in solar masses. 

\begin{figure}[h]
\begin{center}
   \includegraphics[width=0.5\textwidth]{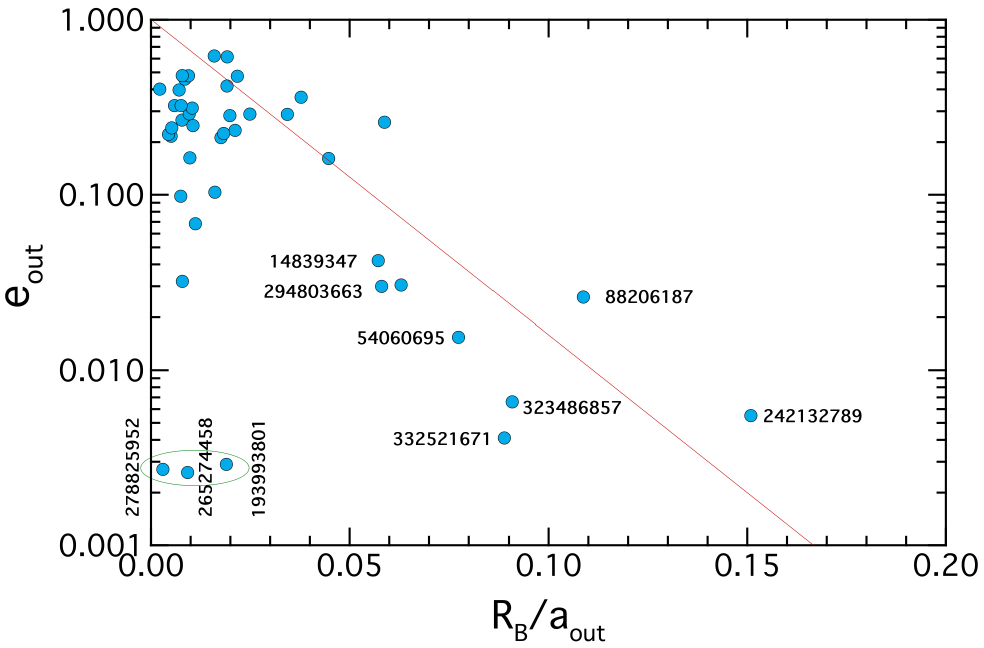}
   \caption{Eccentricity of the outer orbit of compact triply eclipsing triples vs.~the ratio of the tertiary's radius to the outer semimajor axis. With three exceptions (lower left corner), it seems reasonable to infer that tidal circularization of the outer orbit by evolved (i.e., large) tertiaries is responsible for the decaying outer eccentricity with increasing $R_{\rm B}/a_{\rm out}$. }
\label{fig:circularization}
\end{center}
\end{figure}

In Figure \ref{fig:circularization} we show how the outer eccentricities in the sample of compact triples we have studied vary with the ratio of the tertiary star's radius to the outer semimajor axis.  Most of the systems have $e_{\rm out}$ between 0.1 and 0.7.  These same systems also have $R_B/a_{\rm out} \lesssim 0.04$.  However, there are some 9 systems which have considerably smaller $e_{\rm out}$ and also substantially larger values of $R_B/a_{\rm out}$ ranging from $\sim$0.05-0.15, where we postulate that tidal interactions between the tertiary and the inner binary have tended to circularise the outer orbit.  There are three exceptions to this trend where the systems have very small $e_{\rm out}$ and also very low ratios of $R_B/a_{\rm out}$, where tidal circularization should not have played much of a role. In these cases, perhaps the systems were simply born with $e_{\rm out}$ as low as that measured at the current epoch ($\sim$0.003).

\begin{figure}[h]
\begin{center}
   \includegraphics[width=0.5\textwidth]{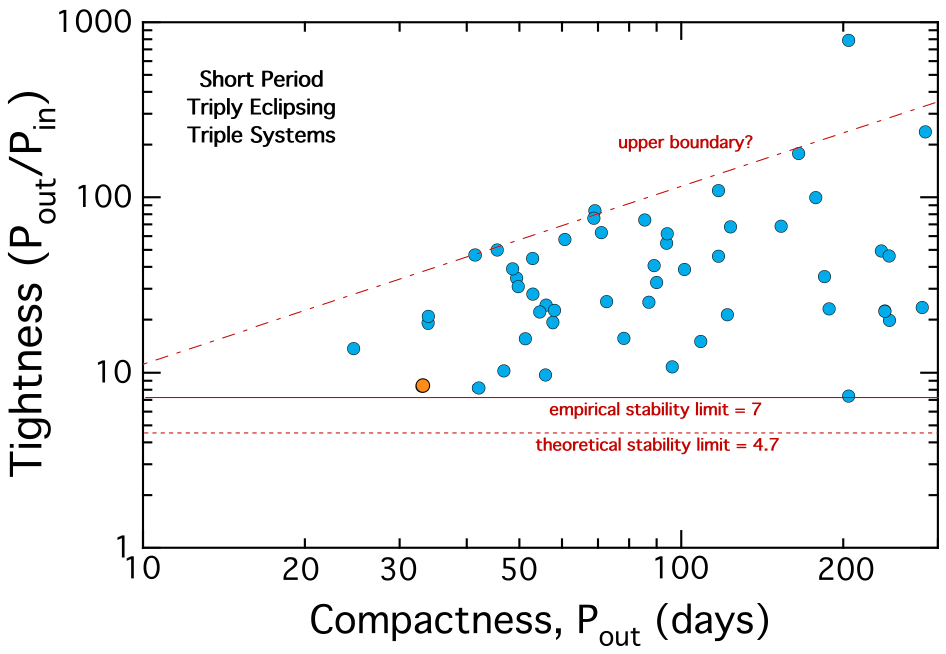}
   \caption{Tightness of triply eclipsing triple systems ($P_{\rm out}/P_{\rm in}$) as a function of the system compactness. $\lambda$ Tau \citep[orange circle,][]{ebbighausenstruve956} is not triply eclipsing, but it is otherwise such a noteworthy benchmark system that we include it for reference. The most compact triple, TIC 290061484, is the leftmost system in the plot. The empirical and theoretical lower limits for dynamical stability are marked with solid and dashed horizontal lines, respectively. A speculative upper boundary is marked with the sloped red dot-dashed line.}
\label{fig:compactness}
\end{center}
\end{figure}

As more and more CHTs become well characterised, a pattern is emerging as to how they populate a tightness-compactness diagram.  We show in Fig.~\ref{fig:compactness} the most compact 50 systems that we have collected (most of them from our analysed systems).  All the systems are triply eclipsing except for $\lambda$ Tau (marked in orange) which is sufficiently exceptional that we show it in spite of having no third-body eclipses. They all have $P_{\rm out}/P_{\rm min} > 7$ which seems to be an empirical limit for long-term dynamical stability. The tightest system is KIC 7668648 with $P_{\rm out} = 205$ d, and $P_{\rm out}/P_{\rm in} = 7.36$ \citep{orosz2023}. The Mardling \& Aarseth relations \citep{mardlingaarseth01} suggests a minimum ratio $P_{\rm out}/P_{\rm min} > 4.7$ for long-term dynamical stability even for coplanar and circular orbits. The most compact triple known to date is TIC 290061484 with an outer period of just 24.5 days \citep{kostovetal24}.  There is an intriguing empirical upper envelope shown by the sloped red line in Fig.~\ref{fig:compactness}.  This line is given mathematically by the condition that $P_{\rm in} \simeq 1$ day.  Such a period, of course, is typical of the shorter periods in our sample.  Interestingly, this line extrapolates to a minimum outer period for UCHTs of $P_{\rm out,min} \sim 7$ days.  Of course, if compact triples can form around contact type binaries, which so far have not been found, then the ultimate theoretical minimum period for UCHT might be as short as just a few days. 

\begin{figure}[h]
\begin{center}
   \includegraphics[width=0.5\textwidth]{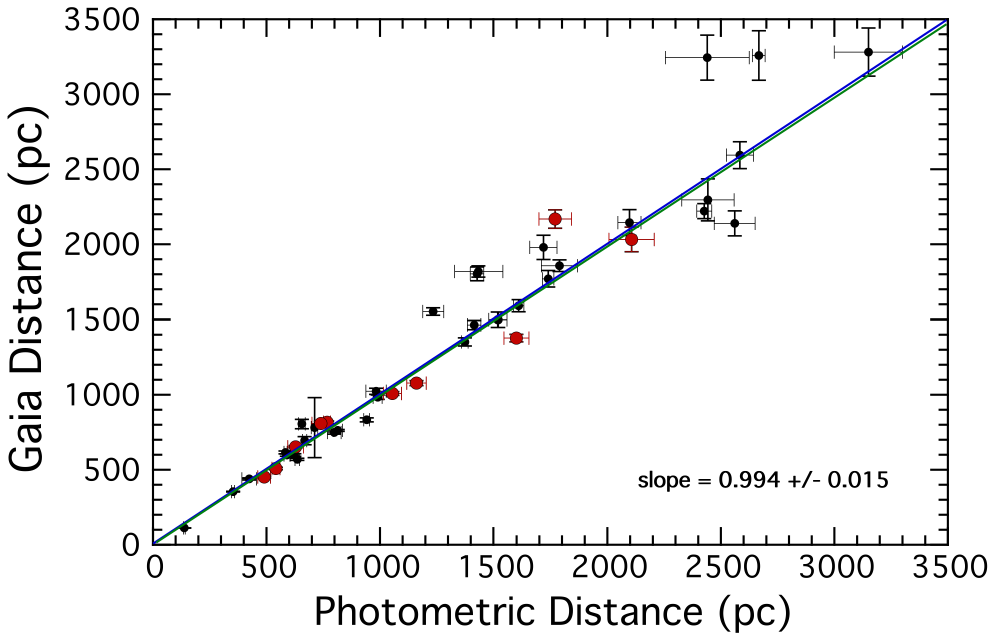}
   \caption{Comparison of Gaia distances \citep{bailer-jonesetal21} to 43 triple systems with distances found from our photodynamical fits to the system parameters. The systems marked in red are the 10 from the current work with fitted distances. The blue curve is the line where the Gaia and our photometric distances would match. The green curve results from a formal orthogonal distance regression with a fitted slope of $0.994 \pm 0.015$. There are three systems not shown.  For TIC 388459317 and TIC 52041148, the Gaia points are off the plot and have rather large uncertainties compared to the photometric ones.  TIC 280883908 has $d_{\rm Gaia} = 3072 \pm 1000$\,pc compared to $d_{\rm photo} = 1183 \pm 40$\,pc, where the Gaia point is quite obviously not very good, and its presence on the plot with such a large error bar would be visually distracting.}
\label{fig:distances}
\end{center}
\end{figure}   

Finally, in regard to our statistical-level results, we mention our photometric distance determinations in comparison with the parallactic distances found by Gaia.   Fig.~\ref{fig:distances} shows how our photometric distances stack up against Gaia's distances \citep{bailer-jonesetal21}.  These are the same triply eclipsing systems discussed above, and augmented by several other triple systems subjected to the same photodynamical analysis, but ones which are not triply eclipsing.  These latter sources are taken from \citet{borkovitsetal20a,borkovitsetal22b,gaulme22}, and \citet{borkovitsmitnyan23}. 

The overall agreement between the our photometric distances and the Gaia results is fairly impressive, and the two sets have broadly comparable error bars. The fitted slope relating them, found from an `orthogonal distance regression' (ODR) is $0.994 \pm 0.0165$.  The value of $\chi^2$ per degree of freedom reaches unity only after the uncertainties on both data sets have been increased by a factor of 2.8. Despite this general agreement, there are a number of points where the two distances differ by more than a few statistical error bars.  It is not obvious to us from an inspection of this plot which distance is manifestly more accurate. It is possible that one or both of the distance sets have underestimated their uncertainties.  We note that most of the outer orbits, range from about 1/6-1 year, and in some cases may be somewhat distorting Gaia's parallax measurement.  For our photometric distances, some of the SED data may have been taken during eclipses, the \texttt{PARSEC} isochrones are based on non-rotating stars, and there are likely to be uncertainties in the wavelength-dependent interstellar extinction that we rely on.

\section*{Data availability}

Tables F.1--F.10 are only available in electronic form at the CDS via anonymous ftp to cdsarc.u-strasbg.fr (130.79.128.5) or via \url{http://cdsweb.u-strasbg.fr/cgi-bin/qcat?J/A+A/}.

\begin{acknowledgements}
This project has received funding from the HUN-REN Hungarian Research Network.

T.\,B., T.\,M., I.\,B.\,B. and A.\,P. acknowledge the financial support of the Hungarian National Research, Development and Innovation Office -- NKFIH Grants K-147131 and K-138962.

V.\,B.\,K. is grateful for financial support from NSF grant AST-2206814.

Z.\,G., T.\,P., and Z.\,D. acknowledge support from the ESA PRODEX project `Observation of Exoplanets with the CHEOPS Space Observatory' (PEA 4000137122), and the ESA PRODEX project ?Hungarian Contribution to ESA's Ariel Space Telescope Mission: `II. High-Precision Photometry with Ariel' between ELTE E\"otv\"os Lor\'and University and the European Space Agency, as well as from the VEGA grant of the Slovak Academy of Sciences (No. 2/0031/22), the Slovak Research and Development Agency contract (No. 
APVV-20-0148), and the support of the city of Szombathely.

This paper makes extensive use of data collected by the \textit{TESS} mission. Funding for the \textit{TESS} mission is provided by the NASA Science Mission directorate. Some of the data presented in this paper were obtained from the Mikulski Archive for Space Telescopes (MAST). STScI is operated by the Association of Universities for Research in Astronomy, Inc., under NASA contract NAS5-26555. Support for MAST for non-HST data is provided by the NASA Office of Space Science via grant NNX09AF08G and by other grants and contracts.

Distances and other astrometric properties for all targets, and outer orbits for two of our sources,  were taken from the archives of European  Space Agency (ESA)  mission {\it Gaia}\footnote{\url{https://www.cosmos.esa.int/gaia}},  processed  by  the {\it   Gaia}   Data   Processing   and  Analysis   Consortium   (DPAC)\footnote{\url{https://www.cosmos.esa.int/web/gaia/dpac/consortium}}.  Funding for the DPAC  has been provided  by national  institutions, in  particular the institutions participating in the {\it Gaia} Multilateral Agreement.

Some of the SED fluxes and magnitudes were obtained with the Wide-field Infrared Survey Explorer, which is a joint project of the University of California, Los Angeles, and the Jet Propulsion Laboratory/California Institute of Technology, funded by the National Aeronautics and Space Administration. 

Additionally, some of the SED fluxes and magnitudes were obtained with the Two Micron All Sky Survey, which is a joint project of the University of Massachusetts and the Infrared Processing and Analysis Center/California Institute of Technology, funded by the National Aeronautics and Space Administration and the National Science Foundation.

We  used the  Simbad  service  operated by  the  Centre des  Donn\'ees Stellaires (Strasbourg,  France). 

This research has also made use of the VizieR catalogue access tool, CDS, Strasbourg, France (DOI : 10.26093/cds/vizier). The original description of the VizieR service was published in \citet{ochsenbein00}.

\end{acknowledgements}

\begin{appendix} 

\onecolumn

\section{Tabulated system parameters}
\label{app:inittables} 

Coordinates, catalogue passband magnitudes, and some other catalogue parameters of our ten target systems.

\begin{table*}[ht]
\hspace{-20px}
\footnotesize
\centering
\caption{Main properties of the first five of the ten triple systems from different catalogues}
\tiny
\begin{tabular}{lccccc}
\hline
\hline
Parameter			       &  198581208	    & 265274458 	 &  283846096	    &  337993842       & 351404069 \\  
\hline
RA J2000 			       &  17:04:25.48	    & 23:50:46.62	 &  21:03:33.20     &  22:31:19.72     & 21:25:46.26  \\
Dec J2000 		               &  46:35:33.53	    & 73:09:24.27	 &  21:51:42.01     &  60:51:25.35     & 38:59:27.77  \\
$T^a$                                  & $13.76 \pm 0.01$   & $12.03 \pm 0.01$   & $13.45 \pm 0.01$ & $13.07 \pm 0.03$ & $12.57 \pm 0.01$ \\
G$^b$ 				       & $14.24 \pm 0.00$   & $12.42 \pm 0.00$   & $14.07 \pm 0.00$ & $13.60 \pm 0.00$ & $13.05 \pm 0.00$ \\ 
G$_{\rm BP}^b$ 		               & $14.62 \pm 0.00$   & $12.73 \pm 0.03$   & $14.60 \pm 0.00$ & $14.07 \pm 0.00$ & $13.44 \pm 0.00$ \\
G$_{\rm RP}^b$ 		               & $13.68 \pm 0.00$   & $11.93 \pm 0.01$   & $13.38 \pm 0.00$ & $12.95 \pm 0.00$ & $12.49 \pm 0.00$ \\
B$^a$                                  & $15.28 \pm 0.01$   & $13.13 \pm 0.52$   & $15.41 \pm NA$   & $14.69 \pm 0.14$   & $13.94 \pm 0.05$ \\
V$^c$                                  & $14.51 \pm 0.02$   & $12.50 \pm 0.05$   & $14.35 \pm 0.08$ & $13.94 \pm 0.14$ & $13.23 \pm 0.06$ \\  
J$^d$				       & $13.01 \pm 0.03$   & $11.34 \pm 0.02$   & $12.50 \pm 0.03$ & $12.09 \pm 0.03$ & $11.93 \pm 0.03$ \\
H$^d$ 				       & $12.61 \pm 0.03$   & $11.20 \pm 0.03$   & $12.04 \pm 0.03$ & $11.87 \pm 0.03$ & $11.56 \pm 0.03$ \\
K$^d$ 			               & $12.51 \pm 0.03$   & $11.10 \pm 0.02$   & $11.93 \pm 0.02$ & $11.77 \pm 0.03$ & $11.51 \pm 0.02$    \\
W1$^e$ 				       & $12.47 \pm 0.02$   & $11.02 \pm 0.02$   & $11.80 \pm 0.02$ & $11.70 \pm 0.02$ & $11.38 \pm 0.02$     \\
W2$^e$ 				       & $12.49 \pm 0.02$   & $11.02 \pm 0.02$   & $11.82 \pm 0.02$ & $11.69 \pm 0.02$ & $11.41 \pm 0.02$    \\
W3$^e$ 				       & $11.98 \pm 0.19$   & $10.30 \pm 0.06$   & $11.91 \pm 0.29$ & $12.44 \pm 0.43$ & $11.31 \pm 0.14$    \\
$T_{\rm eff}$ [K]$^b$ 	               & $5615 \pm 55$      & $9091 \pm 296$	 & $4975 \pm 16$    & NA               &  NA	\\  
$T_{\rm eff}$ [K]$^a$ 	               & $5465 \pm 35$      & $7972 \pm 153$	 & $4923 \pm 122$   & $8968 \pm 123$   & $5757 \pm 125$ \\  
Radius [$R_\odot$]$^a$ 	               & $1.55 \pm NA$      & $2.29 \pm 0.08 $   & $1.04 \pm NA$    & $4.55 \pm NA$    & $2.50 \pm 0.12$ \\
Distance [pc]$^f$		       & $1081\pm 17$	    &  $816 \pm 8$	 &  $452 \pm 6$     &  $2186 \pm 64$   & $1009 \pm 9$ \\  
$E(B-V)^a$                             & $0.03 \pm NA$      & $0.43 \pm 0.01$	 & $0.08 \pm 0.01$  & $0.79 \pm 0.03$  & $0.10 \pm 0.01$  \\ 
$\mu_\alpha$ [mas/yr]$^b$	       & $-8.01 \pm 0.02$   & $-1.59 \pm 0.01$   & $-6.25 \pm 0.02$ & $-2.81 \pm 0.01$ & $-7.11 \pm 0.01$ \\   
$\mu_\delta$ [mas/yr]$^b$  	       & $-6.01 \pm 0.02$   & $-2.60 \pm 0.01$   & $-13.7 \pm 0.02$ & $-1.25 \pm 0.01$ & $-9.93 \pm 0.01$ \\  
RUWE$^{b,g}$                           & 1.05               & 1.16               & 1.38 	    & 1.00	       & 0.96	\\
astr\_ex\_noise [mas]$^{b,h}$          &  0                 &  0.09              &  0.10	    &  0   	       &  0.03        \\
astr\_ex\_noise\_sig$^{b,h}$           &  0                 &   11               &  5.4 	    &  0	       &  1.0	      \\
$P_{\rm binary}$$^i$ [d]               & 2.8769             & 2.9978             & 5.8082	    & 2.6272	       & 8.9113     \\
$P_{\rm triple}$$^i$ [d]               & 72.65              & 57.72              & 55.95	    & 101.38	       & 96.24       \\
\hline
\label{tbl:mags10-1}
\end{tabular}  

\small
\textbf{Notes.} General: ``NA" and ellipses in this table indicate that the value is not available. (a) \textit{TESS} Input Catalog (TIC v8.2) \citep{TIC8}. (b) Gaia EDR3 \citep{GaiaEDR3}; the uncertainty in $T_{\rm eff}$ listed here is 1.5 times the geometric mean of the upper and lower error bars of \verb|teff_gspphot|. Magnitude uncertainties listed as 0.00 are $\lesssim 0.005$. (c) AAVSO Photometric All Sky Survey (APASS) DR9, \citep{APASS}, \url{http://vizier.u-strasbg.fr/viz-bin/VizieR?-source=II/336/apass9}. (d) 2MASS catalogue \citep{2MASS}.  (e) WISE point source catalogue \citep{WISE}. (f) \citet{bailer-jonesetal21}, geometric distances. (g) The Gaia renormalized unit weight error (RUWE) is the square root of the normalized $\chi^2$ of the astrometric fit to the along-scan observations. Values in excess of about unity are sometimes taken to be a sign of stellar multiplicity. (h) Abbreviations for \verb|astrometric_excess_noise| and \verb|astrometric_excess_noise_sig| (\citealt{lindegren21}; \url{https://gea.esac.esa.int/archive/documentation/GDR2/Gaia_archive/chap_datamodel/sec_dm_main_tables/ssec_dm_gaia_source.html}); these are a measure of ``the disagreement, expressed as an angle, between the observations of a source and the best-fitting standard astrometric model.'' Values of \verb|astrometric_excess_noise_sig| $\gtrsim 2$ are considered significant. (i) Binary and outer orbital periods from this work; given here for reference purposes. 
\end{table*}

\begin{table*}
\hspace{-20px}
\footnotesize
\centering
\caption{Main properties of the second five of the ten triple systems from different catalogues}
\tiny
\begin{tabular}{lccccc}
\hline
\hline
Parameter			       &  378270875        &    403792414      &   403916758       &   405789362       &   461500036        \\  
\hline
RA J2000 			       & 20:13:45.58       & 19:16:23.08       & 01:52:48.09       & 21:37:44.55       & 22:19:19.64        \\
Dec J2000 		               & 47:40:39.86       & 26:23:02.86       & 67:21:05.98       & 64:48:20.15       & 85:04:13.37        \\
$T^a$                                  & $13.12 \pm 0.04$  & $13.78 \pm 0.01$  & $11.31 \pm 0.04$  & $14.45 \pm 0.02$  & $13.78 \pm 0.01$   \\
G$^b$ 				       & $13.79 \pm 0.00$  & $14.51 \pm 0.00$  & $12.39 \pm 0.00$  & $15.19 \pm 0.00$  & $14.34 \pm 0.00$       \\ 
G$_{\rm BP}^b$ 		               & $14.38 \pm 0.00$  & $15.16 \pm 0.00$  & $13.57 \pm 0.00$  & $15.85 \pm 0.00$  & $14.79 \pm 0.00$       \\
G$_{\rm RP}^b$ 		               & $13.05 \pm 0.00$  & $13.73 \pm 0.00$  & $11.33 \pm 0.00$  & $14.40 \pm 0.00$  & $13.71 \pm 0.00$        \\
B$^a$                                  & $15.09 \pm 0.23$  & $16.15 \pm 0.06$  & $15.25 \pm NA$    & $16.69 \pm 0.07$  & $15.60 \pm 0.30$     \\
V$^c$                                  & $14.00 \pm 0.03$  & $14.83 \pm 0.07$  & $13.31 \pm 0.07$  & $15.65 \pm 0.06$  & $14.68 \pm 0.03$    \\       
J$^d$				       & $12.01 \pm 0.02$  & $12.64 \pm 0.02$  & $9.73 \pm 0.02 $  & $13.57 \pm 0.03$  & $12.96 \pm 0.03$    \\
H$^d$ 				       & $11.52 \pm 0.03$  & $12.10 \pm 0.02$  & $9.03 \pm 0.03$   & $13.16 \pm 0.03$  & $12.59 \pm 0.03$     \\
K$^d$ 			               & $11.37 \pm 0.03$  & $11.98 \pm 0.03$  & $8.82 \pm 0.02$   & $13.01 \pm 0.03$  & $12.55 \pm 0.03$    \\
W1$^e$ 				       & $11.21 \pm 0.02$  & $11.97 \pm 0.02$  & $8.65 \pm 0.02$   & $12.71 \pm 0.03$  & $12.52 \pm 0.02$     \\
W2$^e$ 				       & $11.27 \pm 0.02$  & $12.01 \pm 0.02$  & $8.68 \pm 0.02$   & $12.72 \pm 0.03$  & $12.52 \pm 0.02$    \\
W3$^e$ 				       & $11.79 \pm 0.19$  & $11.85 \pm 0.25$  & $8.58 \pm 0.03$   & $12.70 \pm NA$    & $12.43 \pm 0.31$    \\
$T_{\rm eff}$ [K]$^b$ 	               & $5015 \pm 43$     & $4903 \pm 21$     & $4871 \pm 12$     & $7194 \pm 73$     & $5858 \pm 39$         \\  
$T_{\rm eff}$ [K]$^a$ 	               & $4705 \pm 122$    & $5018 \pm 123$    & $4211 \pm 122$    & $6129 \pm 124$    & $5561 \pm 122$        \\  
Radius [$R_\odot$]$^a$ 	               & $1.38 \pm NA$     & $1.34 \pm NA$     & $9.74 \pm NA $    & $2.91 \pm NA$     & $2.25 \pm NA$ \\
Distance [pc]$^f$		       & $509 \pm 7$       & $654\pm 7$        &  $804 \pm 8$      & $2036 \pm 187$    &  $1355 \pm 32$  \\  
$E(B-V)^a$                             & $0.07 \pm 0.06$   & $0.26 \pm 0.01$   & $0.51 \pm 0.08$   & $0.58 \pm 0.02$   & $0.15 \pm 0.01$     \\ 
$\mu_\alpha$ [mas/yr]$^b$	       & $3.38 \pm 0.03$   & $5.59 \pm 0.01$   & $3.25 \pm 0.01$   & $-4.49 \pm 0.03$  & $-1.18 \pm 0.02$    \\   
$\mu_\delta$ [mas/yr]$^b$  	       & $2.00 \pm 0.03$   & $7.21 \pm 0.02$   & $-1.36 \pm 0.02$  & $-3.31 \pm 0.02$  & $1.12 \pm 0.02$     \\  
RUWE$^{b,g}$                           &  1.93             & 1.03              & 1.33              & 0.96              & 0.99                    \\
astr\_ex\_noise [mas]$^{b,h}$          &  0.23             &  0                &  0.14             &   0               &  0                        \\
astr\_ex\_noise\_sig$^{b,h}$           &   31              &  0                &   26              &   0               &  0                         \\
$P_{\rm binary}$$^i$ [d]               & 2.5769            & 4.9949            & 1.3371            & 4.5778            & 2.4712                     \\
$P_{\rm triple}$$^i$ [d]               & 58.12             & 78.20             & 71.06             & 46.81             & 54.56                       \\
\hline
\label{tbl:mags10-2}
\end{tabular}  

\small
\tablefoot{See the notes under Table~\ref{tbl:mags10-1}.}
\end{table*}

\begin{table*}
\centering
\caption{TESS observation sectors for the triples}
\small
\begin{tabular}{lcc}
\hline
\hline
Object & Sectors observed & Third body events  \\
\hline
TIC 198581208 & S24--26,51--53,78--80 & 24,26,51,52,79,80 \\ 
TIC 265274458\tablefootmark{a} & 17,19,24--25,52,58--59,78--79,85--86 & 17,19,24,52,58,59,78,79,85,86 \\
TIC 283846096 & 15,41,55,82 & 15,41,55,82 \\
TIC 337993842 & 16--17,24,57,84--85 & 17, 84 \\
TIC 351404069 & 15--16,55--56,75--76,82--83 & 55, 56, 76, 83 \\
TIC 378270875 & 14--15,41,55--56,75--76,81--83 & 14,55,76,82  \\
TIC 403792414 & 40--41,54,80--81 & 40,41,54,80--81 \\
TIC 403916758 & 18,24--25,52,58--59,78--79,85--86 & 24,25,58,79,85 \\
TIC 405789362 & 16--18,24,56--58,76--78,83--85 & 16,17,18,24,56,57,58,76,77,78,83,84,85 \\
TIC 461500036 & 18,20,25--26,40,52--53.58--60,73,78--79,85--86 & 20,25,53,59,73,79,85 \\
\hline
\label{tbl:sectors}  
\end{tabular}

\tablefoot{None of these sources will be observed in further TESS sectors until the end of Cycle 8 observations; \tablefoottext{a} TIC\,265274458 was also observed with 2-min cadence time in Sectors between 24 and 79.}
\end{table*}

%\onecolumn

\FloatBarrier
%\onecolumn

%\newpage

\section{Tabulated results of the photodynamical analyses}
\label{app:resulttables} 

%\FloatBarrier

In this appendix we tabulate the results of the joint photodynamical analyses.

\begin{table*}
\centering
\caption{Definitions of triple system parameters in Tables~\ref{tab:syntheticfit_TIC198581208265274458} -- \ref{tab:syntheticfit_TIC405789362461500036}}
\label{tbl:definitions}
\small
\begin{tabular}{lc}
\hline
\hline
Parameter$^a$ & Definition   \\
\hline
$t_0$ & Epoch time for osculating elements    \\
$P$ & Orbital period  \\ 
$a$ & Orbital semimajor axis  \\
$e$ & Orbital eccentricity \\
$\omega$ & Argument of periastron (of secondary) \\
$i$ & Orbital inclination angle \\
$\mathcal{T}_0^\mathrm{inf/sup}$ & Time of conjunction of the secondary$^b$ \\
$\tau$ & Time of periastron passage  \\
$\Omega$ & Longitude of the node relative to \\
& the node of the inner orbit \\
$i_{\rm mut}$ & Mutual inclination angle$^c$   \\
$q$ & Mass ratio (secondary/primary)  \\ 
$K_\mathrm{pri}$ & ``K'' velocity amplitude of primary \\
$K_\mathrm{sec}$ & ``K'' velocity amplitude of secondary \\
$R/a$ & Stellar radius divided by semimajor axis \\
$T_{\rm eff}/T_{\rm eff,Aa}$ & Temperature relative to EB primary \\
fractional flux  & Stellar contribution in the given band \\
$M$ & Stellar mass  \\
$R$ & Stellar radius   \\
$T_\mathrm{eff}$ & Stellar effective temperature  \\ 
$L_\mathrm{bol}$ & Stellar bolometric luminosity  \\
$M_\mathrm{bol}$ & Stellar absolute bolometric magnitude \\
$M_V$ & Stellar absolute visual magnitude \\
$\log g$ & log surface gravity (cgs units) \\
$[M/H]$ & log metallicity abundance to H, by mass \\
$E(B-V)$ & Colour excess in B-V bands  \\
extra light, $\ell_4$  & Contaminating flux in the given band   \\
$(M_V)_\mathrm{tot}$ & System absolute visual magnitude   \\ 
distance & Distance to the source  \\
\hline   % Table 5
\end{tabular}

\textit{Notes}. (a) The units for the parameters are given in Tables~\ref{tab:syntheticfit_TIC198581208265274458}-- \ref{tab:syntheticfit_TIC405789362461500036}. (b) The superscript of ``inf/sup'' indicates inferior vs.~superior conjunctions. (By default we give inferior conjunctions. Superior conjunctions are indicated by asteriks.) (c) More explicitly, this is the angle between the orbital planes of the inner binary and the outer triple orbit.

\end{table*} 

\begin{table*}
 \centering
\caption{Orbital and astrophysical parameters of TICs\,198581208 and 265274458 from the joint photodynamical light curve, ETV, SED and \texttt{PARSEC} isochrone solution. }
 \label{tab:syntheticfit_TIC198581208265274458}
\scalebox{0.91}{\begin{tabular}{@{}lllllll}
\hline
\hline
 & \multicolumn{3}{c}{TIC\,198581208} &  \multicolumn{3}{c}{TIC\,265274458}\\
\hline
\multicolumn{7}{c}{orbital elements} \\
\hline
   & \multicolumn{6}{c}{subsystem}  \\
   & \multicolumn{2}{c}{Aa--Ab} & A--B & \multicolumn{2}{c}{Aa--Ab} & A--B  \\
  \hline
  $t_0$ [BJD - 2400000]& \multicolumn{3}{c}{$58955.0$} & \multicolumn{3}{c}{$58764.5$} \\
  $P$ [days] & \multicolumn{2}{c}{$2.869542_{-0.00040}^{+0.000039}$} & $72.6507_{-0.0019}^{+0.0017}$ &  \multicolumn{2}{c}{$2.990213_{-0.000065}^{+0.000067}$} & $57.7193_{-0.0015}^{+0.0013}$ \\
  $a$ [R$_\odot$] & \multicolumn{2}{c}{$9.99_{-0.17}^{+0.16}$} & $100.2_{-1.8}^{+1.6}$ & \multicolumn{2}{c}{$11.588_{-0.083}^{+0.055}$} & $93.74_{-0.64}^{+0.41}$ \\
  $e$ & \multicolumn{2}{c}{$0.01413_{-0.00062}^{+0.00046}$} & $0.28945_{-0.00099}^{+0.00098}$ & \multicolumn{2}{c}{$0.00248_{-0.00029}^{+0.00037}$} & $0.0026_{-0.0014}^{+0.0015}$ \\
  $\omega$ [deg]& \multicolumn{2}{c}{$97.4_{-3.0}^{+1.5}$} & $95.600_{-0.070}^{+0.071}$ & \multicolumn{2}{c}{$302_{-10}^{+13}$} & $58_{-18}^{+8}$ \\ 
  $i$ [deg] & \multicolumn{2}{c}{$89.50_{-0.15}^{+0.14}$} & $89.60_{-0.02}^{+0.02}$ & \multicolumn{2}{c}{$90.48_{-0.24}^{+0.25}$} & $90.22_{-0.07}^{+0.06}$ \\
  $\mathcal{T}_0^\mathrm{inf/sup}$ [BJD - 2400000]& \multicolumn{2}{c}{$58957.6102_{-0.0001}^{+0.0001}$} & $58975.6080_{-0.0131}^{+0.0117}$ & \multicolumn{2}{c}{$58766.3499_{-0.0001}^{+0.0001}$} & $58781.1864_{-0.0024}^{+0.0023}$ \\
  $\tau$ [BJD - 2400000]& \multicolumn{2}{c}{$58956.236_{-0.024}^{+0.013}$} & $58941.243_{-0.012}^{+0.013}$ & \multicolumn{2}{c}{$58763.627_{-0.081}^{+0.112}$} & $58746.8_{-3.9}^{+1.3}$ \\
  $\Omega$ [deg] & \multicolumn{2}{c}{$0.0$} & $-0.06_{-0.17}^{+0.18}$ & \multicolumn{2}{c}{$0.0$} & $0.60_{-0.36}^{+0.33}$ \\
  $i_\mathrm{mut}$ [deg] & \multicolumn{3}{c}{$0.22_{-0.11}^{+0.14}$} & \multicolumn{3}{c}{$0.72_{-0.28}^{+0.36}$} \\
  $\varpi^\mathrm{dyn}$ [deg]& \multicolumn{2}{c}{$277.4_{-3.0}^{+1.5}$} & $275.6_{-0.07}^{+0.07}$ & \multicolumn{2}{c}{$122_{-10}^{+13}$} & $236_{-24}^{+8}$ \\
  $i^\mathrm{dyn}$ [deg] & \multicolumn{2}{c}{$0.19_{-0.10}^{+0.12}$} & $0.03_{-0.02}^{+0.02}$ & \multicolumn{2}{c}{$0.62_{-0.25}^{+0.31}$} & $0.09_{-0.04}^{+0.05}$ \\
  $\Omega^\mathrm{dyn}$ [deg] & \multicolumn{2}{c}{$326_{-71}^{+73}$} & $146_{-71}^{+73}$ & \multicolumn{2}{c}{$114_{-24}^{+35}$} & $294_{-24}^{+35}$ \\
  $i_\mathrm{inv}$ [deg] & \multicolumn{3}{c}{$89.59_{-0.03}^{+0.02}$} & \multicolumn{3}{c}{$90.26_{-0.05}^{+0.05}$} \\
  $\Omega_\mathrm{inv}$ [deg] & \multicolumn{3}{c}{$-0.05_{-0.15}^{+0.16}$} & \multicolumn{3}{c}{$0.52_{-0.31}^{+0.29}$} \\
  \hline
  mass ratio $[q=M_\mathrm{sec}/M_\mathrm{pri}]$ & \multicolumn{2}{c}{$0.563_{-0.012}^{+0.010}$} & $0.576_{-0.004}^{+0.004}$ & \multicolumn{2}{c}{$0.229_{-0.003}^{+0.002}$} & $0.421_{-0.005}^{+0.003}$ \\
  $K_\mathrm{pri}$ [km\,s$^{-1}$] & \multicolumn{2}{c}{$63.47_{-0.71}^{+0.78}$} & $26.67_{-0.53}^{+0.56}$ & \multicolumn{2}{c}{$36.53_{-0.36}^{+0.30}$} & $24.32_{-0.20}^{+0.17}$ \\ 
  $K_\mathrm{sec}$ [km\,s$^{-1}$] & \multicolumn{2}{c}{$112.7_{-2.4}^{+2.5}$} & $46.29_{-0.78}^{+0.76}$ & \multicolumn{2}{c}{$159.6_{-1.2}^{+1.0}$} & $57.85_{-0.47}^{+0.35}$ \\ 
  \hline
  \multicolumn{7}{c}{Apsidal and nodal motion related parameters} \\
  \hline
$P_\mathrm{apse}$ [year] & \multicolumn{2}{c}{$15.63_{-0.07}^{+0.07}$} & $97.23_{-0.50}^{+0.66}$ & \multicolumn{2}{c}{$13.56_{-0.09}^{+0.11}$} & $90.93_{-0.57}^{+0.63}$ \\ 
$P_\mathrm{apse}^\mathrm{dyn}$ [year] & \multicolumn{2}{c}{$7.33_{-0.03}^{+0.03}$} & $12.10_{-0.04}^{+0.04}$ & \multicolumn{2}{c}{$6.35_{-0.04}^{+0.05}$} & $10.55_{-0.06}^{+0.07}$ \\ 
$P_\mathrm{node}^\mathrm{dyn}$ [year] & \multicolumn{3}{c}{$13.82_{-0.06}^{+0.06}$} & \multicolumn{3}{c}{$11.94_{-0.08}^{+0.07}$} \\
$\Delta\omega_\mathrm{3b}$ [arcsec/cycle] & \multicolumn{2}{c}{$1368.6_{-6.1}^{+6.4}$} & $21305_{-68}^{+72}$ & \multicolumn{2}{c}{$1661_{-12}^{+11}$} & $19408_{-120}^{+113}$ \\ 
$\Delta\omega_\mathrm{GR}$ [arcsec/cycle] & \multicolumn{2}{c}{$1.343_{-0.046}^{+0.042}$} & $0.230_{-0.008}^{+0.007}$ & \multicolumn{2}{c}{$1.664_{-0.024}^{+0.016}$} & $0.292_{-0.004}^{+0.003}$ \\ 
$\Delta\omega_\mathrm{tide}$ [arcsec/cycle] & \multicolumn{2}{c}{$18.3_{-1.0}^{+1.1}$} & $0.035_{-0.002}^{+0.002}$ & \multicolumn{2}{c}{$8.38_{-0.39}^{+0.37}$} & $0.034_{-0.002}^{+0.001}$  \\ 
  \hline  
\multicolumn{7}{c}{stellar parameters} \\
\hline
   & Aa & Ab &  B & Aa & Ab &  B \\
  \hline
 \multicolumn{7}{c}{Relative quantities} \\
  \hline
 fractional radius [$R/a$]  & $0.1270_{-0.0015}^{+0.0015}$ & $0.0581_{-0.0012}^{+0.0012}$ & $0.0096_{-0.0002}^{+0.0003}$ & $0.1463_{-0.0014}^{+0.0012}$ & $0.0365_{-0.0005}^{+0.0004}$ & $0.0092_{-0.0001}^{+0.0001}$ \\
 temperature relative to $(T_\mathrm{eff})_\mathrm{Aa}$ & $1$ & $0.6502_{-0.0056}^{+0.0057}$ & $0.9531_{-0.0042}^{+0.0042}$ & $1$ & $0.3923_{-0.0042}^{+0.0056}$ & $0.6463_{-0.0074}^{+0.0064}$ \\
 fractional flux [in \textit{TESS}-band] & $0.6487_{-0.0219}^{+0.0154}$ & $0.0245_{-0.0009}^{+0.0009}$ & $0.3194_{-0.030}^{+0.0136}$ & $0.9261_{-0.0108}^{+0.0058}$ & $0.0019_{-0.0001}^{+0.0001}$ & $0.0643_{-0.0020}^{+0.0019}$ \\
 \hline
 \multicolumn{7}{c}{Physical quantities} \\
  \hline 
 $M$ [M$_\odot$] & $1.038_{-0.057}^{+0.054}$ & $0.584_{-0.026}^{+0.026}$ & $0.935_{-0.050}^{+0.048}$ & $1.898_{-0.043}^{+0.029}$ & $0.434_{-0.007}^{+0.007}$ & $0.981_{-0.019}^{+0.012}$ \\
 $R$ [R$_\odot$] & $1.270_{-0.032}^{+0.033}$ & $0.581_{-0.021}^{+0.020}$ & $0.965_{-0.040}^{+0.044}$ & $1.698_{-0.028}^{+0.017}$ & $0.424_{-0.009}^{+0.006}$ & $0.863_{-0.019}^{+0.013}$ \\
 $T_\mathrm{eff}$ [K]& $6095_{-64}^{+99}$ & $3956_{-52}^{+72}$ & $5801_{-65}^{+99}$ & $8537_{-147}^{+165}$ & $3345_{-20}^{+42}$ & $5516_{-63}^{+64}$ \\
 $L_\mathrm{bol}$ [L$_\odot$] & $1.99_{-0.17}^{+0.23}$ & $0.074_{-0.008}^{+0.012}$ & $0.95_{-0.12}^{+0.16}$ & $13.8_{-1.3}^{+1.0}$ & $0.020_{-0.001}^{+0.001}$ & $0.618_{-0.048}^{+0.046}$ \\
 $M_\mathrm{bol}$ & $4.02_{-0.12}^{+0.10}$ & $7.60_{-0.17}^{+0.12}$ & $4.83_{-0.17}^{+0.14}$ & $1.92_{-0.08}^{+0.11}$ & $9.00_{-0.06}^{+0.06}$ & $5.29_{-0.08}^{+0.09}$ \\
 $M_V           $ & $4.05_{-0.13}^{+0.11}$ & $8.68_{-0.23}^{+0.15}$ & $4.88_{-0.19}^{+0.15}$ & $1.91_{-0.06}^{+0.09}$ & $11.03_{-0.14}^{+0.11}$ & $5.40_{-0.09}^{+0.10}$ \\
 $\log g$ [dex] & $4.247_{-0.012}^{+0.014}$ & $4.675_{-0.012}^{+0.013}$ & $4.439_{-0.016}^{+0.014}$ & $4.257_{-0.007}^{+0.007}$ & $4.820_{-0.007}^{+0.010}$ & $4.556_{-0.007}^{+0.010}$ \\
 \hline
\multicolumn{7}{c}{Global system parameters} \\
  \hline
$\log$(age) [dex] &\multicolumn{3}{c}{$9.792_{-0.134}^{+0.103}$} & \multicolumn{3}{c}{$8.420_{-0.042}^{+0.054}$} \\
$[M/H]$  [dex]    &\multicolumn{3}{c}{$-0.105_{-0.058}^{+0.110}$} & \multicolumn{3}{c}{$0.131_{-0.070}^{+0.029}$} \\
$E(B-V)$ [mag]    &\multicolumn{3}{c}{$0.200_{-0.018}^{+0.026}$} & \multicolumn{3}{c}{$0.473_{-0.019}^{+0.018}$} \\
extra light $\ell_4$ [in \textit{TESS}-band] & \multicolumn{3}{c}{$0.006_{-0.004}^{+0.014}$} & \multicolumn{3}{c}{$0.008_{-0.006}^{+0.011}$} \\
$(M_V)_\mathrm{tot}$  &\multicolumn{3}{c}{$3.62_{-0.15}^{+0.12}$} & \multicolumn{3}{c}{$1.86_{-0.06}^{+0.09}$} \\
distance [pc]           & \multicolumn{3}{c}{$1161_{-44}^{+43}$} & \multicolumn{3}{c}{$767_{-17}^{+11}$} \\  
\hline
\end{tabular}}
\end{table*}

\begin{table*}
 \centering
\caption{Orbital and astrophysical parameters of TICs\,283846096 and 337993842 from the joint photodynamical light curve, ETV, SED and \texttt{PARSEC} isochrone solution. }
 \label{tab:syntheticfit_TIC283846096337993842}
\scalebox{0.91}{\begin{tabular}{@{}lllllll}
\hline
\hline
 & \multicolumn{3}{c}{TIC\,283846096} &  \multicolumn{3}{c}{TIC\,337993842}\\
\hline
\multicolumn{7}{c}{orbital elements} \\
\hline
   & \multicolumn{6}{c}{subsystem}  \\
   & \multicolumn{2}{c}{Aa--Ab} & A--B & \multicolumn{2}{c}{Aa--Ab} & A--B \\
  \hline
  $t_0$ [BJD - 2400000]& \multicolumn{3}{c}{$58711.0$} &  \multicolumn{3}{c}{$58738.5$} \\
  $P$ [days] & \multicolumn{2}{c}{$5.76474_{-0.00077}^{+0.00096}$} & $55.9539_{-0.0034}^{+0.0023}$ & \multicolumn{2}{c}{$2.626577_{-0.000046}^{+0.000042}$} & $101.3859_{-0.0028}^{+0.0030}$ \\
  $a$ [R$_\odot$] & \multicolumn{2}{c}{$13.41_{-0.23}^{+0.26}$} & $75.3_{-1.3}^{+1.4}$ & \multicolumn{2}{c}{$11.16_{-0.12}^{+0.09}$} & $156.4_{-1.8}^{+1.9}$ \\
  $e$ & \multicolumn{2}{c}{$0.10579_{-0.00099}^{+0.00080}$} & $0.0684_{-0.0020}^{+0.0021}$ & \multicolumn{2}{c}{$0.00395_{-0.00018}^{+0.00024}$} & $0.2122_{-0.0093}^{+0.0046}$ \\
  $\omega$ [deg]& \multicolumn{2}{c}{$241.4_{-0.9}^{+1.4}$} & $263.9_{-1.0}^{+1.0}$ & \multicolumn{2}{c}{$8.8_{-4.2}^{+4.0}$} & $166.2_{-2.5}^{+2.5}$ \\ 
  $i$ [deg] & \multicolumn{2}{c}{$89.66_{-0.06}^{+0.06}$} & $89.59_{-0.03}^{+0.02}$ & \multicolumn{2}{c}{$88.90_{-0.33}^{+0.30}$} & $89.14_{-0.04}^{+0.04}$ \\
  $\mathcal{T}_0^\mathrm{inf/sup}$ [BJD - 2400000] & \multicolumn{2}{c}{$58712.4342_{-0.0022}^{+0.0041}$} & $58722.9417_{-0.0027}^{+0.0027}$\tablefootmark{*} & \multicolumn{2}{c}{$58740.6730_{-0.0002}^{+0.0003}$} & $58790.0973_{-0.0193}^{+0.0226}$\tablefootmark{*} \\
  $\tau$ [BJD - 2400000] & \multicolumn{2}{c}{$58712.064_{-0.015}^{+0.021}$} & $58693.88_{-0.19}^{+0.19}$ & \multicolumn{2}{c}{$58738.765_{-0.031}^{+0.029}$} & $58703.86_{-0.65}^{+0.68}$ \\
  $\Omega$ [deg] & \multicolumn{2}{c}{$0.0$} & $-0.44_{-0.07}^{+0.06}$ & \multicolumn{2}{c}{$0.0$} & $-1.02_{-0.51}^{+0.41}$ \\
  $i_\mathrm{mut}$ [deg] & \multicolumn{3}{c}{$0.45_{-0.06}^{ +0.07}$}  & \multicolumn{3}{c}{$1.10_{-0.32}^{+0.46}$} \\
  $\varpi^\mathrm{dyn}$ [deg] & \multicolumn{2}{c}{$61.5_{-0.9}^{+1.4}$} & $83.9_{-1.0}^{+1.0}$ & \multicolumn{2}{c}{$188.8_{-4.2}^{+3.9}$} & $346.2_{-2.5}^{+2.5}$ \\
  $i^\mathrm{dyn}$ [deg] & \multicolumn{2}{c}{$0.39_{-0.05}^{+0.06}$} & $0.06_{-0.01}^{+0.02}$ & \multicolumn{2}{c}{$0.99_{-0.29}^{+0.41}$} & $0.11_{-0.03}^{+0.05}$ \\
  $\Omega^\mathrm{dyn}$ [deg] & \multicolumn{2}{c}{$260.4_{-7.6}^{+7.5}$} & $80.4_{-7.6}^{+7.5}$ & \multicolumn{2}{c}{$283_{-16}^{+24}$} & $103_{-16}^{+24}$ \\
  $i_\mathrm{inv}$ [deg] & \multicolumn{3}{c}{$89.60_{-0.03}^{+0.02}$} & \multicolumn{3}{c}{$89.11_{-0.03}^{+0.05}$} \\
  $\Omega_\mathrm{inv}$ [deg] & \multicolumn{3}{c}{$-0.38_{-0.06}^{+0.05}$} & \multicolumn{3}{c}{$-0.92_{-0.46}^{+0.37}$} \\
  \hline
  mass ratio $[q=M_\mathrm{sec}/M_\mathrm{pri}]$ & \multicolumn{2}{c}{$0.636_{-0.004}^{+0.004}$} & $0.880_{-0.002}^{+0.002}$ & \multicolumn{2}{c}{$0.983_{-0.012}^{+0.012}$} & $0.849_{-0.014}^{+0.032}$ \\
  $K_\mathrm{pri}$ [km\,s$^{-1}$] & \multicolumn{2}{c}{$46.05_{-0.80}^{+0.87}$} & $31.95_{-0.55}^{+0.62}$ & \multicolumn{2}{c}{$106.58_{-1.20}^{+1.00}$} & $36.62_{-0.63}^{+1.15}$ \\ 
  $K_\mathrm{sec}$ [km\,s$^{-1}$] & \multicolumn{2}{c}{$72.30_{-1.29}^{+1.48}$} & $36.29_{-0.62}^{+0.70}$ & \multicolumn{2}{c}{$108.44_{-1.46}^{+1.23}$} & $43.10_{-0.37}^{+0.39}$ \\ 
  \hline
  \multicolumn{7}{c}{Apsidal and nodal motion related parameters} \\
  \hline
$P_\mathrm{apse}$ [year] & \multicolumn{2}{c}{$4.228_{-0.005}^{+0.005}$} & $26.41_{-0.05}^{+0.05}$ & \multicolumn{2}{c}{$25.43_{-0.54}^{+0.42}$} & $265.3_{-1.3}^{+4.3}$ \\ 
$P_\mathrm{apse}^\mathrm{dyn}$ [year] & \multicolumn{2}{c}{$1.955_{-0.006}^{+0.005}$} & $3.20_{-0.02}^{+0.01}$ & \multicolumn{2}{c}{$12.90_{-0.21}^{+0.15}$} & $23.81_{-0.22}^{+0.21}$ \\ 
$P_\mathrm{node}^\mathrm{dyn}$ [year] & \multicolumn{3}{c}{$3.64_{-0.01}^{+0.02}$} & \multicolumn{3}{c}{$26.17_{-0.25}^{+0.30}$} \\
$\Delta\omega_\mathrm{3b}$ [arcsec/cycle] & \multicolumn{2}{c}{$10461_{-24}^{+31}$} & $62095_{-212}^{+293}$ & \multicolumn{2}{c}{$676.7_{-6.9}^{+9.2}$} & $15107_{-130}^{+141}$ \\ 
$\Delta\omega_\mathrm{GR}$ [arcsec/cycle] & \multicolumn{2}{c}{$0.61_{-0.02}^{+0.02}$} & $0.202_{-0.007}^{+0.008}$ & \multicolumn{2}{c}{$2.00_{-0.04}^{+0.03}$} & $0.276_{-0.007}^{+0.006}$ \\ 
$\Delta\omega_\mathrm{tide}$ [arcsec/cycle] & \multicolumn{2}{c}{$0.14_{-0.02}^{+0.02}$} & $0.0010_{-0.0001}^{+0.0002}$ & \multicolumn{2}{c}{$44.1_{-4.3}^{+4.7}$} & $0.031_{-0.003}^{+0.003}$ \\ 
  \hline  
\multicolumn{7}{c}{stellar parameters} \\
\hline
   & Aa & Ab &  B & Aa & Ab & B \\
  \hline
 \multicolumn{7}{c}{Relative quantities} \\
  \hline
 fractional radius [$R/a$]  & $0.0441_{-0.0011}^{+0.0013}$ & $0.0279_{-0.0008}^{+0.0010}$ & $0.0112_{-0.0002}^{+0.0004}$ & $0.1214_{-0.0027}^{+0.0026}$ & $0.1186_{-0.0025}^{+0.0025}$ & $0.0176_{-0.0010}^{+0.0011}$ \\
 temperature relative to $(T_\mathrm{eff})_\mathrm{Aa}$ & $1$ & $0.8422_{-0.0210}^{+0.0153}$ & $1.3699_{-0.0092}^{+0.0085}$ & $1$ & $0.9907_{-0.0066}^{+0.0067}$ & $1.3141_{-0.0466}^{+0.0705}$ \\
 fractional flux [in TESS-band] & $0.1095_{-0.0043}^{+0.0053}$ & $0.0197_{-0.0016}^{+0.0010}$ & $0.8311_{-0.0428}^{+0.0214}$ & $0.0906_{-0.0027}^{+0.0029}$ & $0.0838_{-0.0026}^{+0.0030}$ & $0.8010_{-0.0315}^{+0.0147}$ \\
 \hline
 \multicolumn{7}{c}{Physical quantities} \\
  \hline 
 $M$ [M$_\odot$] & $0.594_{-0.031}^{+0.035}$ & $0.378_{-0.019}^{+0.022}$ & $0.856_{-0.044}^{+0.050}$ & $1.361_{-0.045}^{+0.037}$ & $1.338_{-0.042}^{+0.034}$ & $2.282_{-0.084}^{+0.133}$ \\
 $R$ [R$_\odot$] & $0.591_{-0.024}^{+0.028}$ & $0.374_{-0.016}^{+0.021}$ & $0.840_{-0.027}^{+0.047}$ & $1.355_{-0.043}^{+0.040}$ & $1.324_{-0.039}^{+0.039}$ & $2.752_{-0.147}^{+0.163}$ \\
 $T_\mathrm{eff}$ [K] & $3947_{-72}^{+69}$ & $3315_{-22}^{+16}$ & $5410_{-71}^{+60}$ & $6649_{-99}^{+111}$ & $6578_{-80}^{+116}$ & $8736_{-342}^{+544}$ \\
 $L_\mathrm{bol}$ [L$_\odot$] & $0.076_{-0.011}^{+0.014}$ & $0.015_{-0.001}^{+0.001}$ & $0.544_{-0.062}^{+0.083}$ & $3.23_{-0.32}^{+0.34}$ & $2.96_{-0.25}^{+0.30}$ & $40.11_{-4.71}^{+7.67}$ \\
 $M_\mathrm{bol}$ & $7.57_{-0.18}^{+0.17}$ & $9.31_{-0.09}^{+0.10}$ & $5.43_{-0.15}^{+0.13}$ & $3.50_{-0.11}^{+0.11}$ & $3.59_{-0.10}^{+0.10}$ & $0.76_{-0.19}^{+0.14}$ \\
 $M_V           $ & $8.66_{-0.21}^{+0.23}$ & $11.34_{-0.06}^{+0.05}$ & $5.57_{-0.17}^{+0.15}$ & $3.47_{-0.11}^{+0.12}$ & $3.56_{-0.11}^{+0.10}$ & $0.76_{-0.12}^{+0.11}$ \\
 $\log g$ [dex] & $4.666_{-0.017}^{+0.014}$ & $4.867_{-0.021}^{+0.017}$ & $4.519_{-0.022}^{+0.014}$ & $4.307_{-0.015}^{+0.014}$ & $4.320_{-0.015}^{+0.014}$ & $3.916_{-0.055}^{+0.059}$ \\
 \hline
\multicolumn{7}{c}{Global system parameters} \\
  \hline
$\log$(age) [dex] &\multicolumn{3}{c}{$9.860_{-0.113}^{+0.130}$} &\multicolumn{3}{c}{$8.765_{-0.102}^{+0.062}$} \\
$[M/H]$  [dex]    &\multicolumn{3}{c}{$-0.052_{-0.053}^{+0.111}$} &\multicolumn{3}{c}{$0.154_{-0.121}^{+0.066}$} \\
$E(B-V)$ [mag]    &\multicolumn{3}{c}{$0.158_{-0.021}^{+0.021}$} &\multicolumn{3}{c}{$0.764_{-0.038}^{+0.043}$} \\
extra light $\ell_4$ [in TESS-band] &\multicolumn{3}{c}{$0.038_{-0.023}^{+0.050}$} & \multicolumn{3}{c}{$0.024_{-0.016}^{+0.033}$} \\
$(M_V)_\mathrm{tot}$  &\multicolumn{3}{c}{$5.50_{-0.17}^{+0.16}$} &\multicolumn{3}{c}{$0.61_{-0.12}^{+0.11}$} \\
distance [pc]       &\multicolumn{3}{c}{$490_{-21}^{+31}$} & \multicolumn{3}{c}{$1770_{-74}^{+70}$} \\  
\hline
\end{tabular}}
\end{table*}

\begin{table*}
 \centering
\caption{Orbital and astrophysical parameters of TICs\,351404069 and 378270875 from the joint photodynamical light curve, ETV, SED and \texttt{PARSEC} isochrone solution. }
 \label{tab:syntheticfit_TIC351404069378270875}
\scalebox{0.91}{\begin{tabular}{@{}lllllll}
\hline
\hline
 & \multicolumn{3}{c}{TIC\,351404069} &  \multicolumn{3}{c}{TIC\,378270875}\\
\hline
\multicolumn{7}{c}{orbital elements} \\
\hline
   & \multicolumn{6}{c}{subsystem}  \\
   & \multicolumn{2}{c}{Aa--Ab} & A--B & \multicolumn{2}{c}{Aa--Ab} & A--B  \\
  \hline
  $t_0$ [BJD - 2400000]& \multicolumn{3}{c}{$59814.0$} & \multicolumn{3}{c}{$58683.0$} \\
  $P$ [days] & \multicolumn{2}{c}{$8.91637_{-0.00023}^{+0.00024}$} & $96.2431_{-0.0029}^{+0.0033}$ &  \multicolumn{2}{c}{$2.573263_{-0.000042}^{+0.000048}$} & $58.1195_{-0.0015}^{+0.0018}$ \\
  $a$ [R$_\odot$] & \multicolumn{2}{c}{$23.23_{-0.40}^{+0.34}$} & $128.72_{-2.23}^{+1.86}$ & \multicolumn{2}{c}{$9.25_{-0.19}^{+0.24}$} & $85.25_{-1.86}^{+2.17}$ \\
  $e$ & \multicolumn{2}{c}{$0.03890_{-0.00022}^{+0.00019}$} & $0.24879_{-0.00028}^{+0.00029}$ & \multicolumn{2}{c}{$0.00192_{-0.00021}^{+0.00026}$} & $0.3125_{-0.0021}^{+0.0050}$ \\
  $\omega$ [deg]& \multicolumn{2}{c}{$207.93_{-0.19}^{+0.24}$} & $183.74_{-0.54}^{+0.53}$ & \multicolumn{2}{c}{$295.3_{-6.9}^{+6.5}$} & $314.21_{-0.66}^{+0.51}$ \\ 
  $i$ [deg] & \multicolumn{2}{c}{$87.24_{-0.14}^{+0.15}$} & $90.11_{-0.05}^{+0.05}$ & \multicolumn{2}{c}{$88.93_{-0.10}^{+0.10}$} & $89.00_{-0.05}^{+0.05}$ \\
  $\mathcal{T}_0^\mathrm{inf/sup}$ [BJD - 2400000]& \multicolumn{2}{c}{$59823.2857_{-0.0006}^{+0.0006}$} & ${59815.7901_{-0.0119}^{+0.0102}}^*$ & \multicolumn{2}{c}{$58684.2847_{-0.0002}^{+0.0002}$} & $58704.2085_{-0.0200}^{+0.0264}$ \\
  $\tau$ [BJD - 2400000]& \multicolumn{2}{c}{$59821.8437_{-0.0048}^{+0.0061}$} & $59736.988_{-0.128}^{+0.126}$ & \multicolumn{2}{c}{$58681.891_{-0.049}^{+0.047}$} & $58649.808_{-0.081}^{+0.060}$ \\
  $\Omega$ [deg] & \multicolumn{2}{c}{$0.0$} & $1.30_{-0.17}^{+0.22}$ & \multicolumn{2}{c}{$0.0$} & $-0.15_{-0.29}^{+0.26}$ \\
  $i_\mathrm{mut}$ [deg] & \multicolumn{3}{c}{$3.16_{-0.19}^{+0.17}$} & \multicolumn{3}{c}{$0.27_{-0.14}^{+0.19}$} \\
  $\varpi^\mathrm{dyn}$ [deg]& \multicolumn{2}{c}{$27.90_{-0.19}^{+0.24}$} & $3.74_{-0.54}^{+0.53}$ & \multicolumn{2}{c}{$115.3_{-6.9}^{+6.5}$} & $134.21_{-0.66}^{+0.51}$ \\
  $i^\mathrm{dyn}$ [deg] & \multicolumn{2}{c}{$2.46_{-0.15}^{+0.13}$} & $0.70_{-0.04}^{+0.04}$ & \multicolumn{2}{c}{$0.23_{-0.12}^{+0.16}$} & $0.05_{-0.02}^{+0.03}$ \\
  $\Omega^\mathrm{dyn}$ [deg] & \multicolumn{2}{c}{$24.4_{-2.9}^{+3.3}$} & $204.4_{-2.9}^{+3.3}$ & \multicolumn{2}{c}{$290_{-37}^{+101}$} & $110_{-37}^{+101}$ \\
  $i_\mathrm{inv}$ [deg] & \multicolumn{3}{c}{$89.47_{-0.05}^{+0.05}$} & \multicolumn{3}{c}{$88.98_{-0.04}^{+0.04}$} \\
  $\Omega_\mathrm{inv}$ [deg] & \multicolumn{3}{c}{$1.01_{-0.13}^{+0.17}$} & \multicolumn{3}{c}{$-0.12_{-0.24}^{+0.22}$} \\
  \hline
  mass ratio $[q=M_\mathrm{sec}/M_\mathrm{pri}]$ & \multicolumn{2}{c}{$0.826_{-0.005}^{+0.005}$} & $0.460_{-0.002}^{+0.002}$ & \multicolumn{2}{c}{$0.936_{-0.005}^{+0.005}$} & $0.533_{-0.008}^{+0.009}$ \\
  $K_\mathrm{pri}$ [km\,s$^{-1}$] & \multicolumn{2}{c}{$59.55_{-1.03}^{+0.92}$} & $22.03_{-0.40}^{+0.33}$ & \multicolumn{2}{c}{$87.90_{-1.61}^{+2.16}$} & $26.98_{-0.40}^{+1.05}$ \\ 
  $K_\mathrm{sec}$ [km\,s$^{-1}$] & \multicolumn{2}{c}{$72.20_{-1.26}^{+1.13}$} & $47.84_{-0.81}^{+0.71}$ & \multicolumn{2}{c}{$93.94_{-2.04}^{+2.48}$} & $50.93_{-0.89}^{+1.25}$ \\ 
  \hline
  \multicolumn{7}{c}{Apsidal and nodal motion related parameters} \\
  \hline
$P_\mathrm{apse}$ [year] & \multicolumn{2}{c}{$10.92_{-0.03}^{+0.03}$} & $38.41_{-0.05}^{+0.05}$ & \multicolumn{2}{c}{$11.66_{-0.15}^{+0.10}$} & $58.67_{-0.40}^{+0.27}$ \\ 
$P_\mathrm{apse}^\mathrm{dyn}$ [year] & \multicolumn{2}{c}{$4.78_{-0.01}^{+0.01}$} & $6.98_{-0.01}^{+0.01}$ & \multicolumn{2}{c}{$5.32_{-0.06}^{+0.04}$} & $8.42_{-0.08}^{+0.05}$ \\ 
$P_\mathrm{node}^\mathrm{dyn}$ [year] & \multicolumn{3}{c}{$8.53_{-0.02}^{+0.02}$} & \multicolumn{3}{c}{$9.83_{-0.07}^{+0.11}$} \\
$\Delta\omega_\mathrm{3b}$ [arcsec/cycle] & \multicolumn{2}{c}{$6608_{-16}^{+16}$} & $48935_{-74}^{+70}$ & \multicolumn{2}{c}{$1702_{-13}^{+20}$} & $24498_{-140}^{+235}$ \\ 
$\Delta\omega_\mathrm{GR}$ [arcsec/cycle] & \multicolumn{2}{c}{$0.753_{-0.026}^{+0.022}$} & $0.211_{-0.007}^{+0.006}$ & \multicolumn{2}{c}{$1.431_{-0.059}^{+0.074}$} & $0.264_{-0.011}^{+0.014}$ \\ 
$\Delta\omega_\mathrm{tide}$ [arcsec/cycle] & \multicolumn{2}{c}{$3.69_{-0.23}^{+0.25}$} & $0.021_{-0.001}^{+0.001}$ & \multicolumn{2}{c}{$8.35_{-0.51}^{+0.53}$} & $0.015_{-0.001}^{+0.001}$  \\ 
  \hline  
\multicolumn{7}{c}{stellar parameters} \\
\hline
   & Aa & Ab &  B & Aa & Ab &  B \\
  \hline
 \multicolumn{7}{c}{Relative quantities} \\
  \hline
 fractional radius [$R/a$]  & $0.0859_{-0.0011}^{+0.0012}$ & $0.0421_{-0.0016}^{+0.0014}$ & $0.0079_{-0.0003}^{+0.0002}$ & $0.0901_{-0.0013}^{+0.0012}$ & $0.0818_{-0.0012}^{+0.0010}$ & $0.0104_{-0.0003}^{+0.0003}$ \\
 temperature relative to $(T_\mathrm{eff})_\mathrm{Aa}$ & $1$ & $0.9942_{-0.0069}^{+0.0071}$ & $1.0045_{-0.0067}^{+0.0068}$ & $1$ & $0.9486_{-0.0039}^{+0.0036}$ & $1.0242_{-0.0109}^{+0.0104}$ \\
 fractional flux [in \textit{TESS}-band] & $0.4718_{-0.0260}^{+0.0274}$ & $0.1103_{-0.0030}^{+0.0034}$ & $0.1230_{-0.0044}^{+0.0041}$ & $0.3263_{-0.0062}^{+0.0068}$ & $0.2198_{-0.0053}^{+0.0051}$ & $0.4040_{-0.0340}^{+0.0288}$ \\
 \hline
 \multicolumn{7}{c}{Physical quantities} \\
  \hline 
 $M$ [M$_\odot$] & $1.157_{-0.058}^{+0.052}$ & $0.954_{-0.048}^{+0.042}$ & $0.973_{-0.051}^{+0.043}$ & $0.827_{-0.052}^{+0.066}$ & $0.773_{-0.046}^{+0.060}$ & $0.851_{-0.050}^{+0.076}$ \\
 $R$ [R$_\odot$] & $1.993_{-0.051}^{+0.045}$ & $0.977_{-0.052}^{+0.045}$ & $1.018_{-0.060}^{+0.042}$ & $0.834_{-0.020}^{+0.022}$ & $0.756_{-0.025}^{+0.026}$ & $0.884_{-0.034}^{+0.031}$ \\
 $T_\mathrm{eff}$ [K]& $5902_{-100}^{+91}$ & $5863_{-78}^{+75}$ & $5929_{-79}^{+67}$ & $5487_{-72}^{+82}$ & $5205_{-65}^{+82}$ & $5621_{-51}^{+58}$ \\
 $L_\mathrm{bol}$ [L$_\odot$] & $4.34_{-0.46}^{+0.38}$ & $1.02_{-0.15}^{+0.14}$ & $1.14_{-0.18}^{+0.14}$ & $0.571_{-0.044}^{+0.046}$ & $0.379_{-0.025}^{+0.031}$ & $0.696_{-0.055}^{+0.066}$ \\
 $M_\mathrm{bol}$ & $3.18_{-0.09}^{+0.12}$ & $4.75_{-0.14}^{+0.18}$ & $4.62_{-0.12}^{+0.19}$ & $5.38_{-0.08}^{+0.09}$ & $5.82_{-0.09}^{+0.08}$ & $5.16_{-0.10}^{+0.09}$ \\
 $M_V           $ & $3.22_{-0.10}^{+0.14}$ & $4.80_{-0.15}^{+0.19}$ & $4.66_{-0.13}^{+0.20}$ & $5.50_{-0.10}^{+0.10}$ & $6.02_{-0.09}^{+0.10}$ & $5.25_{-0.11}^{+0.20}$ \\
 $\log g$ [dex] & $3.900_{-0.012}^{+0.012}$ & $4.437_{-0.021}^{+0.026}$ & $4.410_{-0.018}^{+0.029}$ & $4.516_{-0.017}^{+0.012}$ & $4.570_{-0.007}^{+0.008}$ & $4.477_{-0.020}^{+0.021}$ \\
 \hline
\multicolumn{7}{c}{Global system parameters} \\
  \hline
$\log$(age) [dex] &\multicolumn{3}{c}{$9.721_{-0.044}^{+0.083}$} & \multicolumn{3}{c}{$9.972_{-0.269}^{+0.084}$} \\
$[M/H]$  [dex]    &\multicolumn{3}{c}{$-0.129_{-0.073}^{+0.132}$} & \multicolumn{3}{c}{$-0.166_{-0.149}^{+0.166}$} \\
$E(B-V)$ [mag]    &\multicolumn{3}{c}{$0.132_{-0.026}^{+0.024}$} & \multicolumn{3}{c}{$0.402_{-0.018}^{+0.018}$} \\
extra light $\ell_4$ [in \textit{TESS}-band] & \multicolumn{3}{c}{$0.294_{-0.027}^{+0.028}$} & \multicolumn{3}{c}{$0.051_{-0.030}^{+0.035}$} \\
$(M_V)_\mathrm{tot}$  &\multicolumn{3}{c}{$2.78_{-0.11}^{+0.16}$} & \multicolumn{3}{c}{$4.35_{-0.09}^{+0.08}$} \\
distance [pc]           & \multicolumn{3}{c}{$1055_{-44}^{+33}$} & \multicolumn{3}{c}{$542_{-15}^{+18}$} \\  
\hline
\end{tabular}}
\end{table*}

\begin{table*}
 \centering
\caption{Orbital and astrophysical parameters of TICs\,403792414 and 403916758 from the joint photodynamical light curve, ETV, SED and \texttt{PARSEC} isochrone solution. }
 \label{tab:syntheticfit_TIC403792414403916758}
\scalebox{0.91}{\begin{tabular}{@{}lllllll}
\hline
 & \multicolumn{3}{c}{TIC\,403792414} &  \multicolumn{3}{c}{TIC\,403916758}\\
\hline
\multicolumn{7}{c}{orbital elements} \\
\hline
   & \multicolumn{6}{c}{subsystem}  \\
   & \multicolumn{2}{c}{Aa--Ab} & A--B & \multicolumn{2}{c}{Aa--Ab} & A--B  \\
  \hline
  $t_0$ [BJD - 2400000]& \multicolumn{3}{c}{$59388.0$} & \multicolumn{3}{c}{$58790.5$} \\
  $P$ [days] & \multicolumn{2}{c}{$4.97416_{-0.00027}^{+0.00026}$} & $78.197_{-0.012}^{+0.012}$ &  \multicolumn{2}{c}{$1.1337956_{-0.0000079}^{+0.0000087}$} & $71.0600_{-0.0018}^{+0.0020}$ \\
  $a$ [R$_\odot$] & \multicolumn{2}{c}{$13.83_{-0.38}^{+0.26}$} & $100.6_{-2.7}^{+2.1}$ & \multicolumn{2}{c}{$5.531_{-0.057}^{+0.067}$} & $110.07_{-1.60}^{+0.77}$ \\
  $e$ & \multicolumn{2}{c}{$0.02263_{-0.00055}^{+0.00081}$} & $0.3232_{-0.0026}^{+0.0026}$ & \multicolumn{2}{c}{$0.00075_{-0.00035}^{+0.00087}$} & $0.0305_{-0.0155}^{+0.0067}$ \\
  $\omega$ [deg]& \multicolumn{2}{c}{$80.90_{-0.56}^{+0.54}$} & $82.364_{-0.057}^{+0.058}$ & \multicolumn{2}{c}{$173_{-41}^{+69}$} & $270.1_{-1.4}^{+1.9}$ \\ 
  $i$ [deg] & \multicolumn{2}{c}{$90.14_{-0.34}^{+0.27}$} & $90.01_{-0.06}^{+0.07}$ & \multicolumn{2}{c}{$86.57_{-0.57}^{+1.16}$} & $87.88_{-0.22}^{+0.19}$ \\
  $\mathcal{T}_0^\mathrm{inf/sup}$ [BJD - 2400000]& \multicolumn{2}{c}{$59391.3367_{-0.0005}^{+0.0004}$} & ${59390.340_{-0.018}^{+0.016}}*$ & \multicolumn{2}{c}{$58792.1781_{-0.0003}^{+0.0002}$} & $58967.071_{-0.036}^{+0.034}$ \\
  $\tau$ [BJD - 2400000]& \multicolumn{2}{c}{$59388.7175_{-0.0076}^{+0.0073}$} & $59389.536_{-0.017}^{+0.017}$ & \multicolumn{2}{c}{$58791.86_{-0.18}^{+0.18}$} & $58897.50_{-1.36}^{+1.54}$ \\
  $\Omega$ [deg] & \multicolumn{2}{c}{$0.0$} & $-0.40_{-0.63}^{+0.56}$ & \multicolumn{2}{c}{$0.0$} & $3.76_{-0.69}^{+1.00}$ \\
  $i_\mathrm{mut}$ [deg] & \multicolumn{3}{c}{$0.60_{-0.28}^{+0.50}$} & \multicolumn{3}{c}{$4.09_{-0.76}^{+0.77}$} \\
  $\varpi^\mathrm{dyn}$ [deg]& \multicolumn{2}{c}{$260.89_{-0.56}^{+0.54}$} & $262.36_{-0.06}^{+0.06}$ & \multicolumn{2}{c}{$359_{-37}^{+74}$} & $90.1_{-1.4}^{+1.9}$ \\
  $i^\mathrm{dyn}$ [deg] & \multicolumn{2}{c}{$0.49_{-0.23}^{+0.41}$} & $0.10_{-0.05}^{+0.09}$ & \multicolumn{2}{c}{$3.78_{-0.70}^{+0.72}$} & $0.30_{-0.06}^{+0.06}$ \\
  $\Omega^\mathrm{dyn}$ [deg] & \multicolumn{2}{c}{$248_{-79}^{+34}$} & $68_{-79}^{+34}$ & \multicolumn{2}{c}{$69_{-5}^{+17}$} & $249_{-5}^{+17}$ \\
  $i_\mathrm{inv}$ [deg] & \multicolumn{3}{c}{$90.04_{-0.07}^{+0.06}$} & \multicolumn{3}{c}{$87.79_{-0.25}^{+0.21}$} \\
  $\Omega_\mathrm{inv}$ [deg] & \multicolumn{3}{c}{$-0.33_{-0.52}^{+0.46}$} & \multicolumn{3}{c}{$3.48_{-0.63}^{+0.92}$} \\
  \hline
  mass ratio $[q=M_\mathrm{sec}/M_\mathrm{pri}]$ & \multicolumn{2}{c}{$0.689_{-0.016}^{+0.015}$} & $0.558_{-0.011}^{+0.010}$ & \multicolumn{2}{c}{$0.982_{-0.028}^{+0.022}$} & $0.983_{-0.021}^{+0.022}$ \\
  $K_\mathrm{pri}$ [km\,s$^{-1}$] & \multicolumn{2}{c}{$57.3_{-1.6}^{+1.8}$} & $24.65_{-0.72}^{+0.68}$ & \multicolumn{2}{c}{$122.1_{-1.9}^{+1.9}$} & $38.65_{-0.48}^{+0.69}$ \\ 
  $K_\mathrm{sec}$ [km\,s$^{-1}$] & \multicolumn{2}{c}{$83.8_{-2.5}^{+1.1}$} & $44.28_{-1.26}^{+0.69}$ & \multicolumn{2}{c}{$124.6_{-2.3}^{+2.3}$} & $39.37_{-0.68}^{+0.62}$ \\ 
  \hline
  \multicolumn{7}{c}{Apsidal and nodal motion related parameters} \\
  \hline
$P_\mathrm{apse}$ [year] & \multicolumn{2}{c}{$10.61_{-0.12}^{+0.13}$} & $50.14_{-0.19}^{+0.19}$ & \multicolumn{2}{c}{$17.30_{-0.95}^{+0.79}$} & $409.6_{-2.9}^{+3.7}$ \\ 
$P_\mathrm{apse}^\mathrm{dyn}$ [year] & \multicolumn{2}{c}{$4.80_{-0.05}^{+0.05}$} & $7.46_{-0.05}^{+0.06}$ & \multicolumn{2}{c}{$10.97_{-0.32}^{+0.37}$} & $28.31_{-0.26}^{+0.24}$ \\ 
$P_\mathrm{node}^\mathrm{dyn}$ [year] & \multicolumn{3}{c}{$8.759_{-0.082}^{+0.073}$} & \multicolumn{3}{c}{$30.41_{-0.30}^{+0.31}$} \\
$\Delta\omega_\mathrm{3b}$ [arcsec/cycle] & \multicolumn{2}{c}{$3677_{-39}^{+35}$} & $37209_{-277}^{+257}$ & \multicolumn{2}{c}{$253.7_{-2.8}^{+3.3}$} & $8906_{-75}^{+82}$ \\ 
$\Delta\omega_\mathrm{GR}$ [arcsec/cycle] & \multicolumn{2}{c}{$0.857_{-0.047}^{+0.033}$} & $0.205_{-0.011}^{+0.008}$ & \multicolumn{2}{c}{$2.637_{-0.054}^{+0.064}$} & $0.266_{-0.008}^{+0.004}$ \\ 
$\Delta\omega_\mathrm{tide}$ [arcsec/cycle] & \multicolumn{2}{c}{$0.67_{-0.12}^{+0.07}$} & $0.0024_{-0.0004}^{+0.0003}$ & \multicolumn{2}{c}{$107_{-10}^{+15}$} & $0.94_{-0.14}^{+0.21}$  \\ 
  \hline  
\multicolumn{7}{c}{stellar parameters} \\
\hline
   & Aa & Ab &  B & Aa & Ab &  B \\
  \hline
 \multicolumn{7}{c}{Relative quantities} \\
  \hline
 fractional radius [$R/a$]  & $0.0599_{-0.0022}^{+0.0019}$ & $0.0421_{-0.0020}^{+0.0015}$ & $0.0076_{-0.0003}^{+0.0003}$ & $0.1456_{-0.0044}^{+0.0039}$ & $0.1421_{-0.0030}^{+0.0037}$ & $0.0631_{-0.0021}^{+0.0026}$ \\
 temperature relative to $(T_\mathrm{eff})_\mathrm{Aa}$ & $1$ & $0.7276_{-0.0123}^{+0.0151}$ & $0.9542_{-0.0192}^{+0.0189}$ & $1$ & $0.9870_{-0.0202}^{+0.0155}$ & $1.0349_{-0.0183}^{+0.0165}$ \\
 fractional flux [in \textit{TESS}-band] & $0.5117_{-0.0300}^{+0.0333}$ & $0.0678_{-0.0050}^{+0.0074}$ & $0.3647_{-0.0320}^{+0.0335}$ & $0.0107_{-0.0008}^{+0.0009}$ & $0.0097_{-0.0008}^{+0.0006}$ & $0.9308_{-0.0493}^{+0.0290}$ \\
 \hline
 \multicolumn{7}{c}{Physical quantities} \\
  \hline 
 $M$ [M$_\odot$] & $0.854_{-0.072}^{+0.042}$ & $0.584_{-0.044}^{+0.040}$ & $0.799_{-0.061}^{+0.059}$ & $0.897_{-0.039}^{+0.030}$ & $0.873_{-0.026}^{+0.032}$ & $1.753_{-0.068}^{+0.053}$ \\
 $R$ [R$_\odot$] & $0.835_{-0.055}^{+0.026}$ & $0.582_{-0.041}^{+0.032}$ & $0.763_{-0.042}^{+0.041}$ & $0.807_{-0.033}^{+0.028}$ & $0.786_{-0.023}^{+0.029}$ & $6.932_{-0.293}^{+0.340}$ \\
 $T_\mathrm{eff}$ [K]& $5440_{-191}^{+157}$ & $3945_{-84}^{+87}$ & $5169_{-119}^{+147}$ & $5185_{-111}^{+110}$ & $5115_{-114}^{+100}$ & $5372_{-149}^{+111}$ \\
 $L_\mathrm{bol}$ [L$_\odot$] & $0.546_{-0.114}^{+0.100}$ & $0.073_{-0.013}^{+0.011}$ & $0.373_{-0.064}^{+0.077}$ & $0.422_{-0.065}^{+0.070}$ & $0.379_{-0.53}^{+0.061}$ & $35.6_{-5.2}^{+6.6}$ \\
 $M_\mathrm{bol}$ & $5.43_{-0.18}^{+0.26}$ & $7.61_{-0.16}^{+0.21}$ & $5.84_{-0.20}^{+0.20}$ & $5.71_{-0.17}^{+0.18}$ & $5.82_{-0.16}^{+0.16}$ & $0.89_{-0.19}^{+0.17}$ \\
 $M_V           $ & $5.56_{-0.22}^{+0.31}$ & $8.72_{-0.23}^{+0.26}$ & $6.06_{-0.25}^{+0.25}$ & $5.92_{-0.21}^{+0.23}$ & $6.06_{-0.20}^{+0.22}$ & $1.19_{-0.13}^{+0.17}$ \\
 $\log g$ [dex] & $4.533_{-0.024}^{+0.021}$ & $4.673_{-0.016}^{+0.029}$ & $4.576_{-0.018}^{+0.018}$ & $4.576_{-0.016}^{+0.017}$ & $4.587_{-0.015}^{+0.013}$ & $2.998_{-0.034}^{+0.025}$ \\
 \hline
\multicolumn{7}{c}{Global system parameters} \\
  \hline
$\log$(age) [dex] &\multicolumn{3}{c}{$9.802_{-0.113}^{+0.182}$} & \multicolumn{3}{c}{$9.275_{-0.036}^{+0.033}$} \\
$[M/H]$  [dex]    &\multicolumn{3}{c}{$-0.072_{-0.169}^{+0.122}$} & \multicolumn{3}{c}{$0.174_{-0.067}^{+0.023}$} \\
$E(B-V)$ [mag]    &\multicolumn{3}{c}{$0.334_{-0.060}^{+0.048}$} & \multicolumn{3}{c}{$0.977_{-0.023}^{+0.029}$} \\
extra light $\ell_4$ [in \textit{TESS}-band] & \multicolumn{3}{c}{$0.054_{-0.022}^{+0.020}$} & \multicolumn{3}{c}{$0.049_{-0.029}^{+0.050}$} \\
$(M_V)_\mathrm{tot}$  &\multicolumn{3}{c}{$4.96_{-0.17}^{+0.29}$} & \multicolumn{3}{c}{$1.17_{-0.13}^{+0.17}$} \\
distance [pc]           & \multicolumn{3}{c}{$629_{-48}^{+26}$} & \multicolumn{3}{c}{$740_{-34}^{+41}$} \\  
\hline
\end{tabular}}
\end{table*}

\begin{table*}
 \centering
\caption{Orbital and astrophysical parameters of TICs\,405789362 and 461500036 from the joint photodynamical light curve, ETV, SED and \texttt{PARSEC} isochrone solution. }
 \label{tab:syntheticfit_TIC405789362461500036}
\scalebox{0.91}{\begin{tabular}{@{}lllllll}
\hline
 & \multicolumn{3}{c}{TIC\,405789362} &  \multicolumn{3}{c}{TIC\,461500036}\\
\hline
\multicolumn{7}{c}{orbital elements} \\
\hline
   & \multicolumn{6}{c}{subsystem}  \\
   & \multicolumn{2}{c}{Aa--Ab} & A--B & \multicolumn{2}{c}{Aa--Ab} & A--B  \\
  \hline
  $t_0$ [BJD - 2400000]& \multicolumn{3}{c}{$58738.0$} & \multicolumn{3}{c}{$58790.5$} \\
  $P$ [days] & \multicolumn{2}{c}{$4.56619_{-0.00021}^{+0.00025}$} & $46.8103_{-0.0025}^{+0.0032}$ &  \multicolumn{2}{c}{$2.469953_{-0.000022}^{+0.000019}$} & $54.56390_{-0.00068}^{+0.00070}$ \\
  $a$ [R$_\odot$] & \multicolumn{2}{c}{$16.34_{-0.23}^{+0.35}$} & $84.0_{-1.2}^{+1.8}$ & \multicolumn{2}{c}{$10.59_{-0.30}^{+0.10}$} & $91.21_{-2.53}^{+0.85}$ \\
  $e$ & \multicolumn{2}{c}{$0.00719_{-0.00027}^{+0.00022}$} & $0.1623_{-0.0021}^{+0.0029}$ & \multicolumn{2}{c}{$0.00129_{-0.00012}^{+0.00014}$} & $0.0320_{-0.0009}^{+0.0011}$ \\
  $\omega$ [deg]& \multicolumn{2}{c}{$329.8_{-2.7}^{+2.3}$} & $351.24_{-0.84}^{+0.78}$ & \multicolumn{2}{c}{$139_{-15}^{+14}$} & $188.00_{-0.26}^{+0.31}$ \\ 
  $i$ [deg] & \multicolumn{2}{c}{$89.60_{-0.56}^{+0.57}$} & $89.72_{-0.10}^{+0.13}$ & \multicolumn{2}{c}{$88.71_{-0.11}^{+0.11}$} & $89.70_{-0.07}^{+0.07}$ \\
  $\mathcal{T}_0^\mathrm{inf/sup}$ [BJD - 2400000]& \multicolumn{2}{c}{$58741.1492_{-0.0005}^{+0.0004}$} & $58753.797_{-0.035}^{+0.023}$ & \multicolumn{2}{c}{$58791.3937_{-0.0002}^{+0.0002}$} & $58842.8333_{-0.0086}^{+0.0081}$ \\
  $\tau$ [BJD - 2400000]& \multicolumn{2}{c}{$58737.333_{-0.034}^{+0.029}$} & $58715.211_{-0.102}^{+0.084}$ & \multicolumn{2}{c}{$58790.49_{-0.11}^{+0.10}$} & $58835.647_{-0.042}^{+0.051}$ \\
  $\Omega$ [deg] & \multicolumn{2}{c}{$0.0$} & $-1.26_{-0.40}^{+0.55}$ & \multicolumn{2}{c}{$0.0$} & $0.92_{-0.20}^{+0.15}$ \\
  $i_\mathrm{mut}$ [deg] & \multicolumn{3}{c}{$1.36_{-0.37}^{+0.43}$} & \multicolumn{3}{c}{$1.36_{-0.17}^{+0.16}$} \\
  $\varpi^\mathrm{dyn}$ [deg]& \multicolumn{2}{c}{$149.8_{-2.7}^{+2.3}$} & $171.23_{-0.84}^{+0.78}$ & \multicolumn{2}{c}{$319_{-15}^{+14}$} & $8.00_{-0.26}^{+0.31}$ \\
  $i^\mathrm{dyn}$ [deg] & \multicolumn{2}{c}{$0.95_{-0.26}^{+0.30}$} & $0.41_{-0.11}^{+0.13}$ & \multicolumn{2}{c}{$1.02_{-0.13}^{+0.12}$} & $0.34_{-0.04}^{+0.04}$ \\
  $\Omega^\mathrm{dyn}$ [deg] & \multicolumn{2}{c}{$276_{-28}^{+28}$} & $96_{-28}^{+28}$ & \multicolumn{2}{c}{$42.7_{-7.1}^{+6.3}$} & $222.7_{-7.1}^{+6.3}$ \\
  $i_\mathrm{inv}$ [deg] & \multicolumn{3}{c}{$89.72_{-0.22}^{+0.14}$} & \multicolumn{3}{c}{$89.45_{-0.05}^{+0.05}$} \\
  $\Omega_\mathrm{inv}$ [deg] & \multicolumn{3}{c}{$-0.88_{-0.28}^{+0.39}$} & \multicolumn{3}{c}{$0.69_{-0.15}^{+0.11}$} \\
  \hline
  mass ratio $[q=M_\mathrm{sec}/M_\mathrm{pri}]$ & \multicolumn{2}{c}{$0.837_{-0.006}^{+0.006}$} & $0.293_{-0.007}^{+0.007}$ & \multicolumn{2}{c}{$0.962_{-0.007}^{+0.008}$} & $0.309_{-0.005}^{+0.006}$ \\
  $K_\mathrm{pri}$ [km\,s$^{-1}$] & \multicolumn{2}{c}{$82.5_{-1.4}^{+1.8}$} & $20.90_{-0.58}^{+0.61}$ & \multicolumn{2}{c}{$106.1_{-2.4}^{+1.2}$} & $20.97_{-0.29}^{+0.33}$ \\ 
  $K_\mathrm{sec}$ [km\,s$^{-1}$] & \multicolumn{2}{c}{$98.7_{-1.6}^{+1.8}$} & $71.25_{-1.09}^{+1.36}$ & \multicolumn{2}{c}{$110.9_{-4.0}^{+1.1}$} & $68.28_{-2.20}^{+0.79}$ \\ 
  \hline
  \multicolumn{7}{c}{Apsidal and nodal motion related parameters} \\
  \hline
$P_\mathrm{apse}$ [year] & \multicolumn{2}{c}{$7.34_{-0.14}^{+0.14}$} & $17.28_{-0.07}^{+0.07}$ & \multicolumn{2}{c}{$13.12_{-0.16}^{+0.15}$} & $47.64_{-0.15}^{+0.16}$ \\ 
$P_\mathrm{apse}^\mathrm{dyn}$ [year] & \multicolumn{2}{c}{$3.04_{-0.05}^{+0.05}$} & $3.99_{-0.04}^{+0.04}$ & \multicolumn{2}{c}{$6.24_{-0.07}^{+0.06}$} & $9.51_{-0.07}^{+0.06}$ \\ 
$P_\mathrm{node}^\mathrm{dyn}$ [year] & \multicolumn{3}{c}{$5.20_{-0.07}^{+0.06}$} & \multicolumn{3}{c}{$11.89_{-0.11}^{+0.12}$} \\
$\Delta\omega_\mathrm{3b}$ [arcsec/cycle] & \multicolumn{2}{c}{$5296_{-80}^{+82}$} & $_{-}^{+}$ & \multicolumn{2}{c}{$1289_{-14}^{+16}$} & $8906_{-75}^{+82}$ \\ 
$\Delta\omega_\mathrm{GR}$ [arcsec/cycle] & \multicolumn{2}{c}{$1.419_{-0.039}^{+0.061}$} & $0.366_{-0.010}^{+0.016}$ & \multicolumn{2}{c}{$2.038_{-0.116}^{+0.040}$} & $0.345_{-0.019}^{+0.006}$ \\ 
$\Delta\omega_\mathrm{tide}$ [arcsec/cycle] & \multicolumn{2}{c}{$29.1_{-2.7}^{+3.3}$} & $0.154_{-0.014}^{+0.017}$ & \multicolumn{2}{c}{$113.7_{-2.1}^{+2.1}$} & $0.228_{-0.004}^{+0.004}$  \\ 
  \hline  
\multicolumn{7}{c}{stellar parameters} \\
\hline
   & Aa & Ab &  B & Aa & Ab &  B \\
  \hline
 \multicolumn{7}{c}{Relative quantities} \\
  \hline
 fractional radius [$R/a$]  & $0.1269_{-0.0024}^{+0.0025}$ & $0.0841_{-0.0032}^{+0.0039}$ & $0.00881_{-0.00032}^{+0.00036}$ & $0.1520_{-0.0012}^{+0.0010}$ & $0.1372_{-0.0014}^{+0.0016}$ & $0.00794_{-0.00014}^{+0.00017}$ \\
 temperature relative to $(T_\mathrm{eff})_\mathrm{Aa}$ & $1$ & $0.9776_{-0.0137}^{+0.0104}$ & $0.7522_{-0.0134}^{+0.0137}$ & $1$ & $0.9942_{-0.0027}^{+0.0026}$ & $0.7624_{-0.0099}^{+0.0085}$ \\
 fractional flux [in \textit{TESS}-band] & $0.6505_{-0.0286}^{+0.0331}$ & $0.2667_{-0.0056}^{+0.0060}$ & $0.0289_{-0.0038}^{+0.0041}$ & $0.5326_{-0.0098}^{+0.0080}$ & $0.4262_{-0.0052}^{+0.0065}$ & $0.0380_{-0.0034}^{+0.0036}$ \\
 fractional flux [in Sloan $r'$-band] & $0.6604_{-0.0227}^{0.0231}$ & $0.2693_{-0.0094}^{0.0084}$ & $0.0254_{-0.0039}^{+0.0042}$ & $0.5107_{-0.0135}^{+0.0154}$ & $0.4078_{-0.0116}^{+0.0176}$ & $0.0319_{-0.0032}^{+0.0035}$ \\
 \hline
 \multicolumn{7}{c}{Physical Quantities} \\
  \hline 
 $M$ [M$_\odot$] & $1.529_{-0.068}^{+0.094}$ & $1.276_{-0.054}^{+0.086}$ & $0.824_{-0.046}^{+0.051}$ & $1.334_{-0.120}^{+0.039}$ & $1.277_{-0.100}^{+0.039}$ & $0.802_{-0.056}^{+0.028}$ \\
 $R$ [R$_\odot$] & $2.074_{-0.060}^{+0.072}$ & $1.373_{-0.069}^{+0.094}$ & $0.740_{-0.037}^{+0.042}$ & $1.609_{-0.044}^{+0.016}$ & $1.453_{-0.056}^{+0.028}$ & $0.724_{-0.035}^{+0.022}$ \\
 $T_\mathrm{eff}$ [K]& $6292_{-103}^{+235}$ & $6151_{-51}^{+134}$ & $4755_{-138}^{+177}$ & $6148_{-39}^{+75}$ & $6114_{-37}^{+64}$ & $4693_{-80}^{+79}$ \\
 $L_\mathrm{bol}$ [L$_\odot$] & $5.99_{-0.55}^{+0.25}$ & $2.40_{-0.26}^{+0.58}$ & $0.251_{-0.048}^{+0.074}$ & $3.30_{-0.20}^{+0.20}$ & $2.63_{-0.22}^{+0.20}$ & $0.227_{-0.031}^{+0.032}$ \\
 $M_\mathrm{bol}$ & $2.83_{-0.21}^{+0.11}$ & $3.82_{-0.24}^{+0.12}$ & $6.27_{-0.28}^{+0.23}$ & $3.47_{-0.06}^{+0.07}$ & $3.72_{-0.08}^{+0.09}$ & $6.38_{-0.14}^{+0.16}$ \\
 $M_V           $ & $2.79_{-0.23}^{+0.12}$ & $3.80_{-0.26}^{+0.13}$ & $6.71_{-0.38}^{+0.33}$ & $3.46_{-0.08}^{+0.08}$ & $3.71_{-0.09}^{+0.11}$ & $6.86_{-0.19}^{+0.21}$ \\
 $\log g$ [dex] & $3.988_{-0.014}^{+0.014}$ & $4.267_{-0.030}^{+0.026}$ & $4.614_{-0.022}^{+0.019}$ & $4.148_{-0.019}^{+0.009}$ & $4.216_{-0.005}^{+0.005}$ & $4.621_{-0.011}^{+0.010}$ \\
 \hline
\multicolumn{7}{c}{Global system parameters} \\
  \hline
$\log$(age) [dex] &\multicolumn{3}{c}{$9.331_{-0.170}^{+0.078}$} & \multicolumn{3}{c}{$9.451_{-0.087}^{+0.168}$} \\
$[M/H]$  [dex]    &\multicolumn{3}{c}{$0.356_{-0.116}^{+0.123}$} & \multicolumn{3}{c}{$0.332_{-0.230}^{+0.104}$} \\
$E(B-V)$ [mag]    &\multicolumn{3}{c}{$0.573_{-0.028}^{+0.052}$} & \multicolumn{3}{c}{$0.294_{-0.013}^{+0.021}$} \\
extra light $\ell_4$ [in \textit{TESS}-band] & \multicolumn{3}{c}{$0.0539_{-0.355}^{+0.0282}$} & \multicolumn{3}{c}{$0.0025_{-0.0018}^{+0.0034}$} \\
extra light $\ell_4$ [in Sloan $r'$-band] & \multicolumn{3}{c}{$0.0437_{-0.0237}^{+0.0244}$} & \multicolumn{3}{c}{$0.050_{-0.034}^{+0.022}$} \\
$(M_V)_\mathrm{tot}$  &\multicolumn{3}{c}{$2.42_{-0.24}^{+0.12}$} & \multicolumn{3}{c}{$2.80_{-0.08}^{+0.09}$} \\
distance [pc]           & \multicolumn{3}{c}{$2107_{-86}^{+112}$} & \multicolumn{3}{c}{$1599_{-63}^{+27}$} \\  
\hline
\end{tabular}}
\end{table*}

\FloatBarrier

\onecolumn
\section{Light curve sections around the third-body eclipses}
\label{app:lightcurve}

In this appendix we plot characteristic light curve sections of nine of the ten investigated systems.

\begin{figure*}[ht]
   \centering
\includegraphics[width=0.30\textwidth]{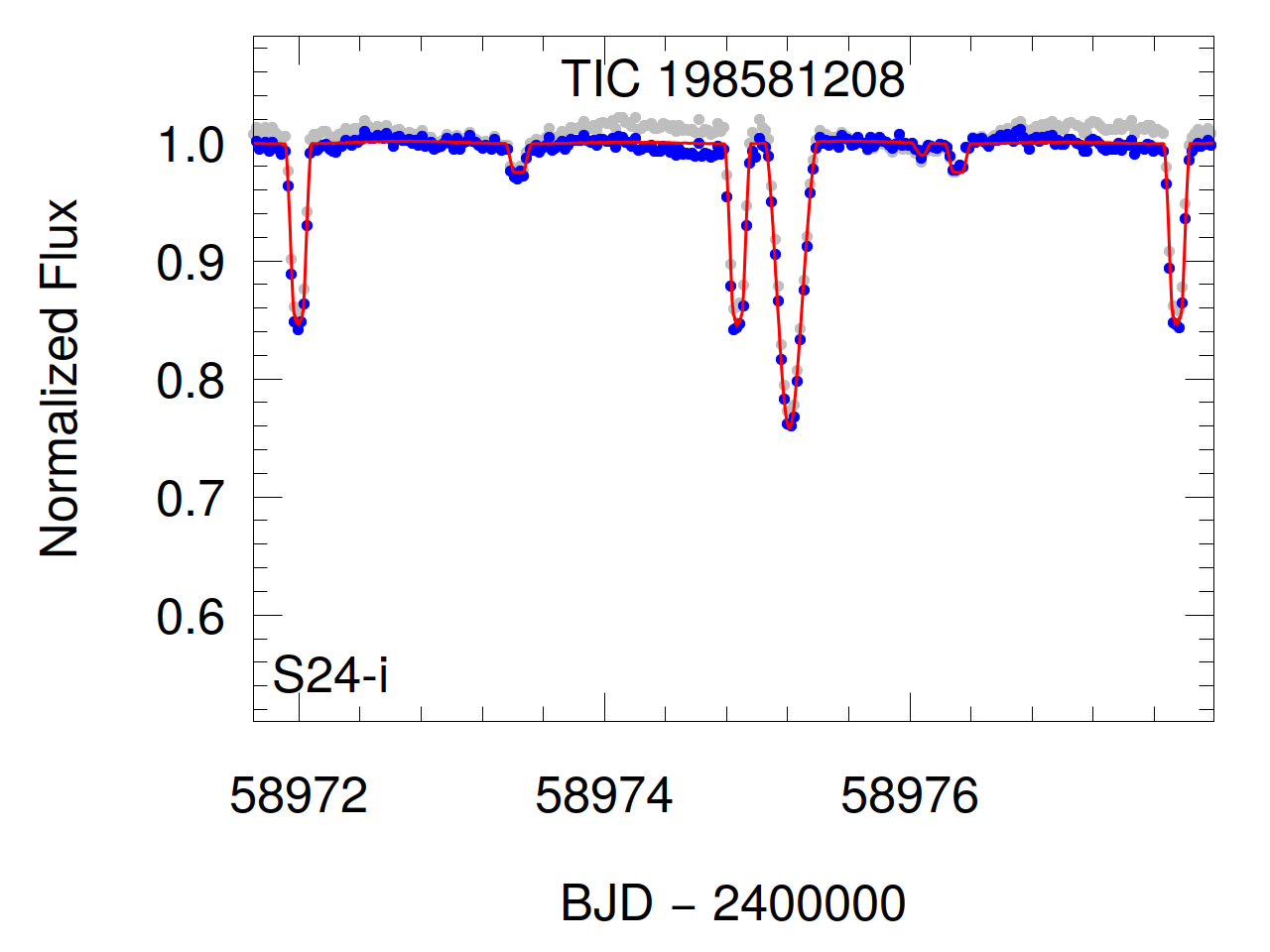}\includegraphics[width=0.30\textwidth]{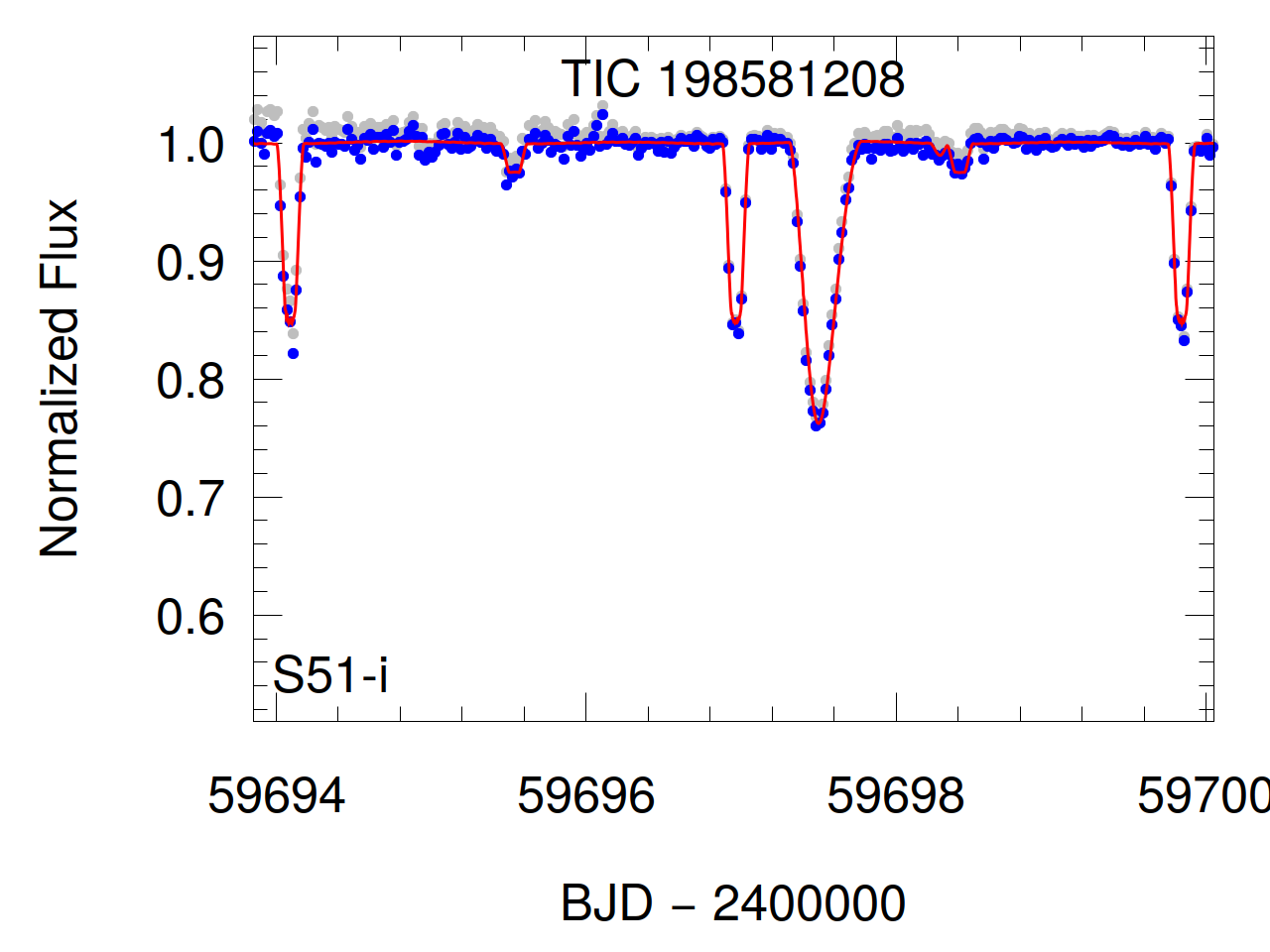}\includegraphics[width=0.30\textwidth]{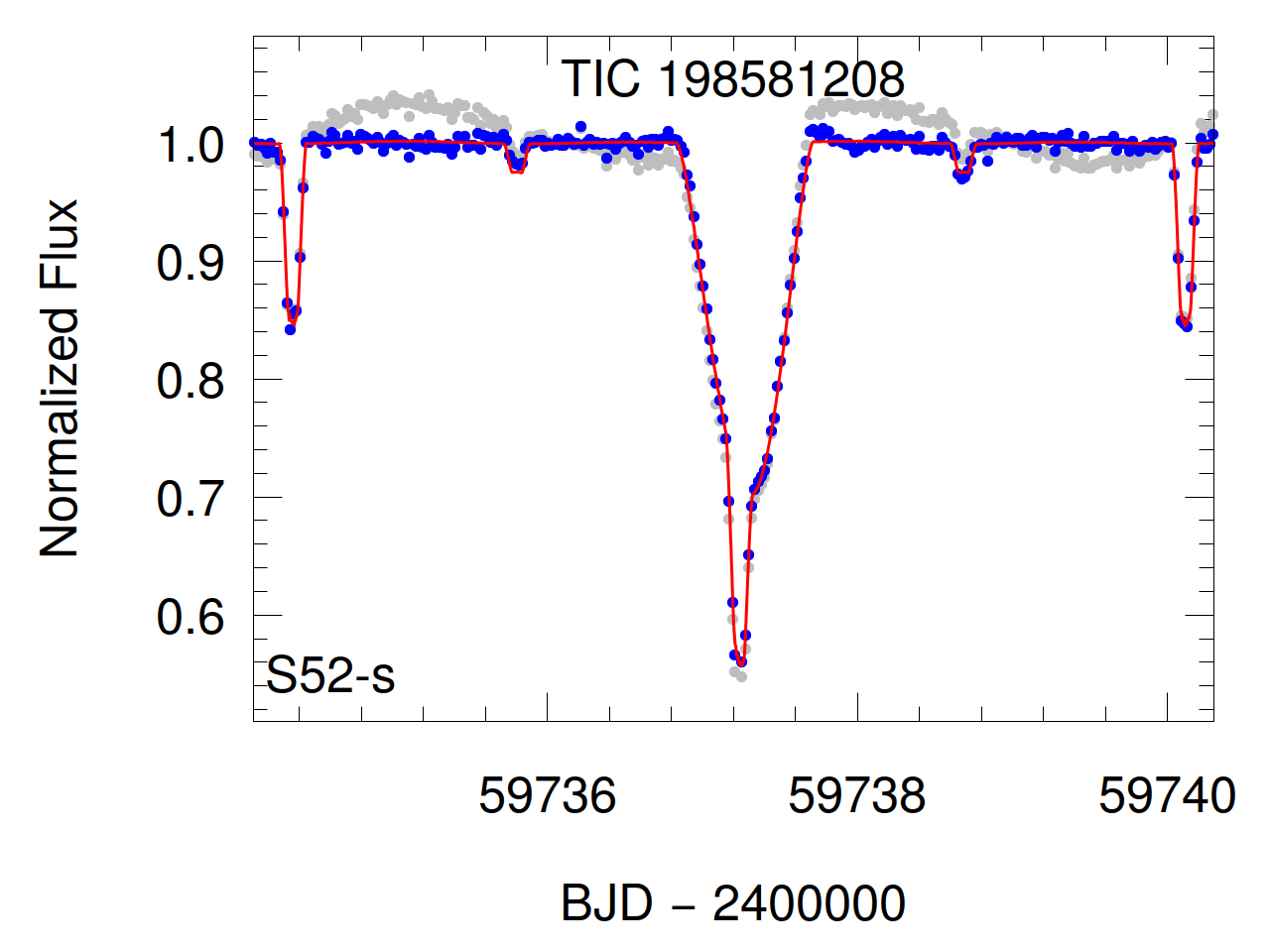}
\includegraphics[width=0.32\textwidth]{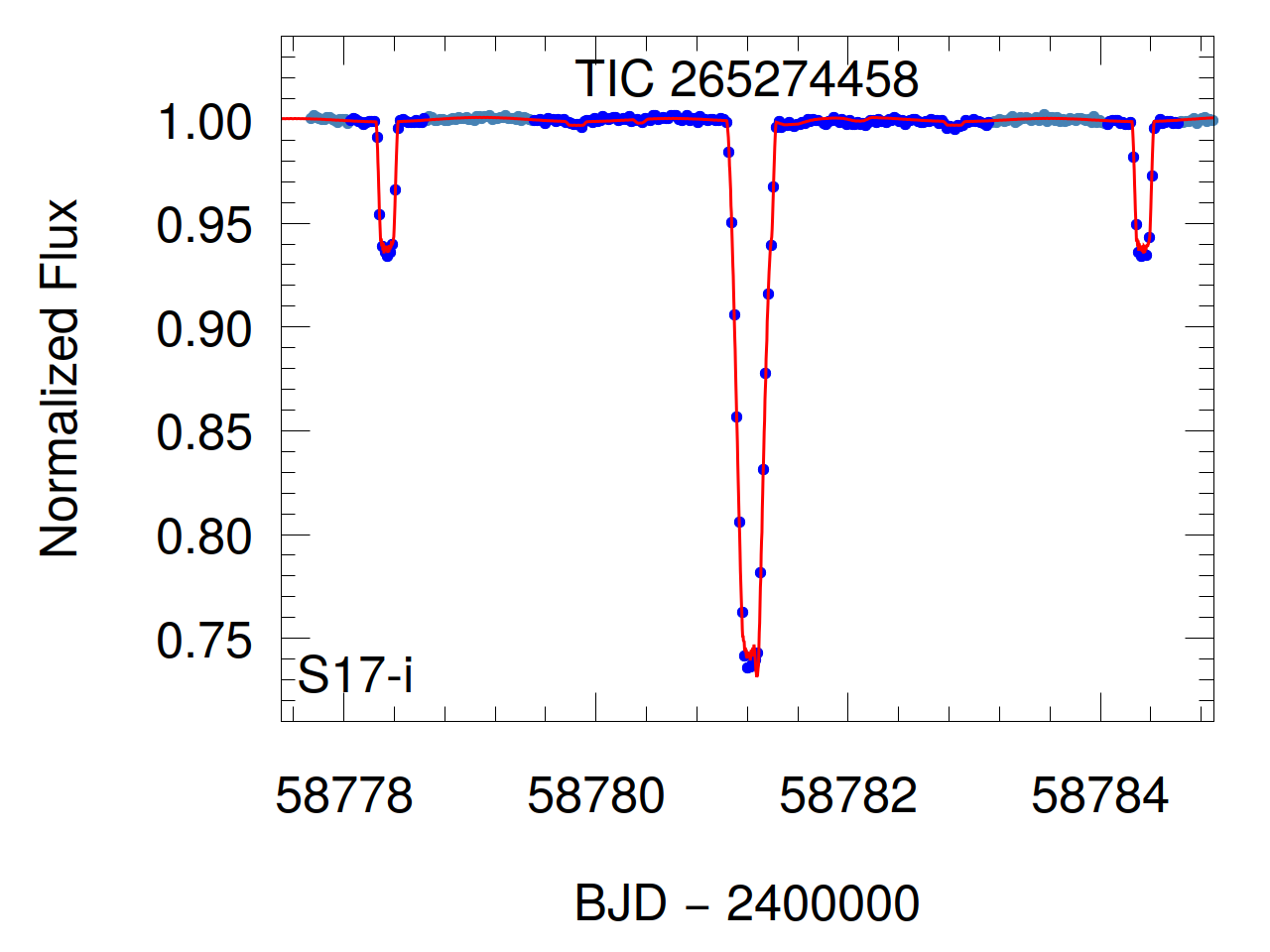}\includegraphics[width=0.32\textwidth]{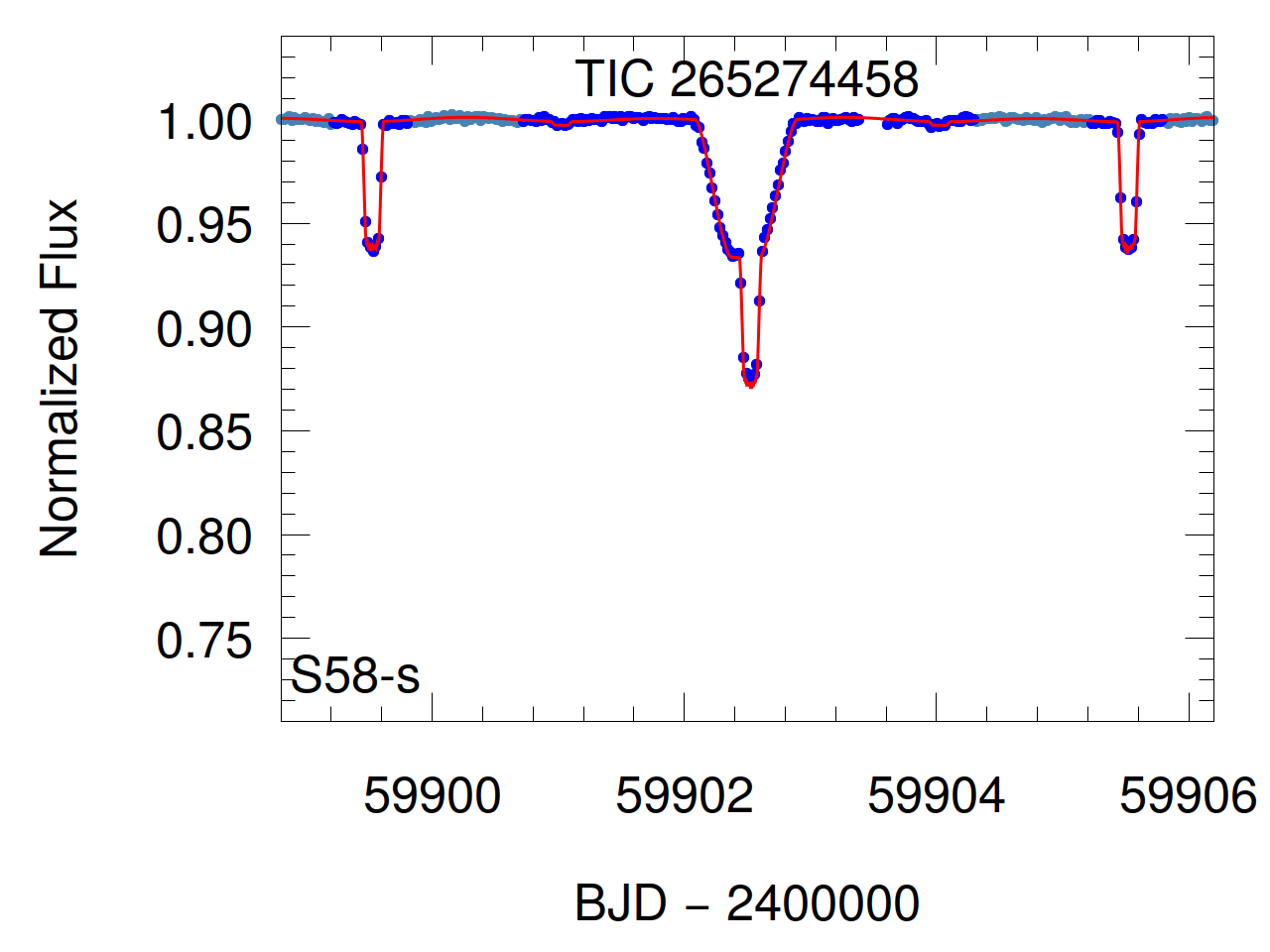}\includegraphics[width=0.32\textwidth]{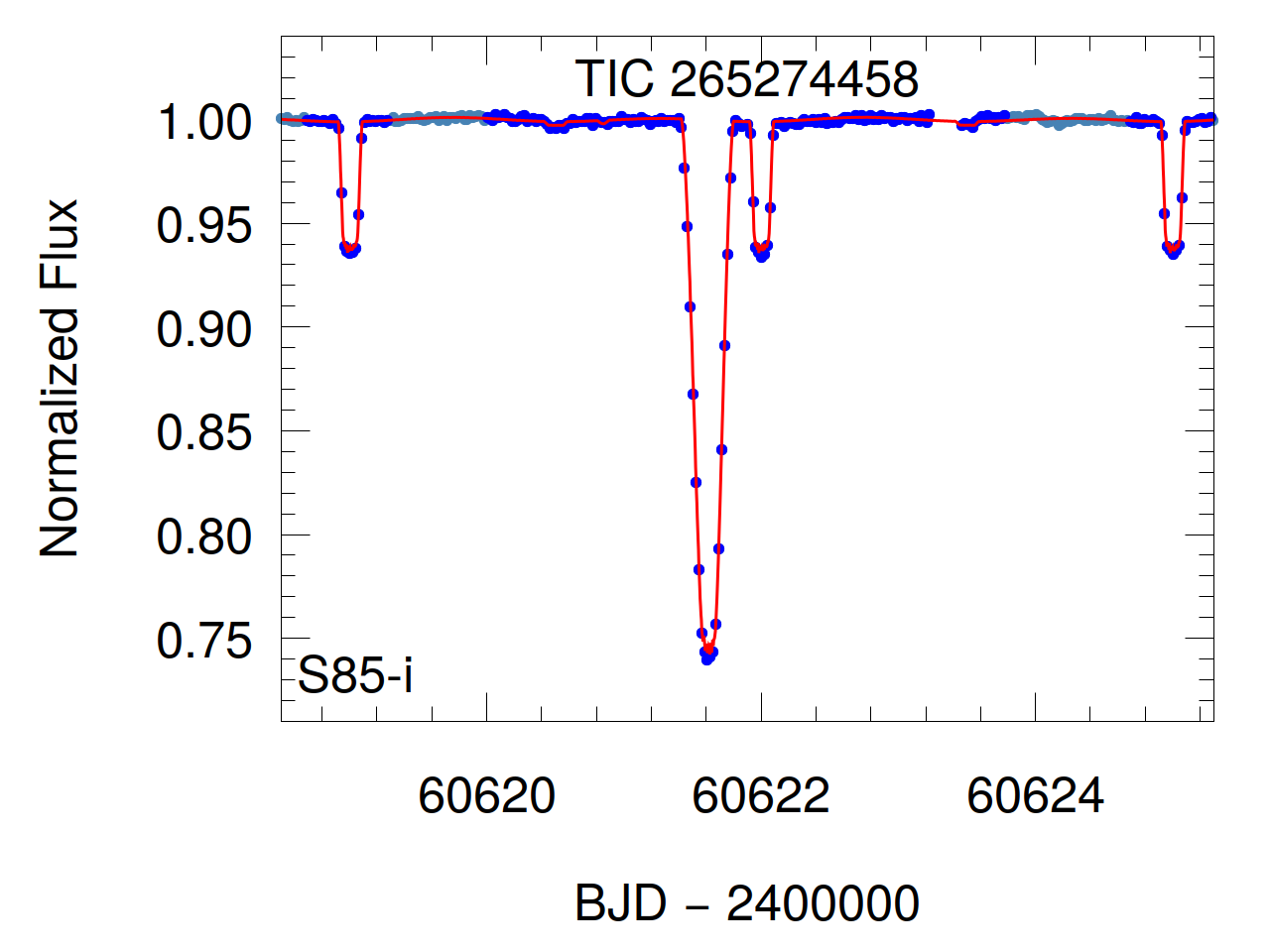}   
\includegraphics[width=0.32\textwidth]{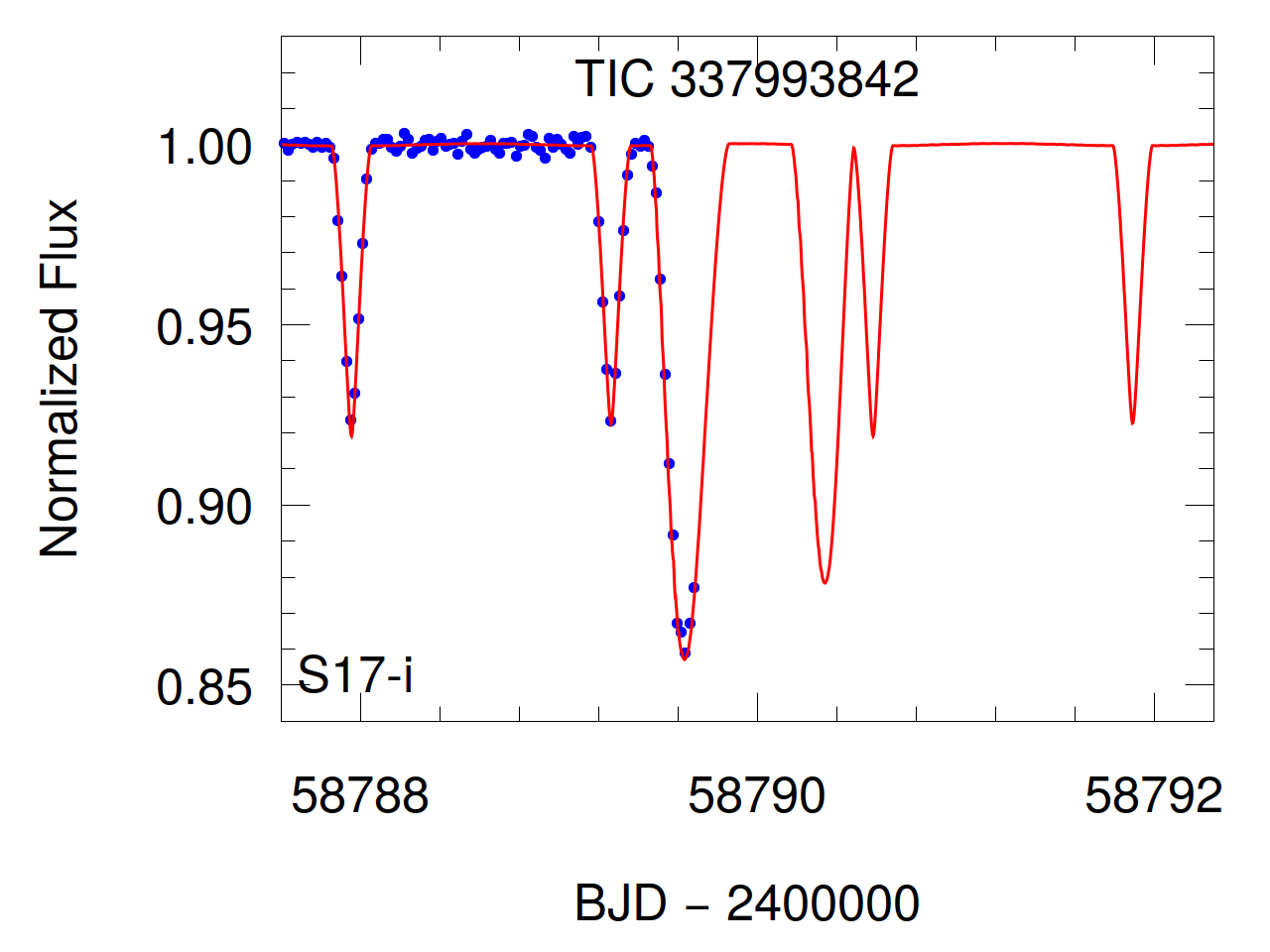}\includegraphics[width=0.32\textwidth]{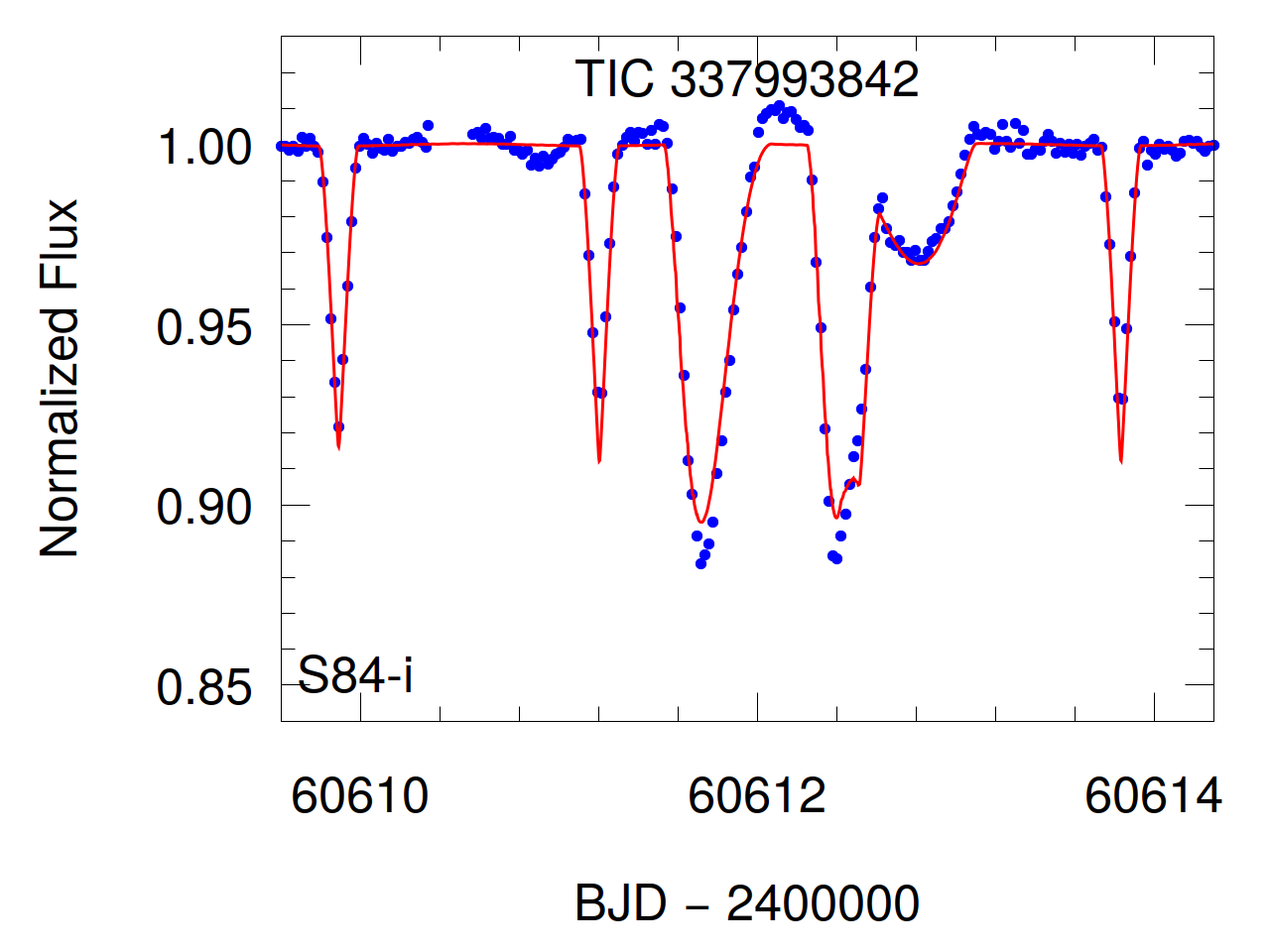}\\
\includegraphics[width=0.32\textwidth]{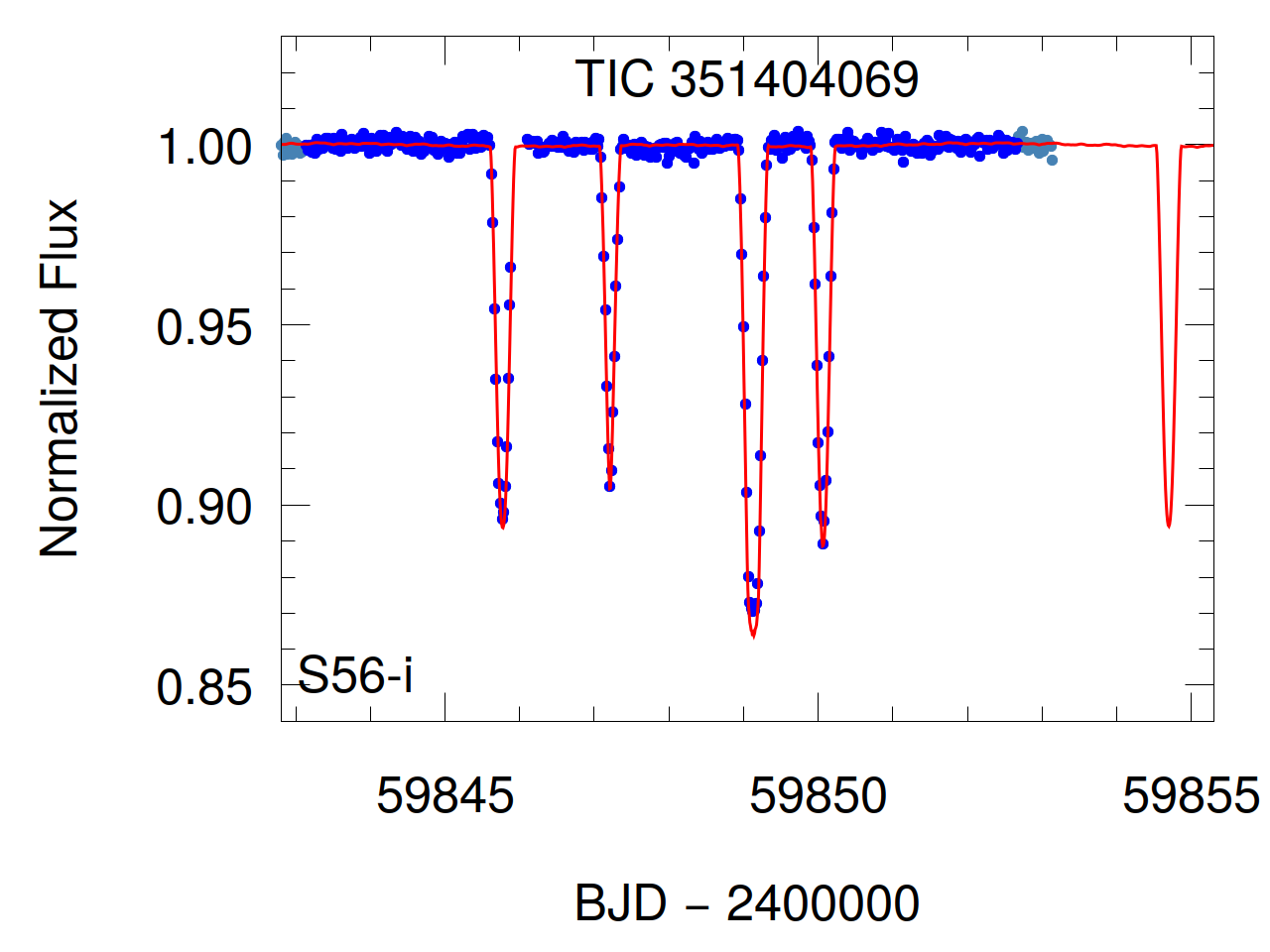}\includegraphics[width=0.32\textwidth]{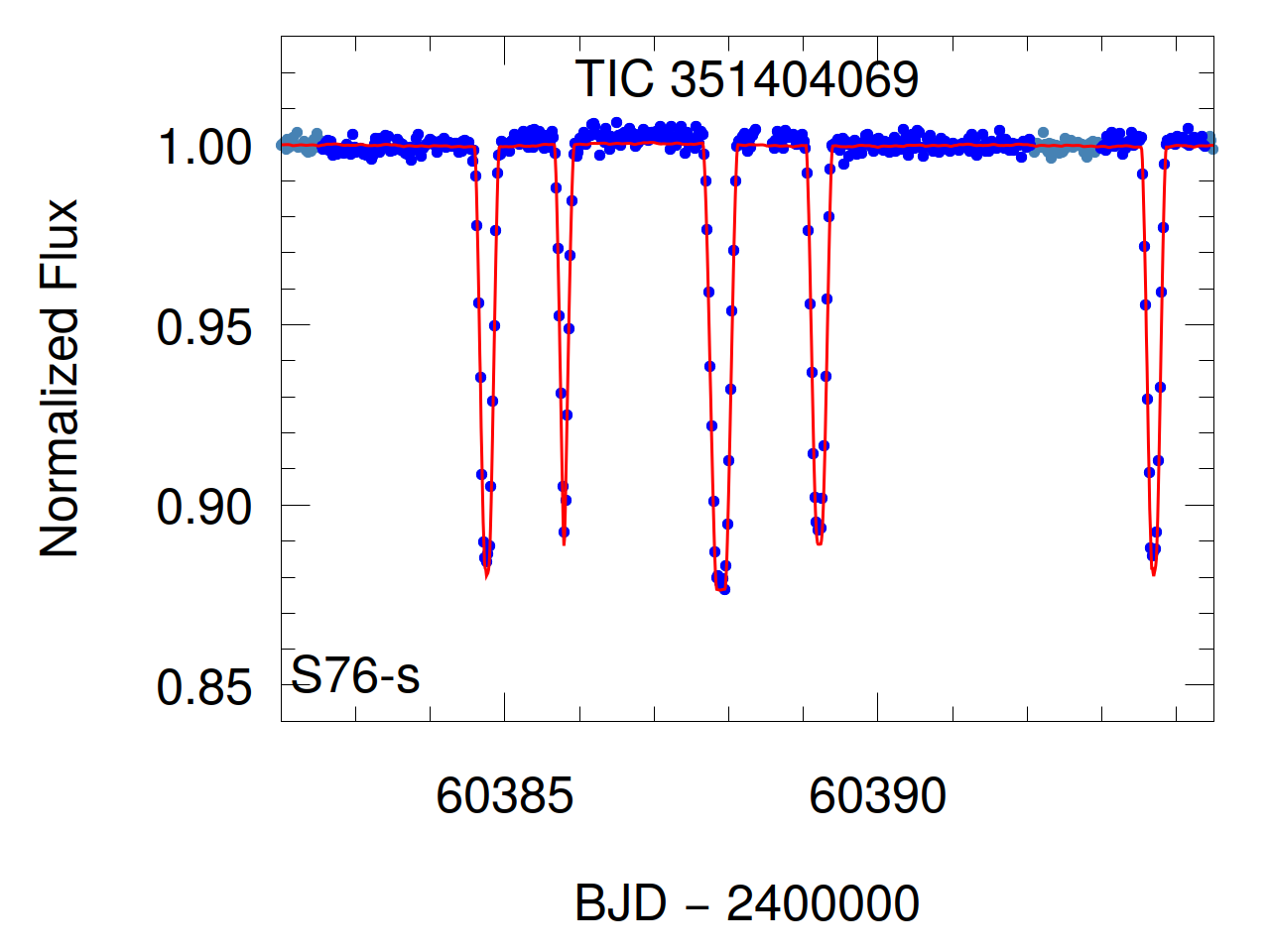}\includegraphics[width=0.32\textwidth]{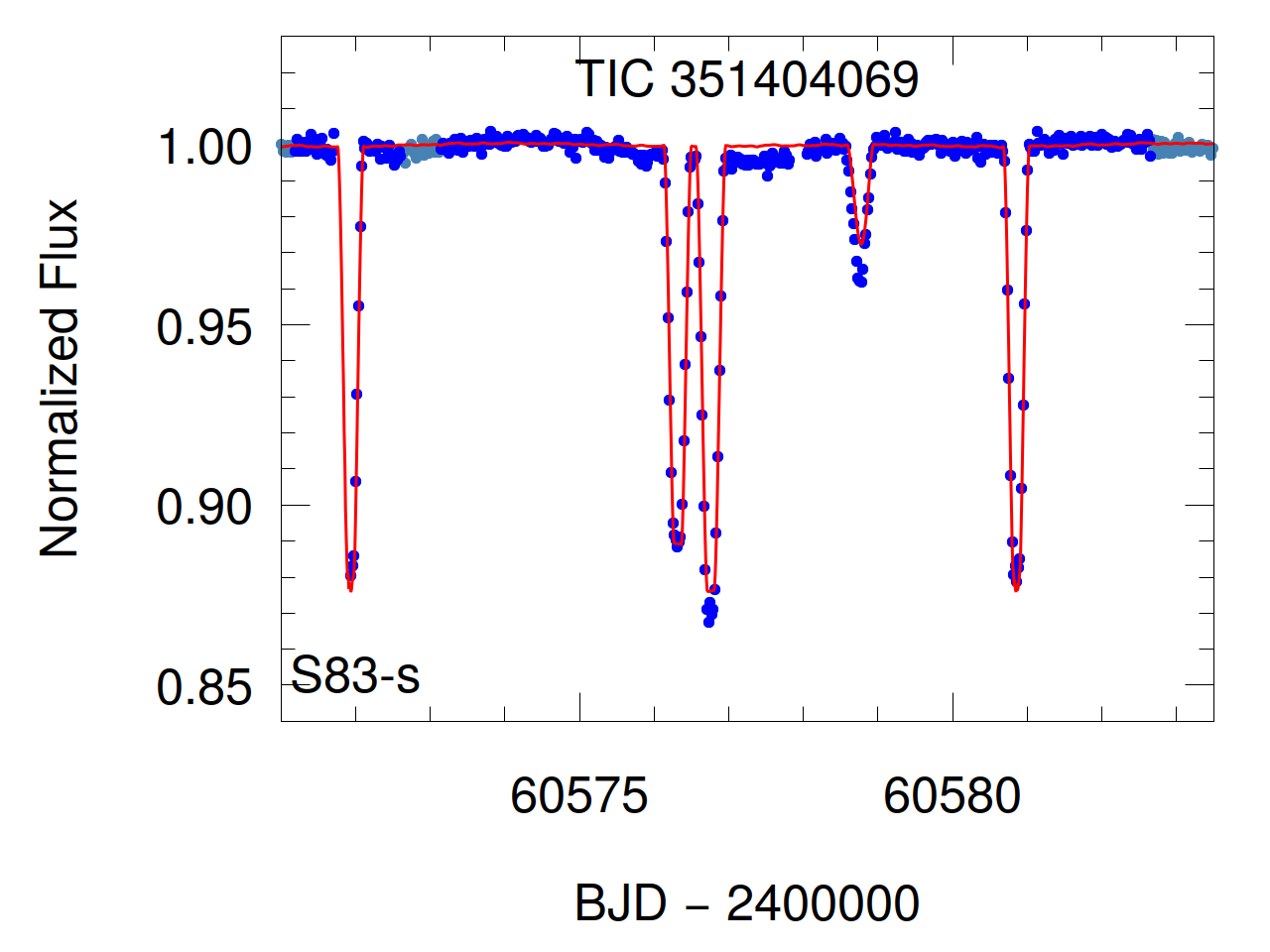}
   \caption{Light curves (blue points) and model fits (smooth red curves) near the third-body eclipses of four targets. {\it From top to bottom panels:} 198581208, 265274458, 337993842, 351404069. The grey points represent original, unsmoothed, but 1800-sec-binned TESS light curves. Dark/pale blue points are for those light curve sections which were used/not-used for the photodynamical solution, after the removal of the likely effects of stellar activity. The sector numbers are indicated in the lower left corner of each panel. Letters `i' or `s' after the sector numbers refer to the inferior or superior conjunction of the third star, respectively.}
   \label{fig:lcs1}
   \end{figure*}  

\begin{figure*}
   \centering
\includegraphics[width=0.32\textwidth]{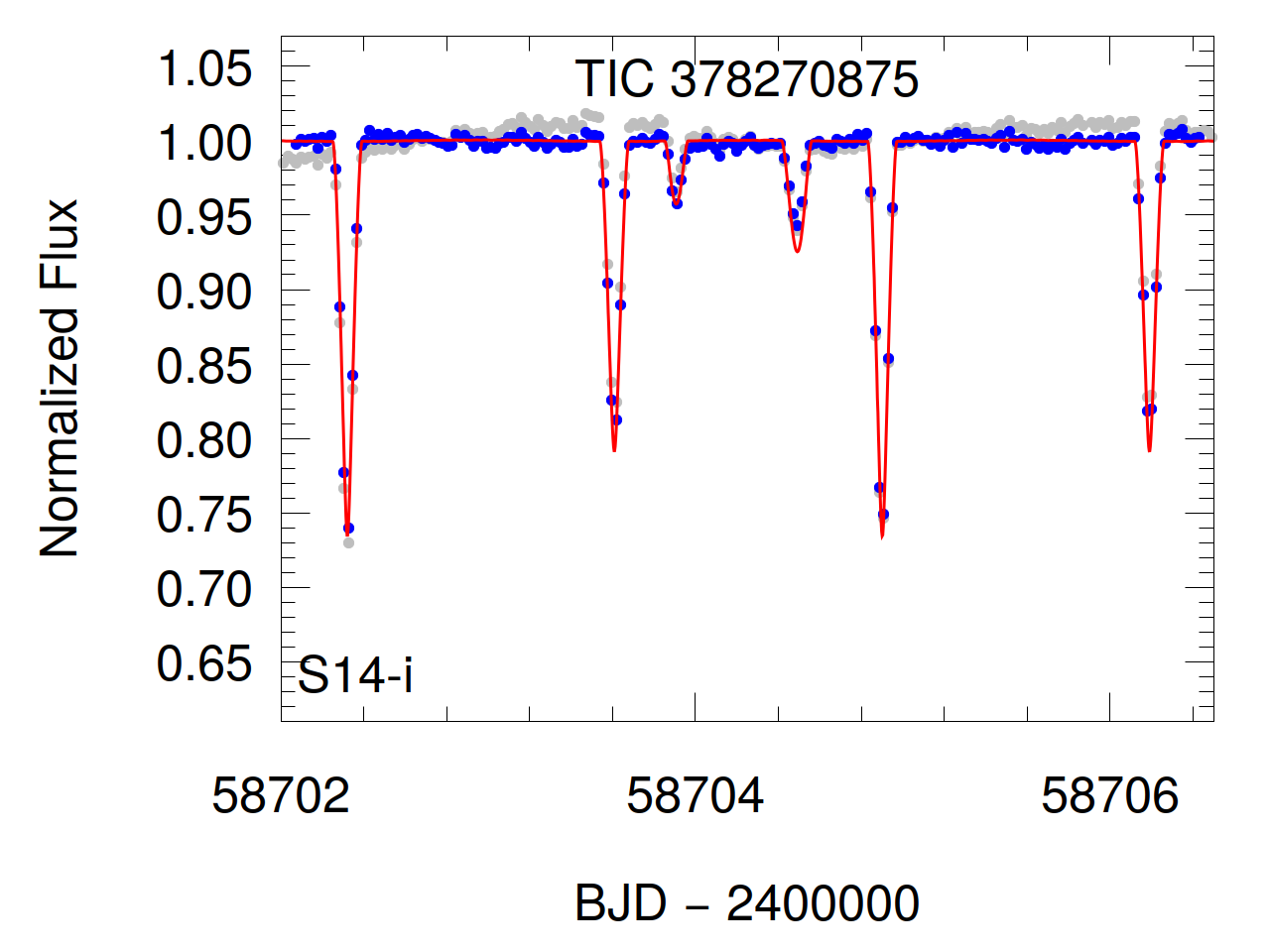}\includegraphics[width=0.32\textwidth]{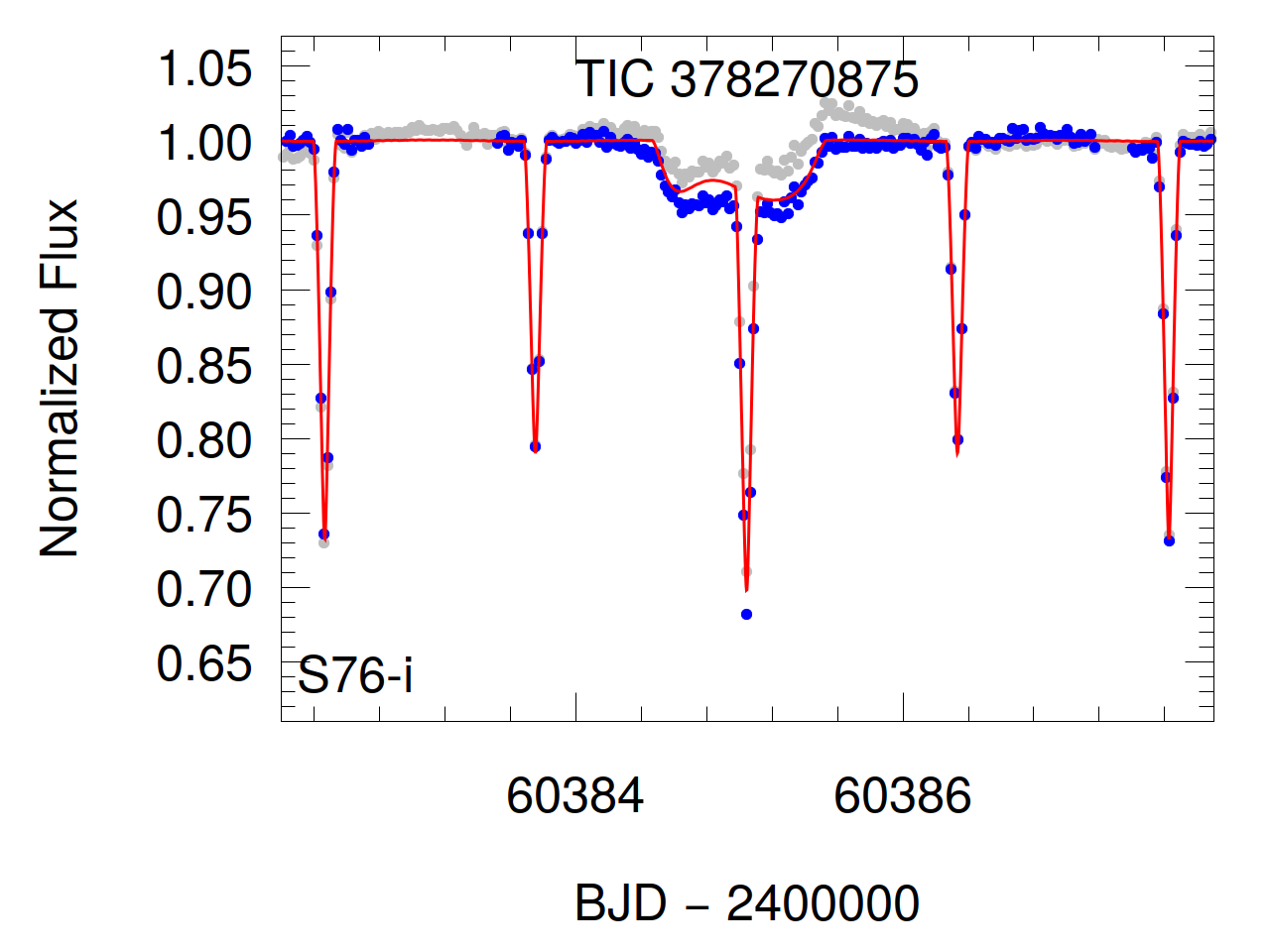}\includegraphics[width=0.32\textwidth]{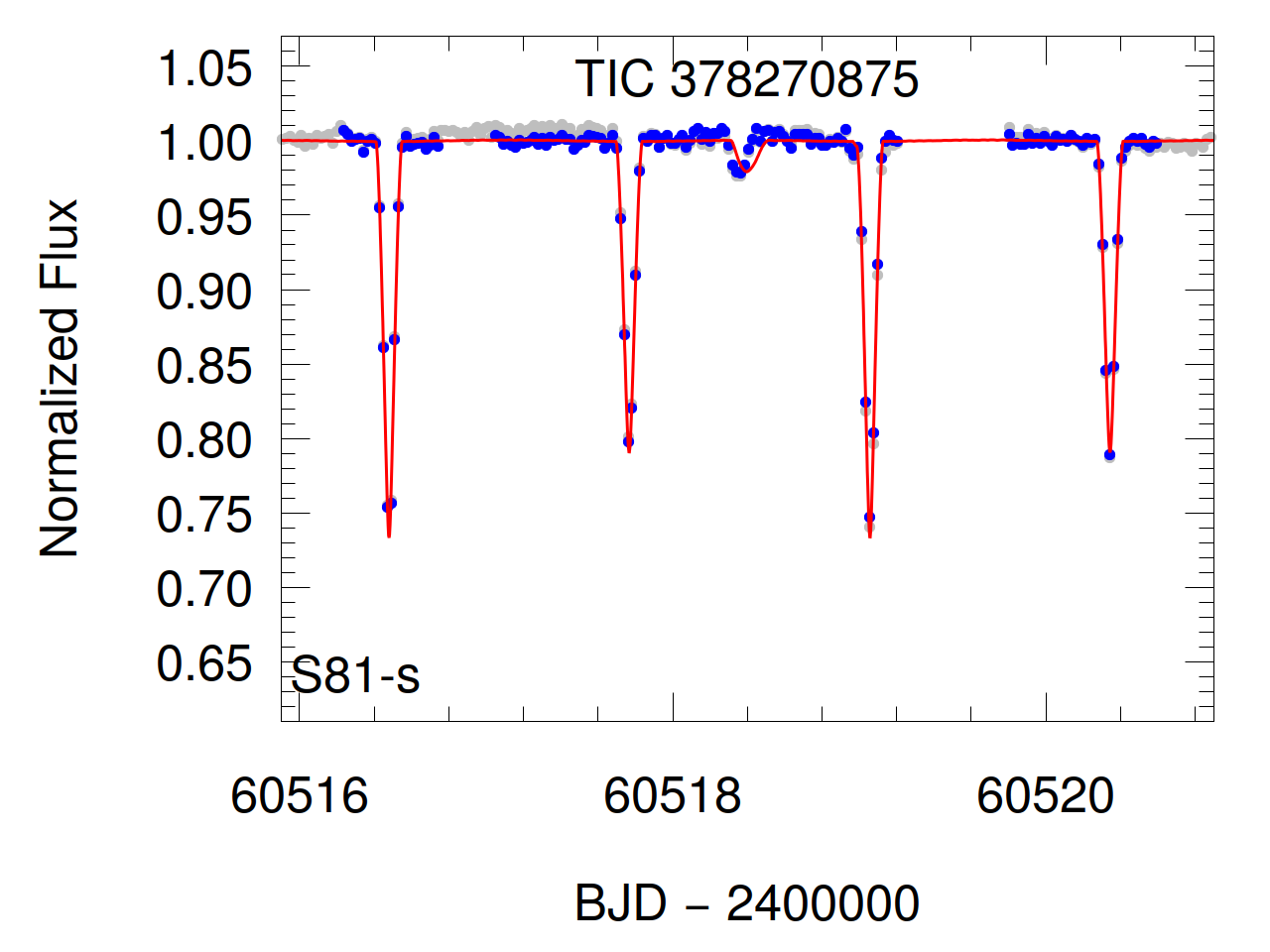}
\includegraphics[width=0.32\textwidth]{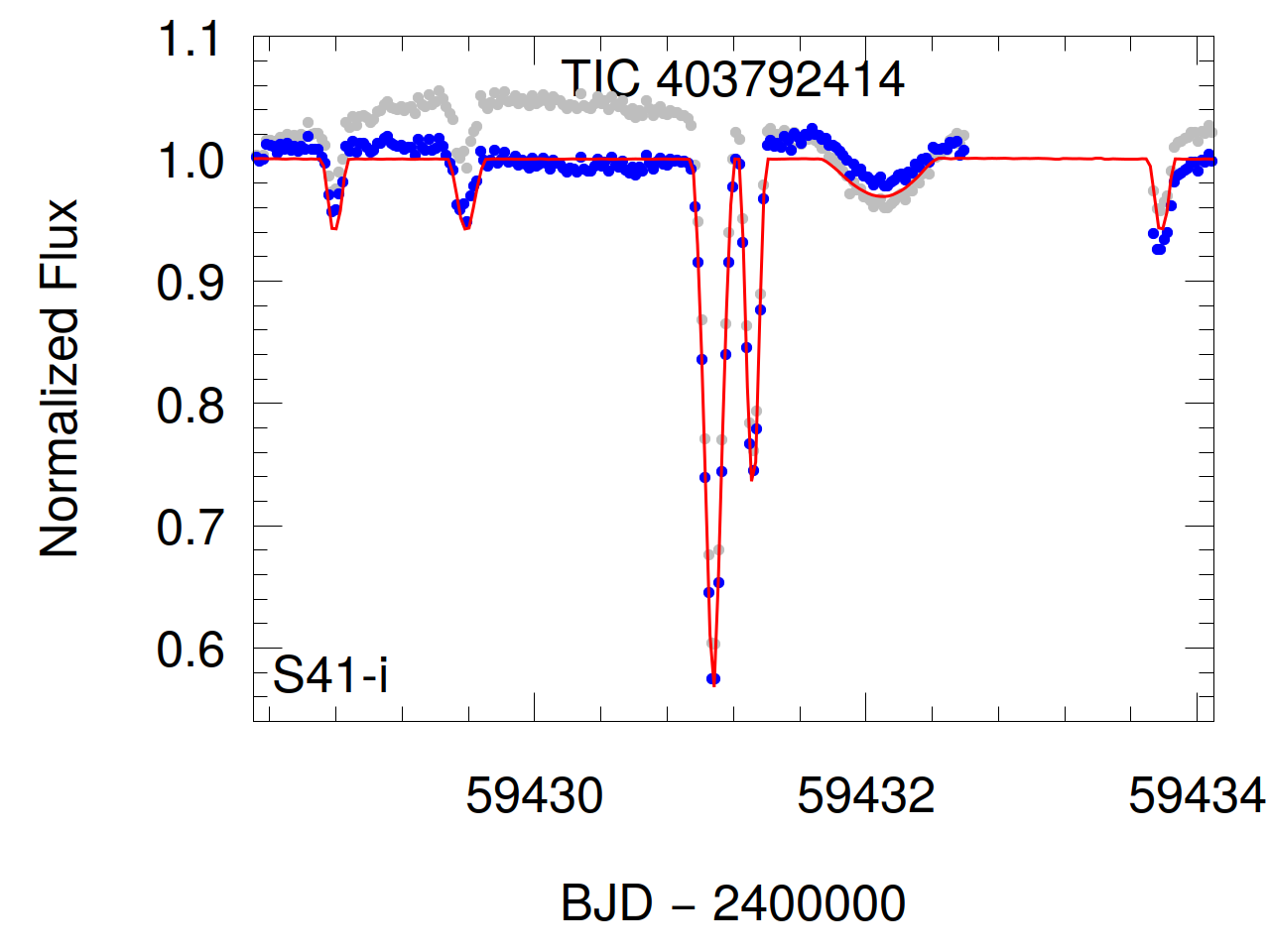}\includegraphics[width=0.32\textwidth]{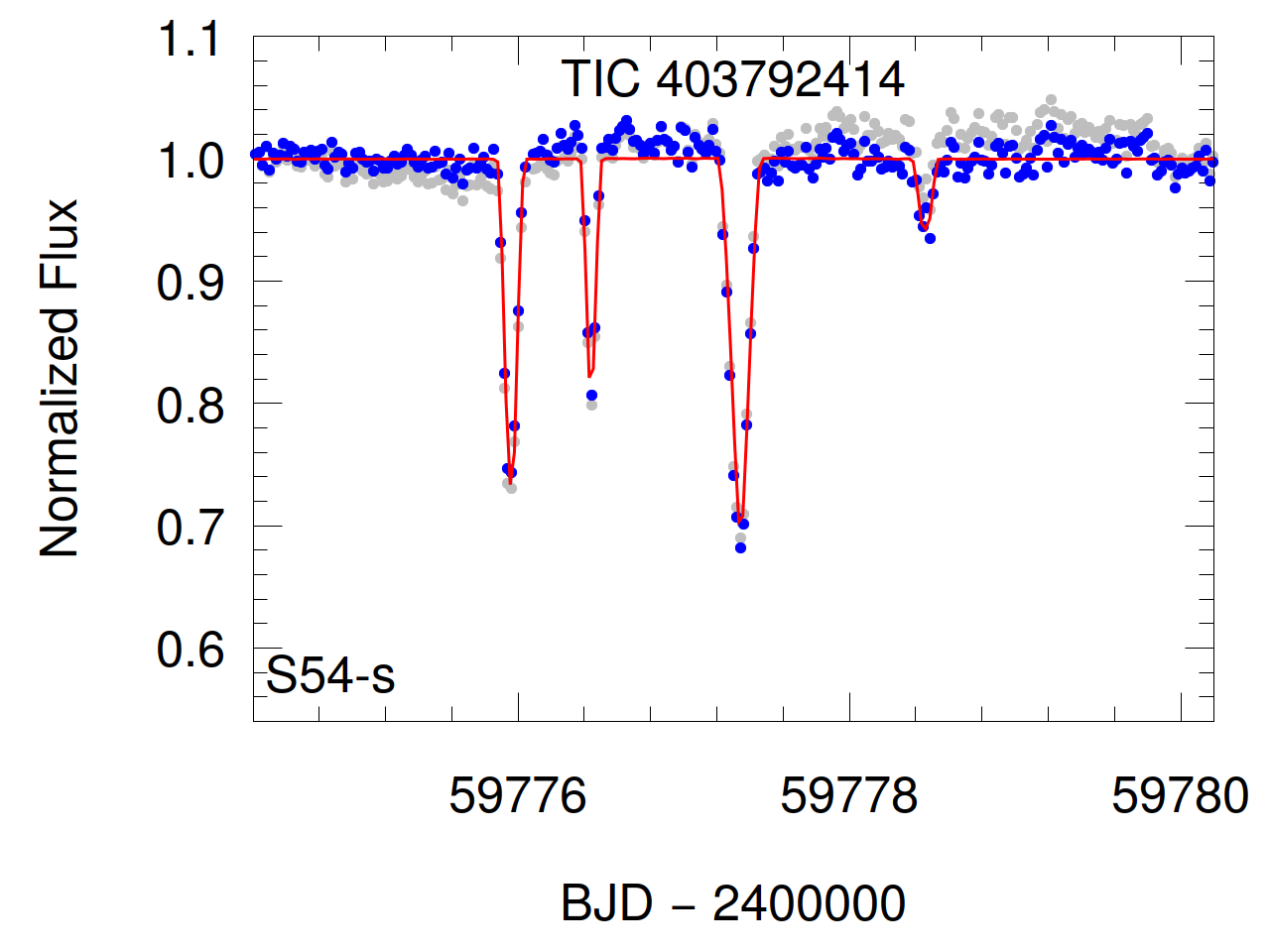}\includegraphics[width=0.32\textwidth]{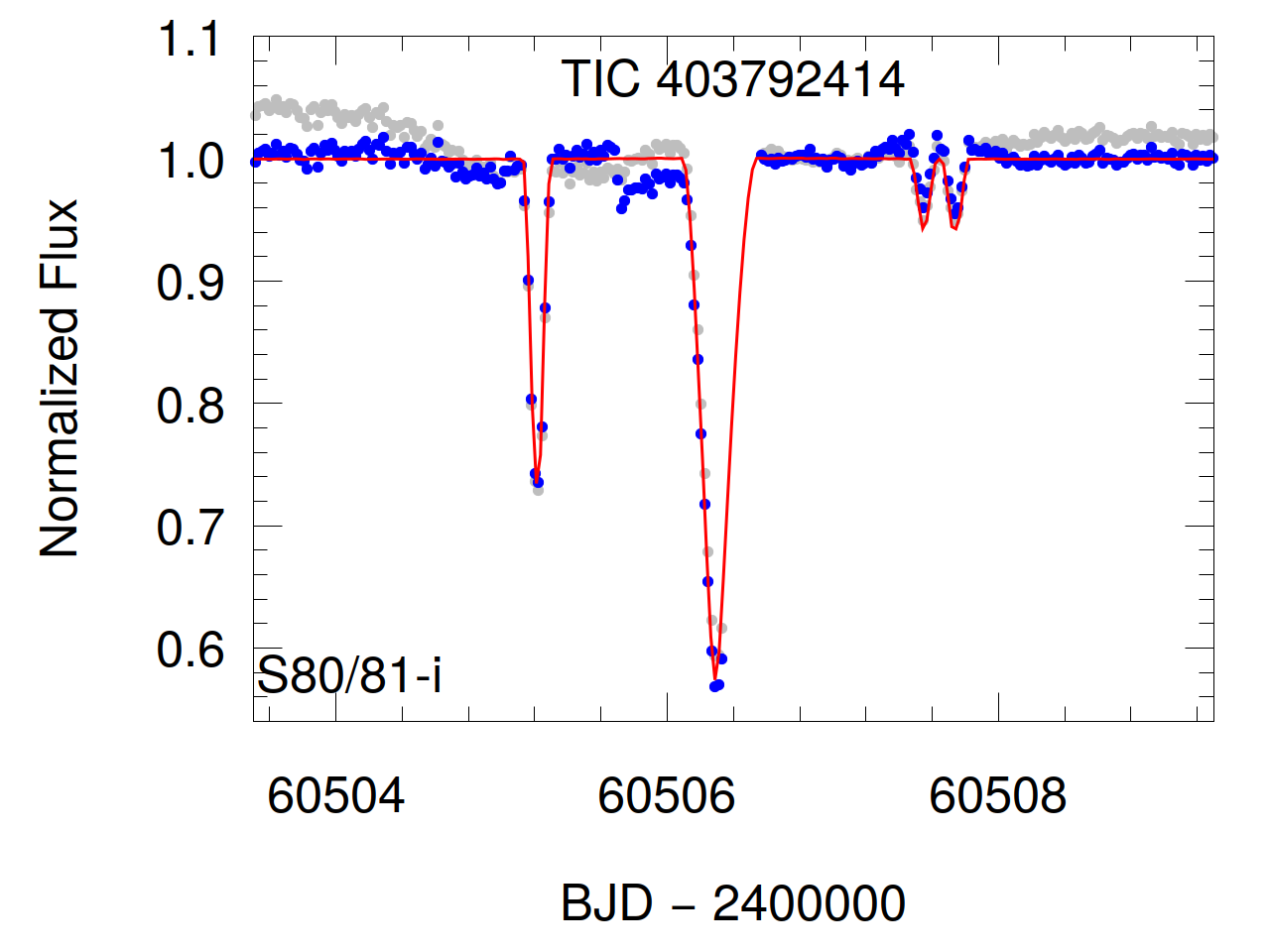}   
\includegraphics[width=0.32\textwidth]{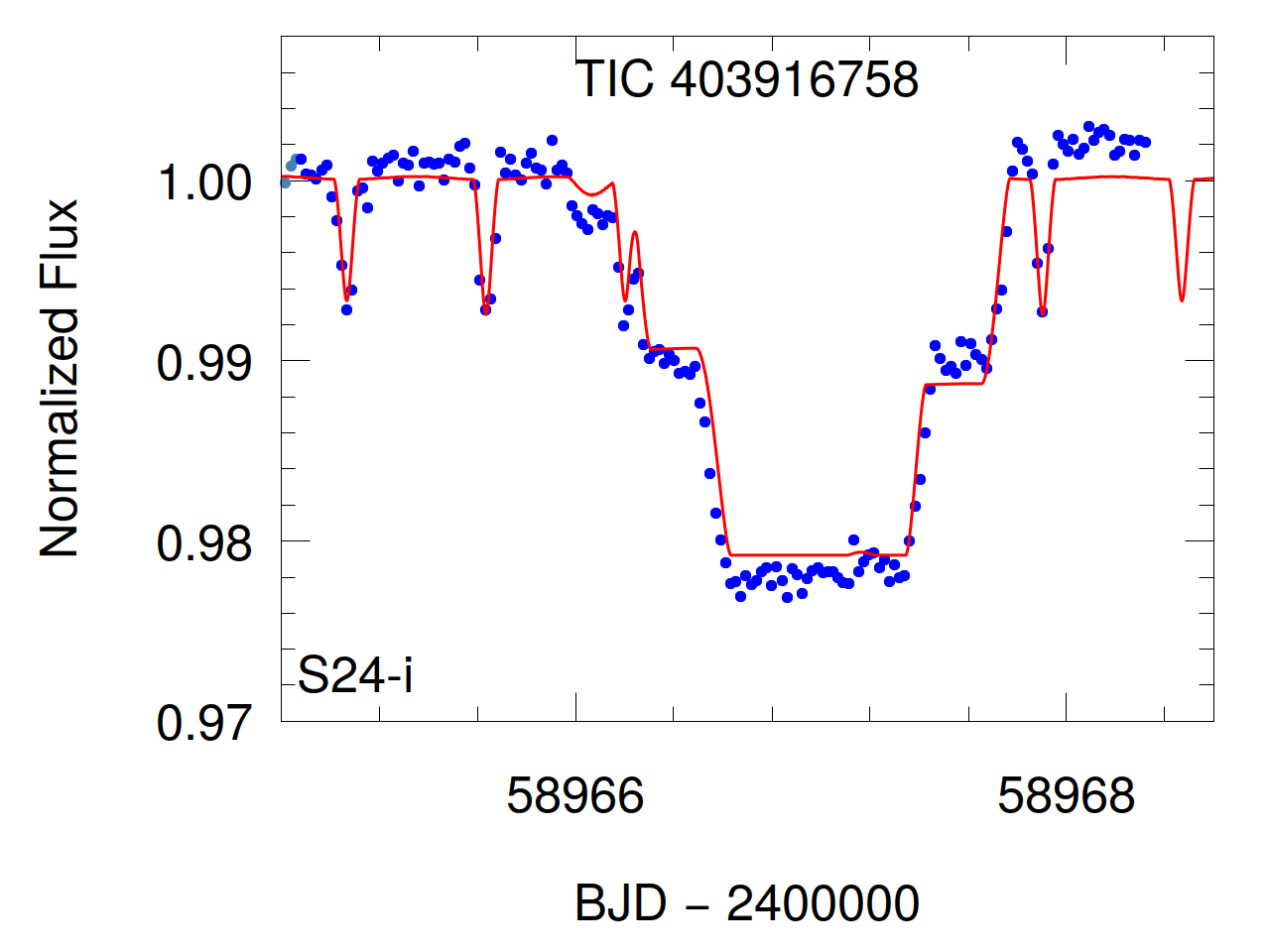}\includegraphics[width=0.32\textwidth]{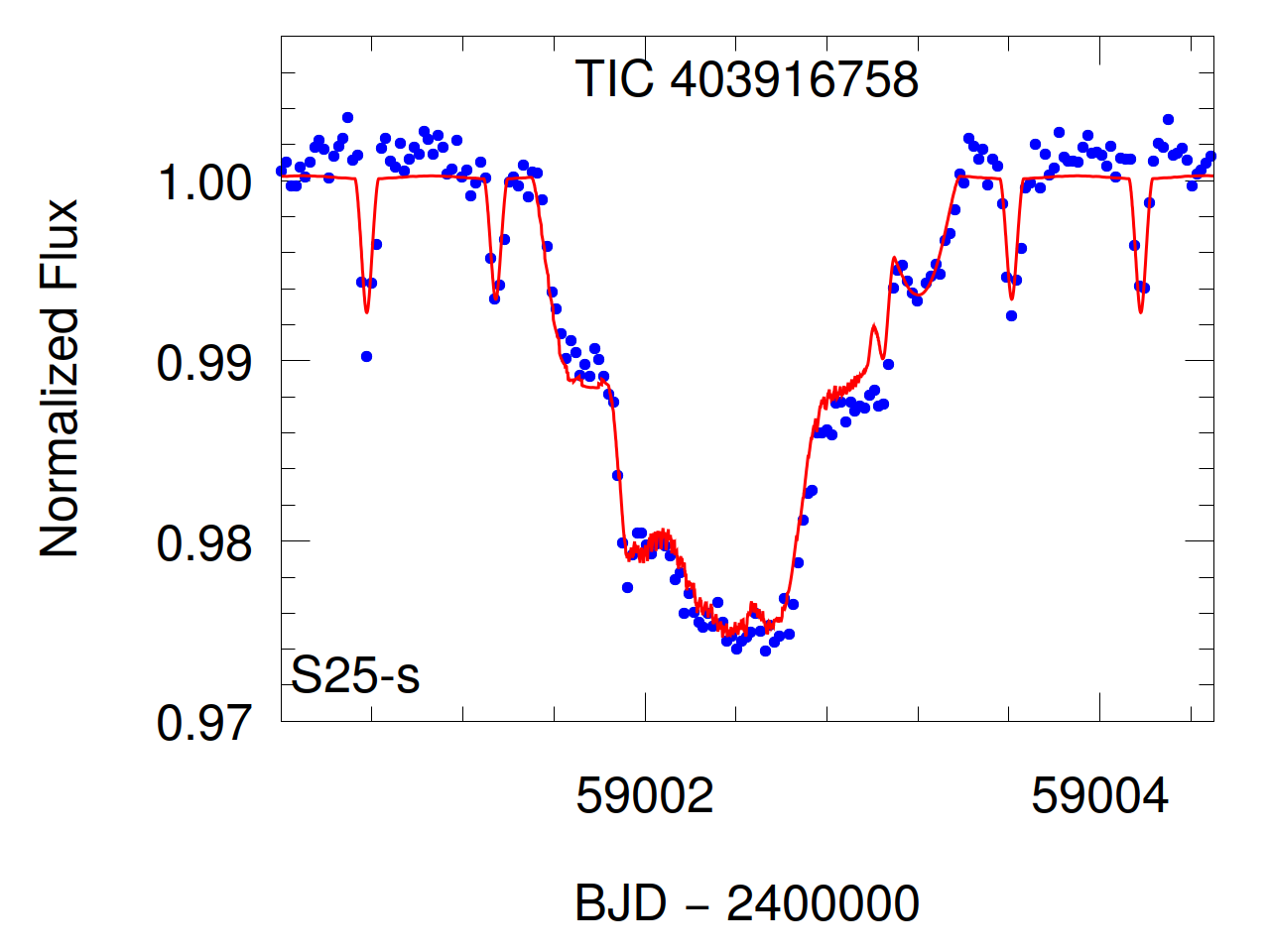}\includegraphics[width=0.32\textwidth]{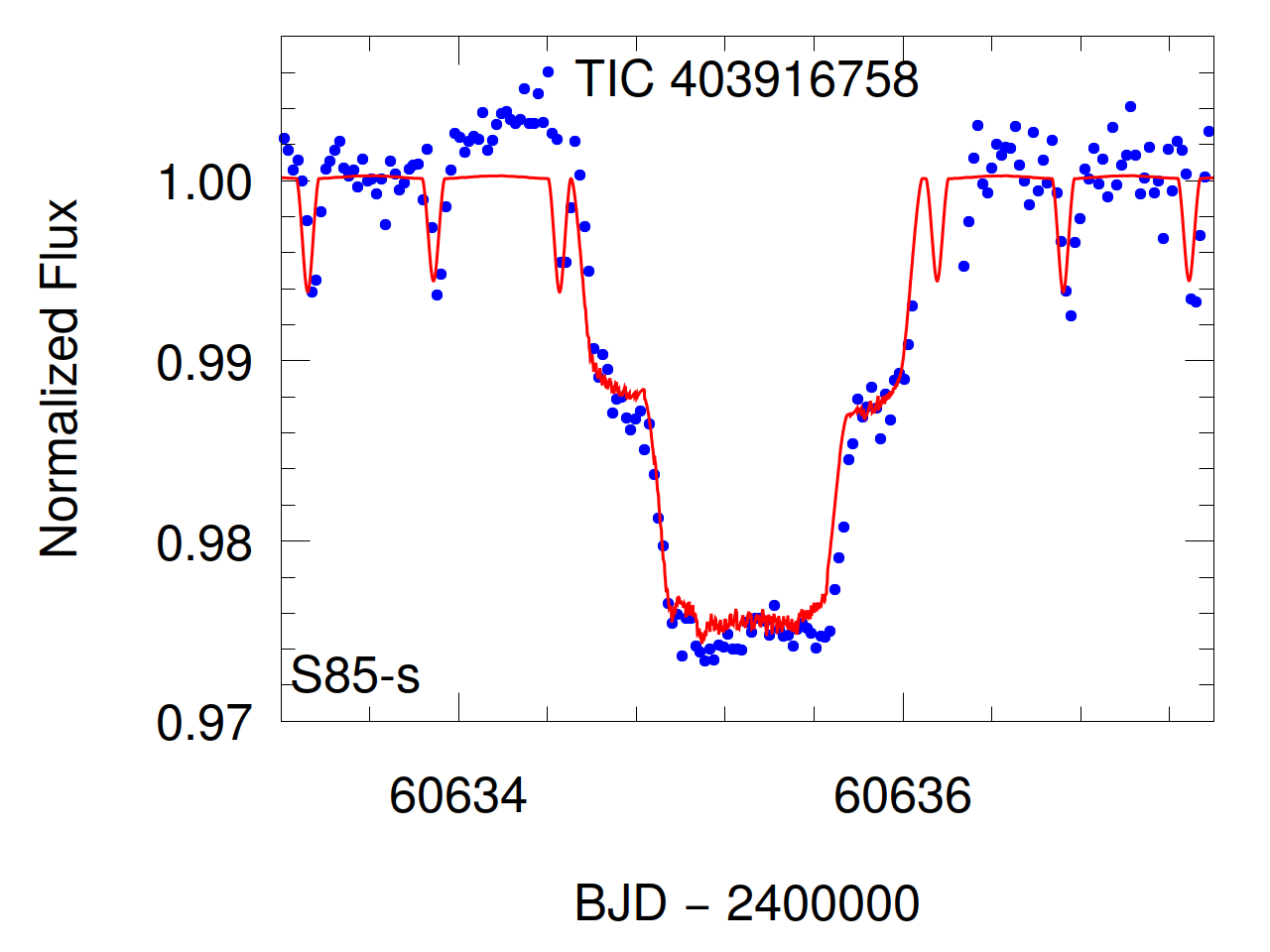}
\includegraphics[width=0.32\textwidth]{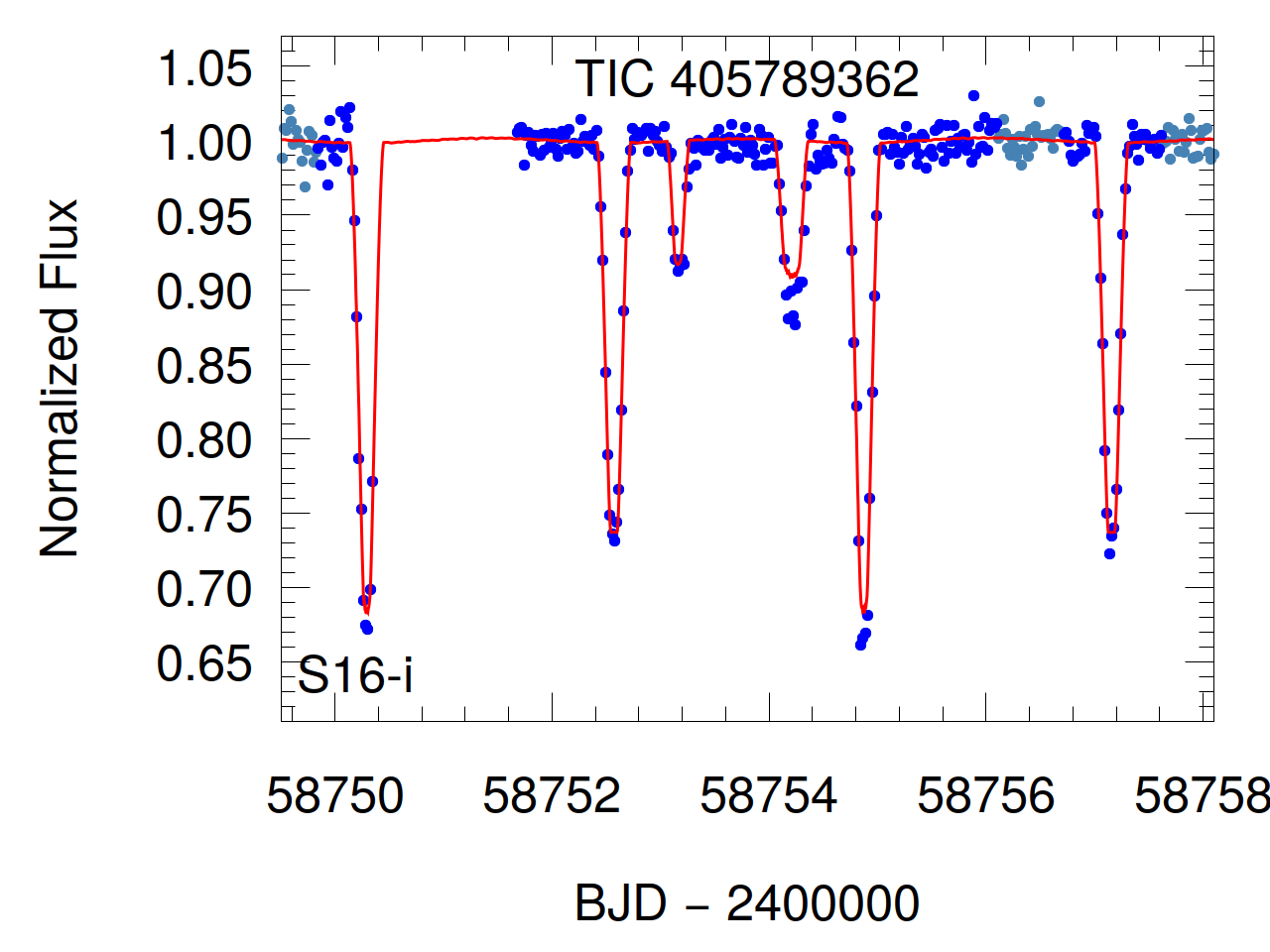}\includegraphics[width=0.32\textwidth]{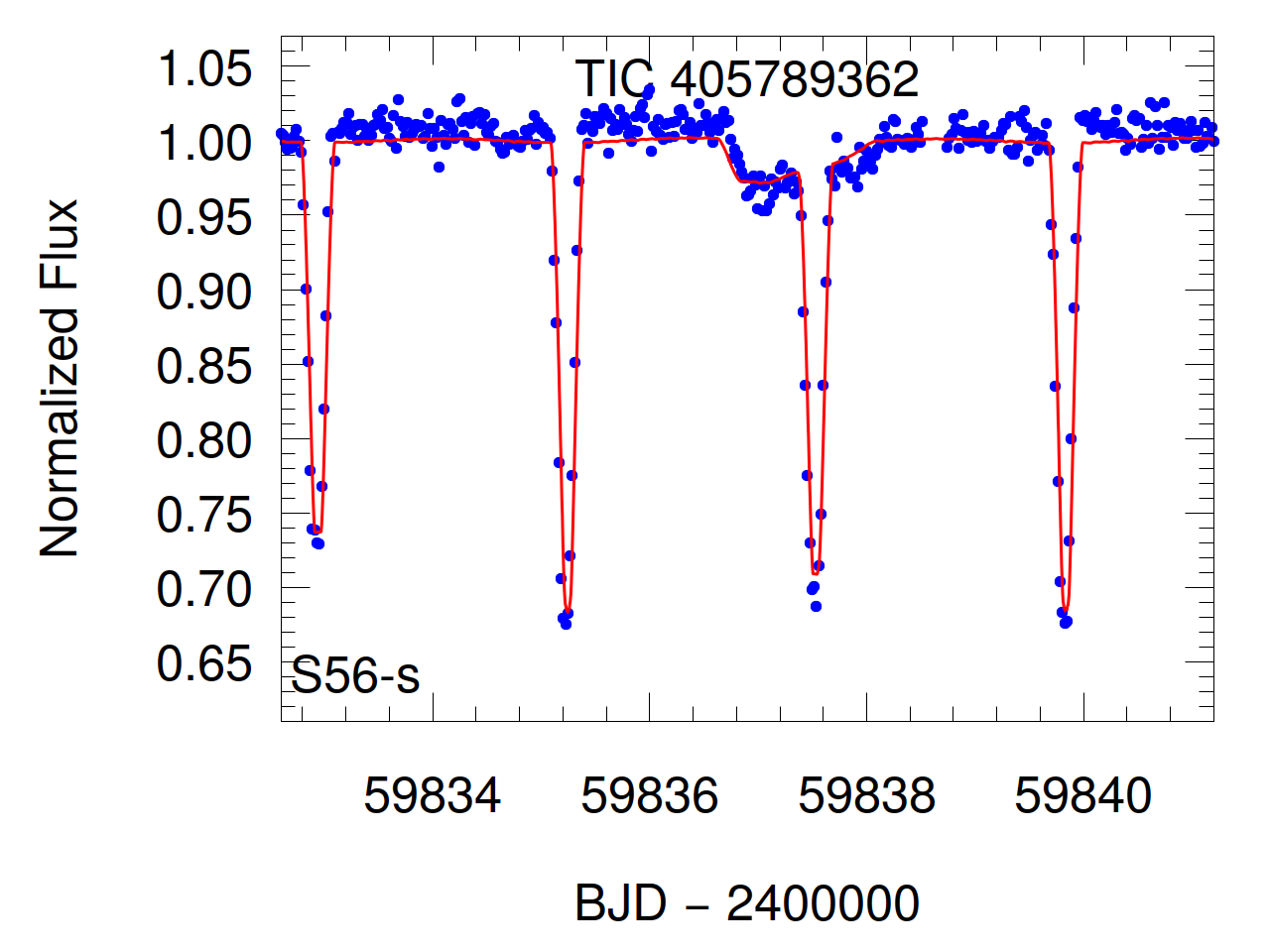}\includegraphics[width=0.32\textwidth]{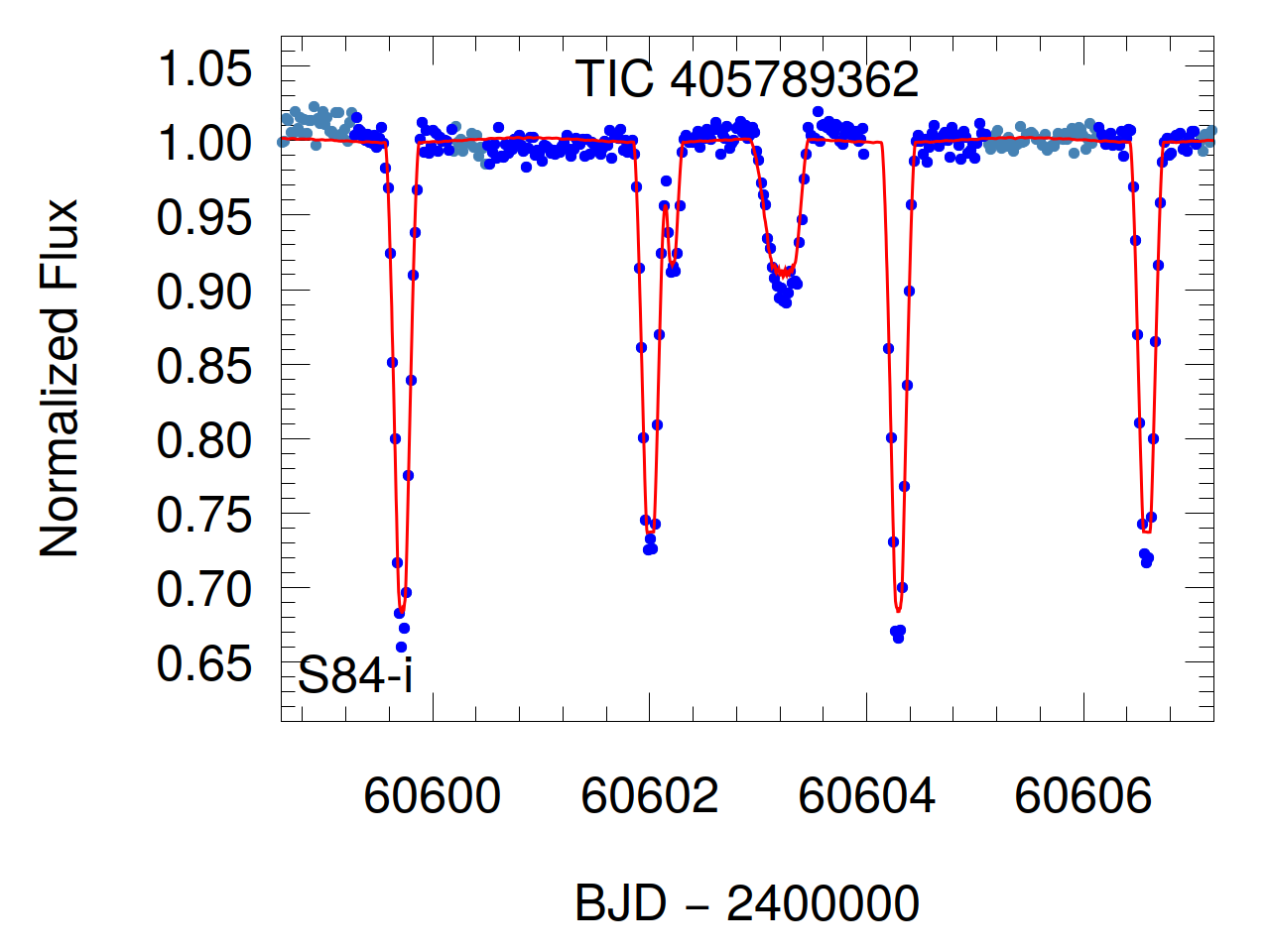}
\includegraphics[width=0.32\textwidth]{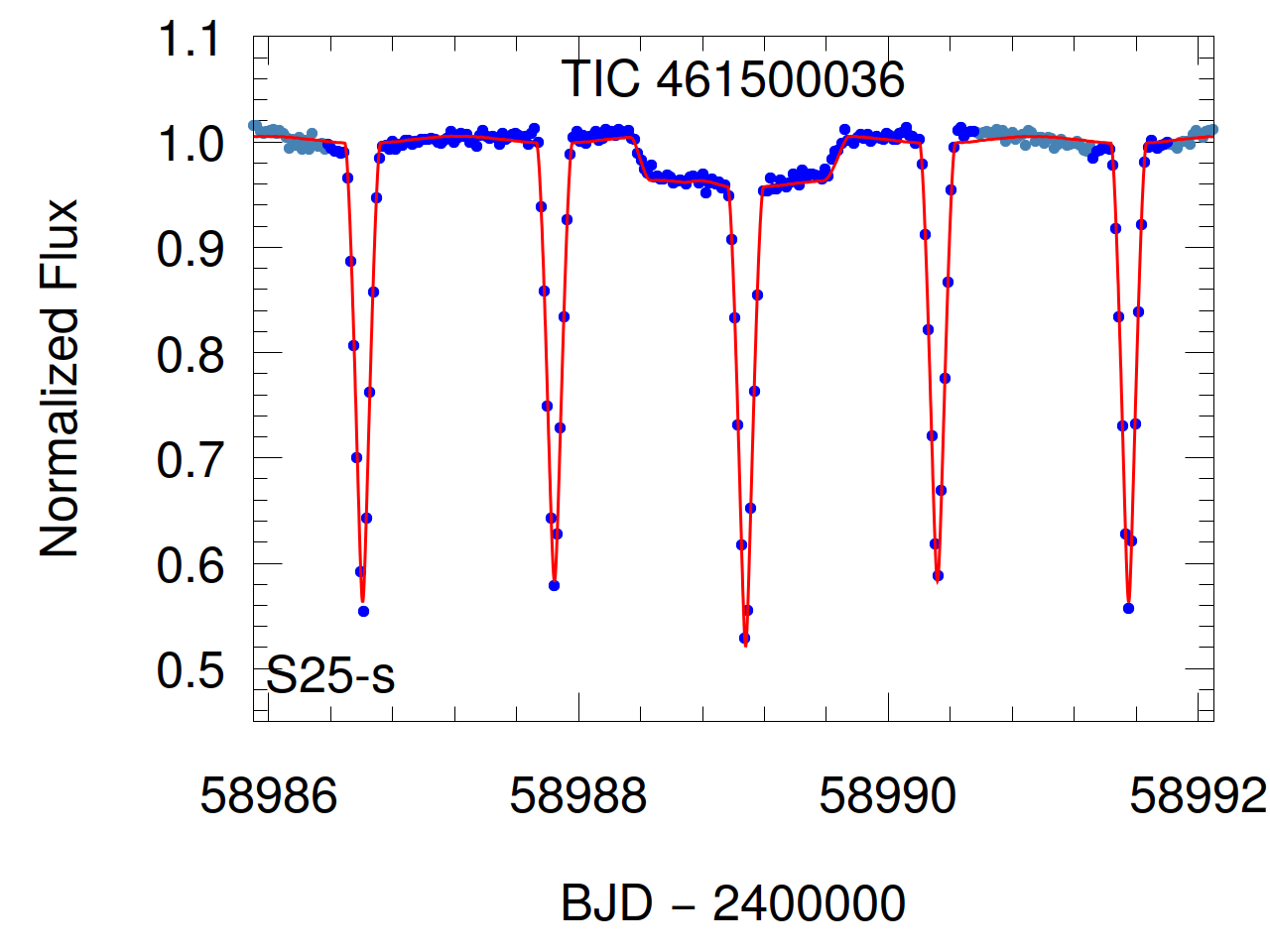}\includegraphics[width=0.32\textwidth]{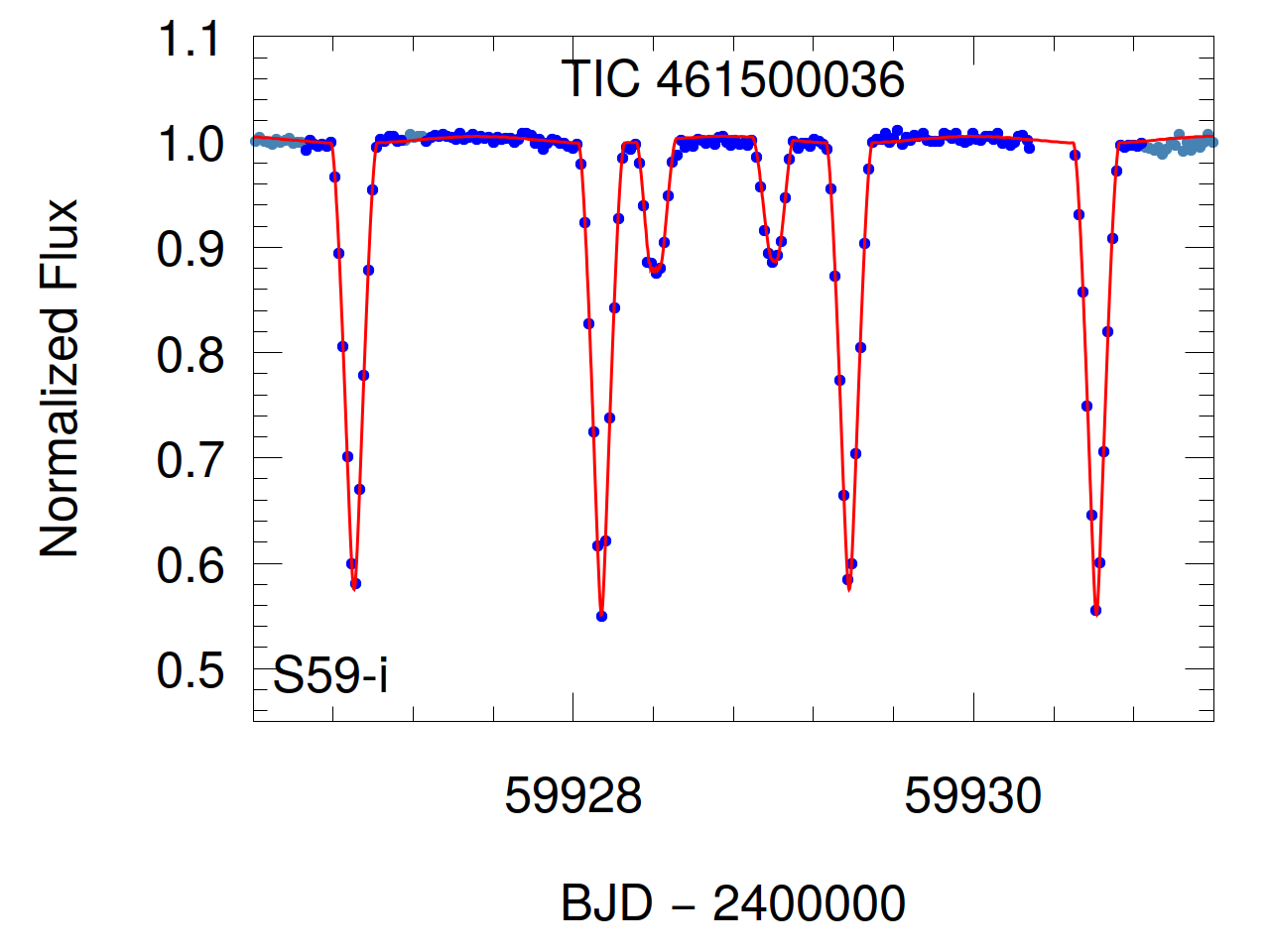}\includegraphics[width=0.32\textwidth]{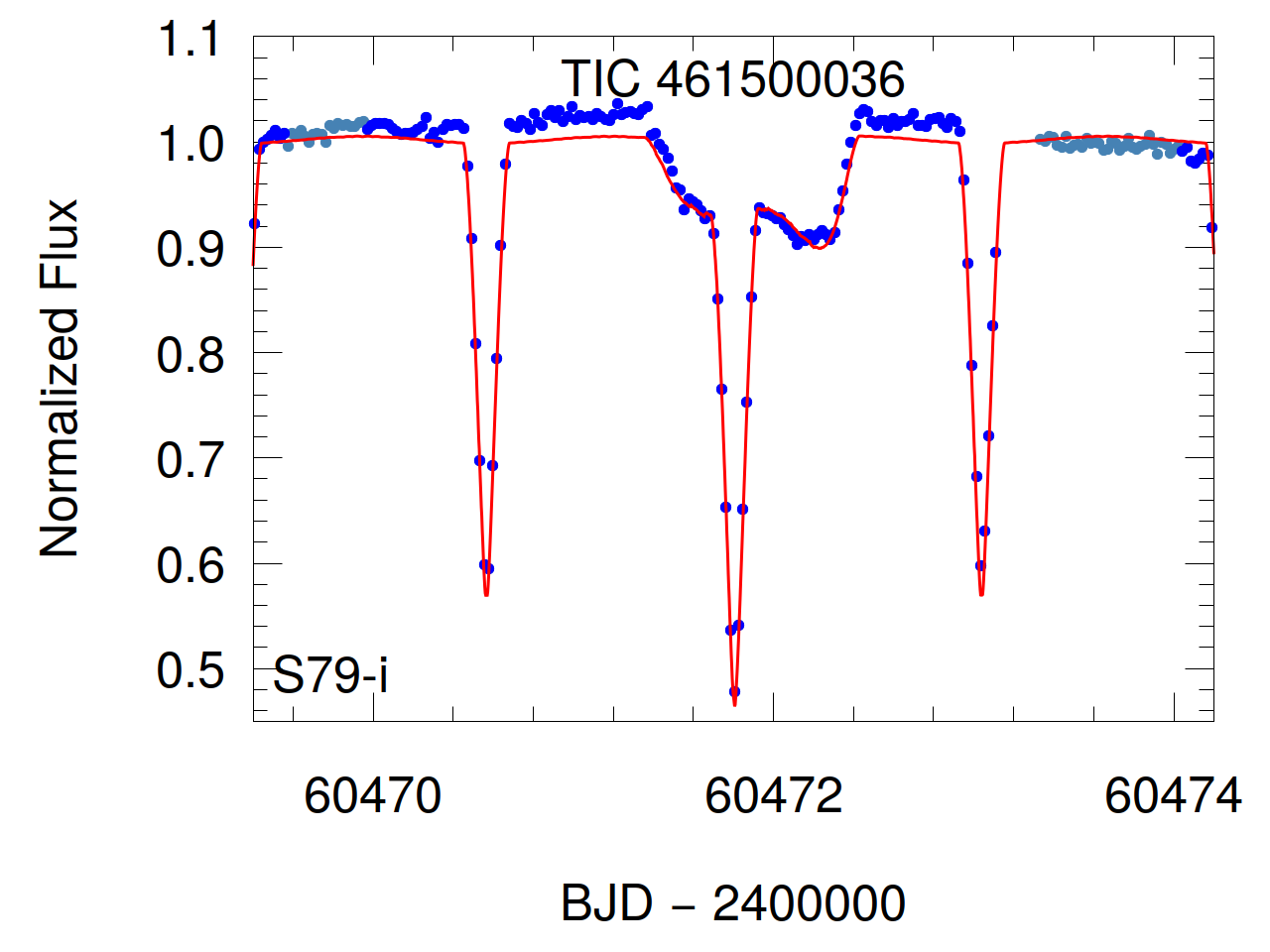}
   \caption{Light curves (blue points) and model fits (smooth red curves) near to the third-body eclipses of the second five targets. {\it From top to bottom panels:} TICs 378270875, 403792414, 403916758, 405789362, 461500036. See Fig.\,\ref{fig:lcs1} for details.}
   \label{fig:lcs2}
   \end{figure*}

\FloatBarrier

\section{Eclipse Timing Variations curves}
\label{app:ETVcurve}

In this appendix we plot the ETV curves of nine of the ten investigated triple systems.

\begin{figure*}[ht]
\centering
     \includegraphics[width=0.32\textwidth]{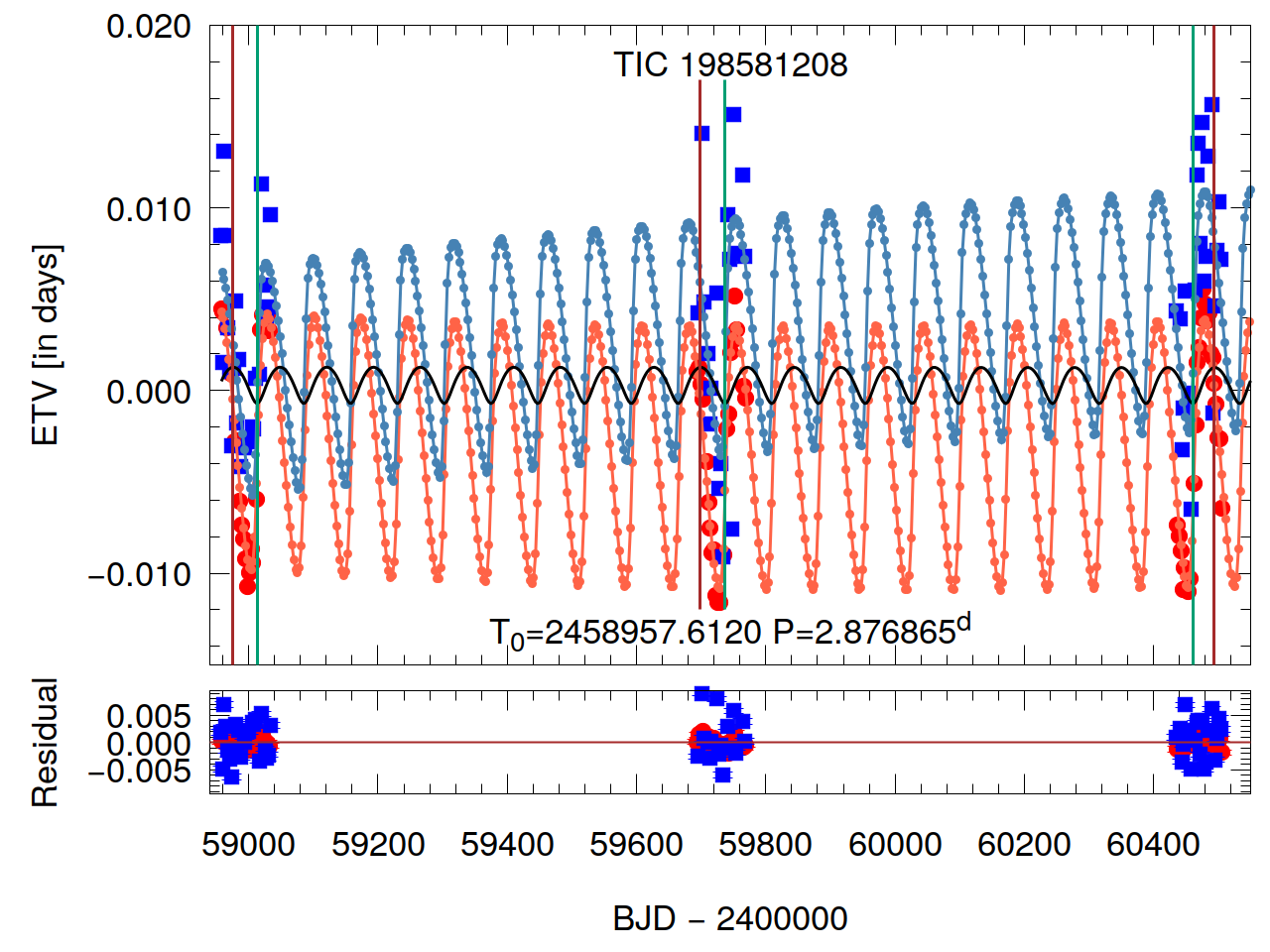}\includegraphics[width=0.32\textwidth]{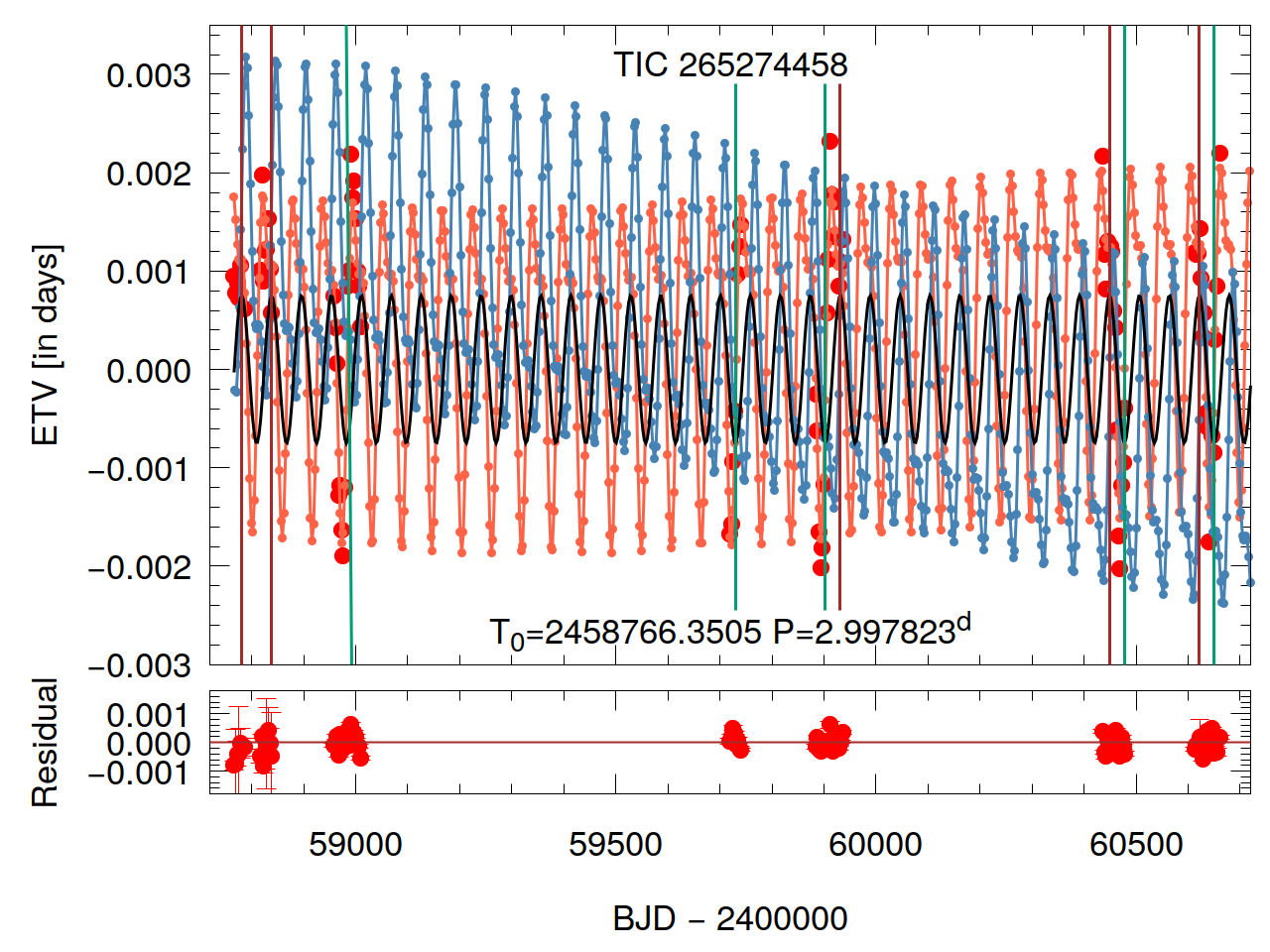}\includegraphics[width=0.32\textwidth]{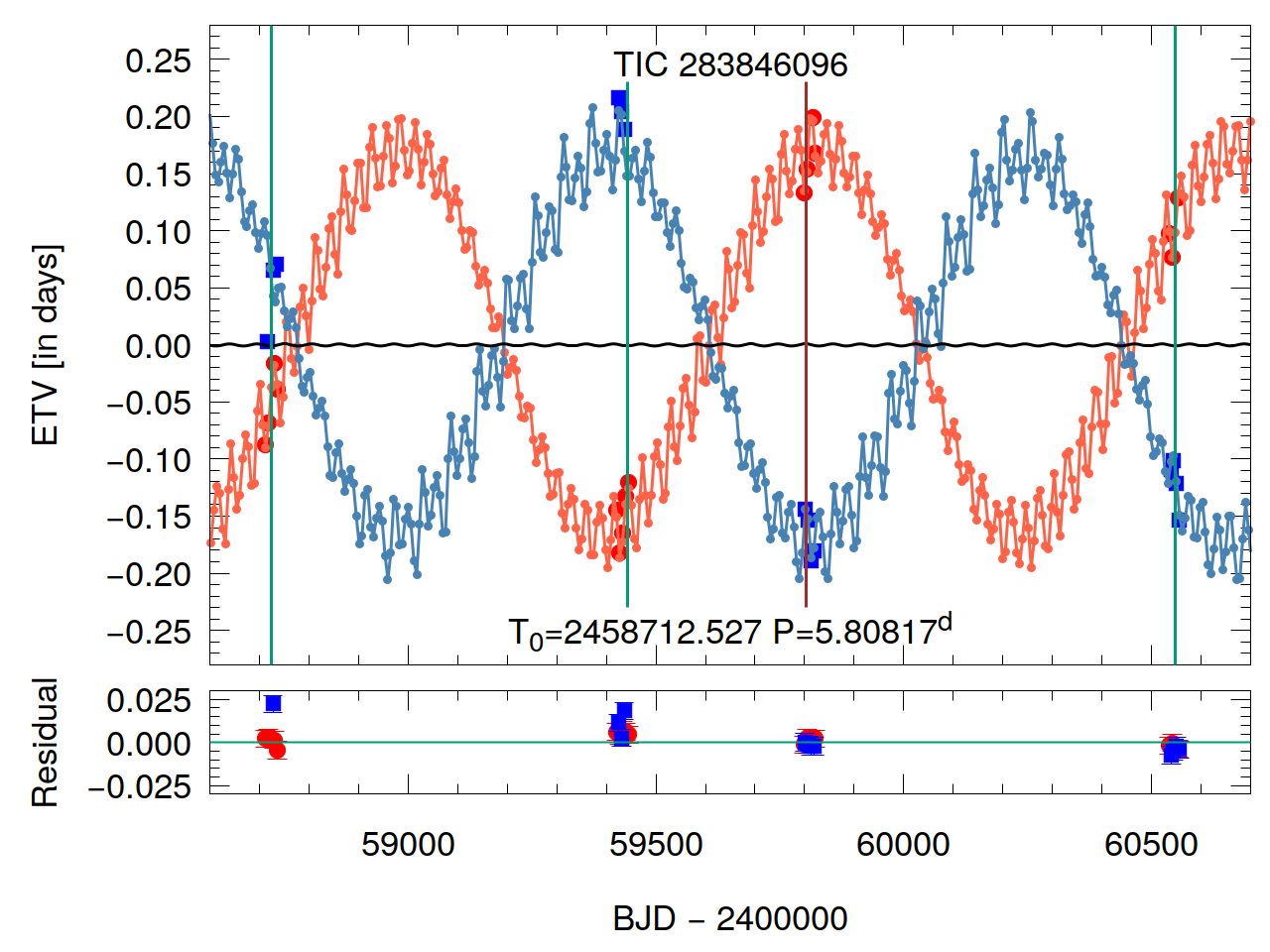}
     \includegraphics[width=0.32\textwidth]{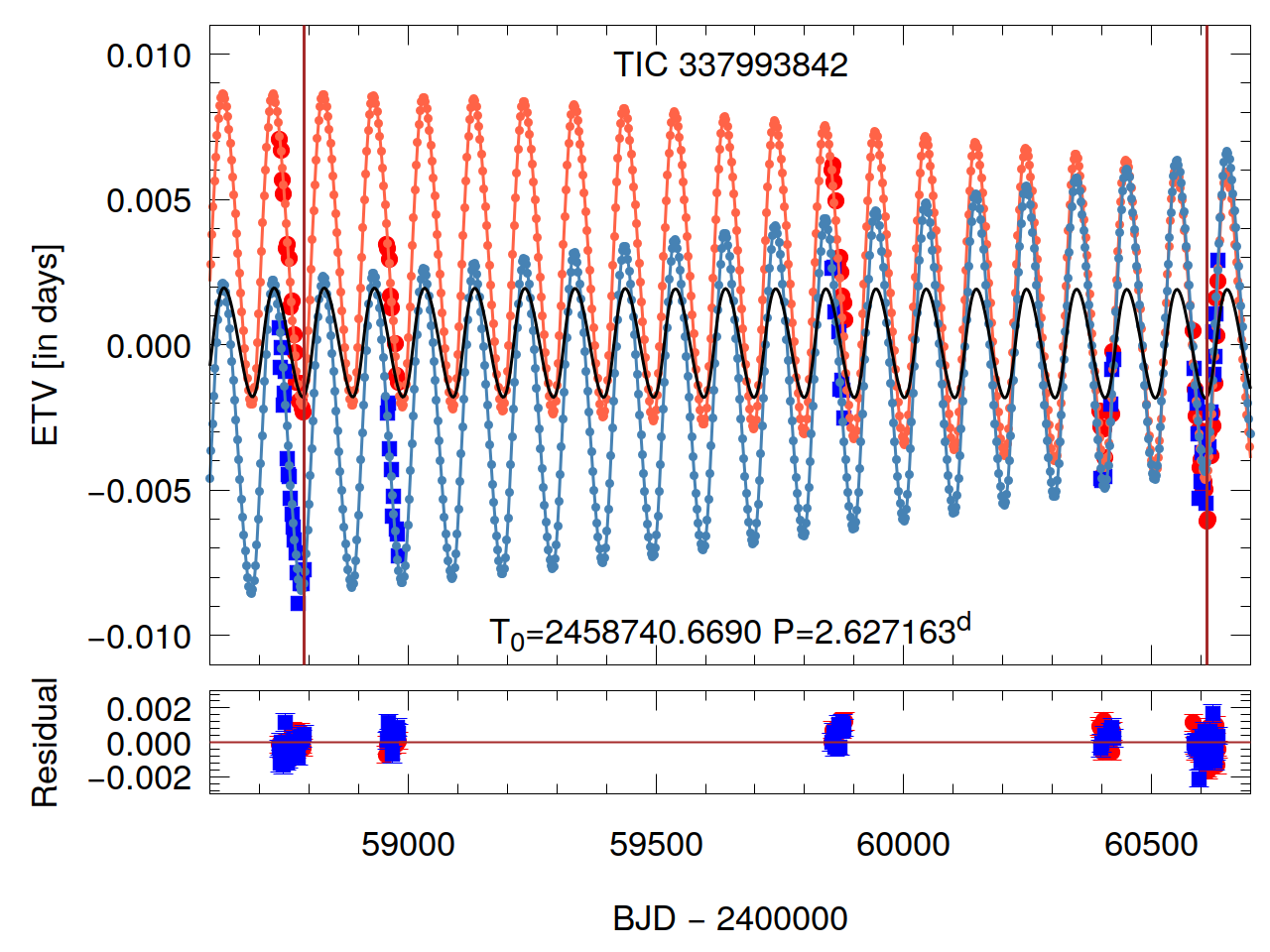}\includegraphics[width=0.32\textwidth]{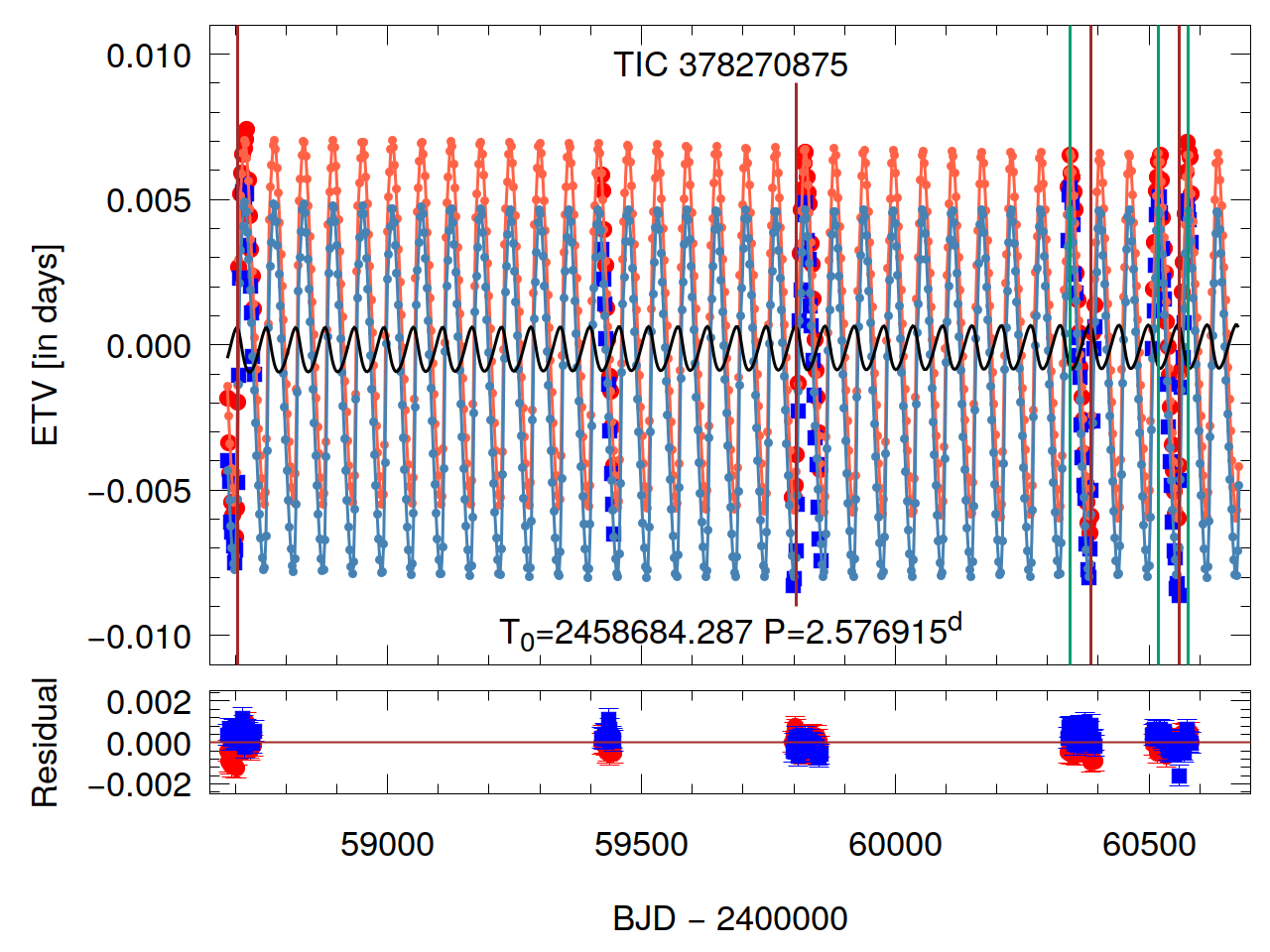}\includegraphics[width=0.32\textwidth]{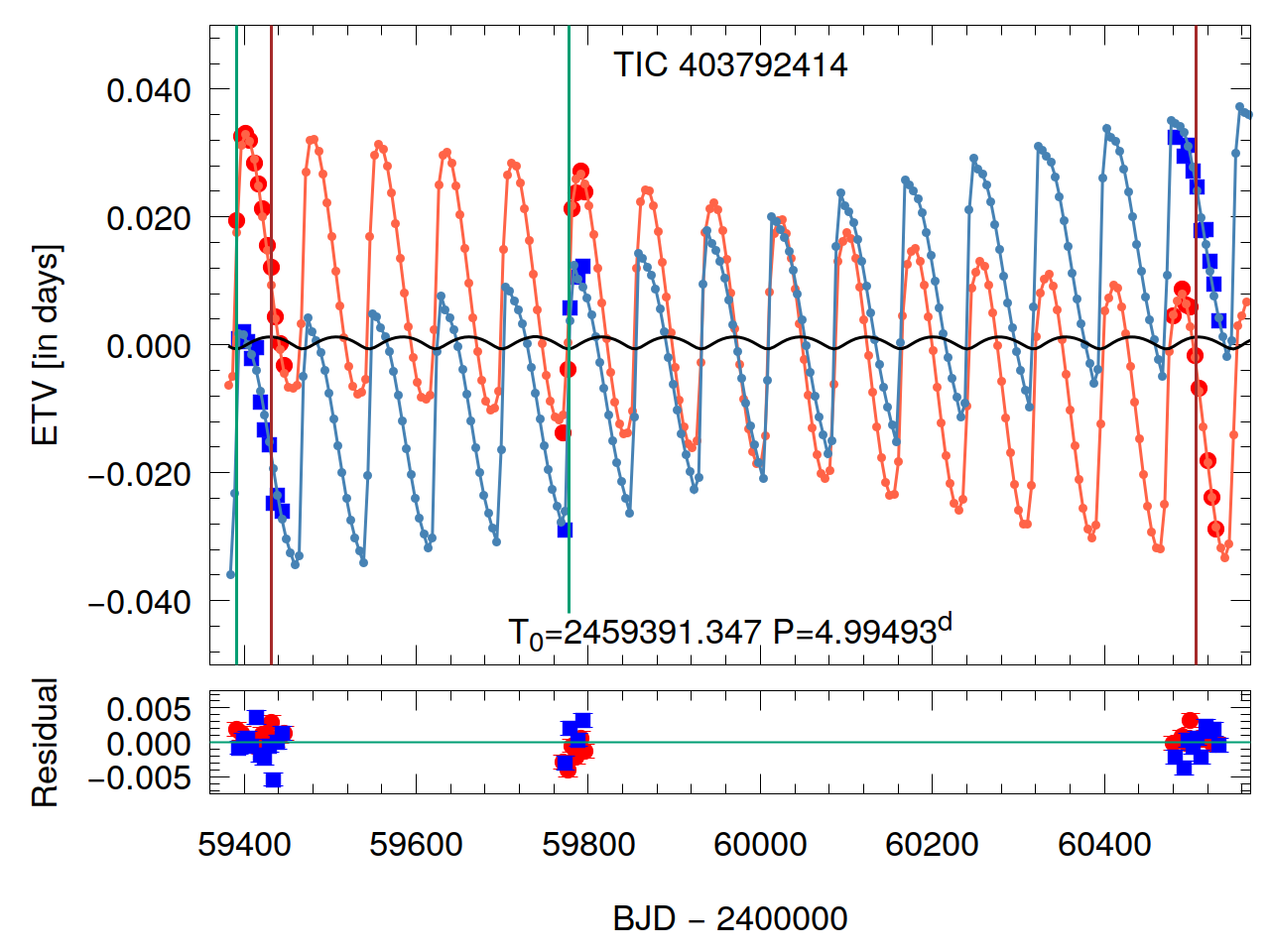}
     \includegraphics[width=0.32\textwidth]{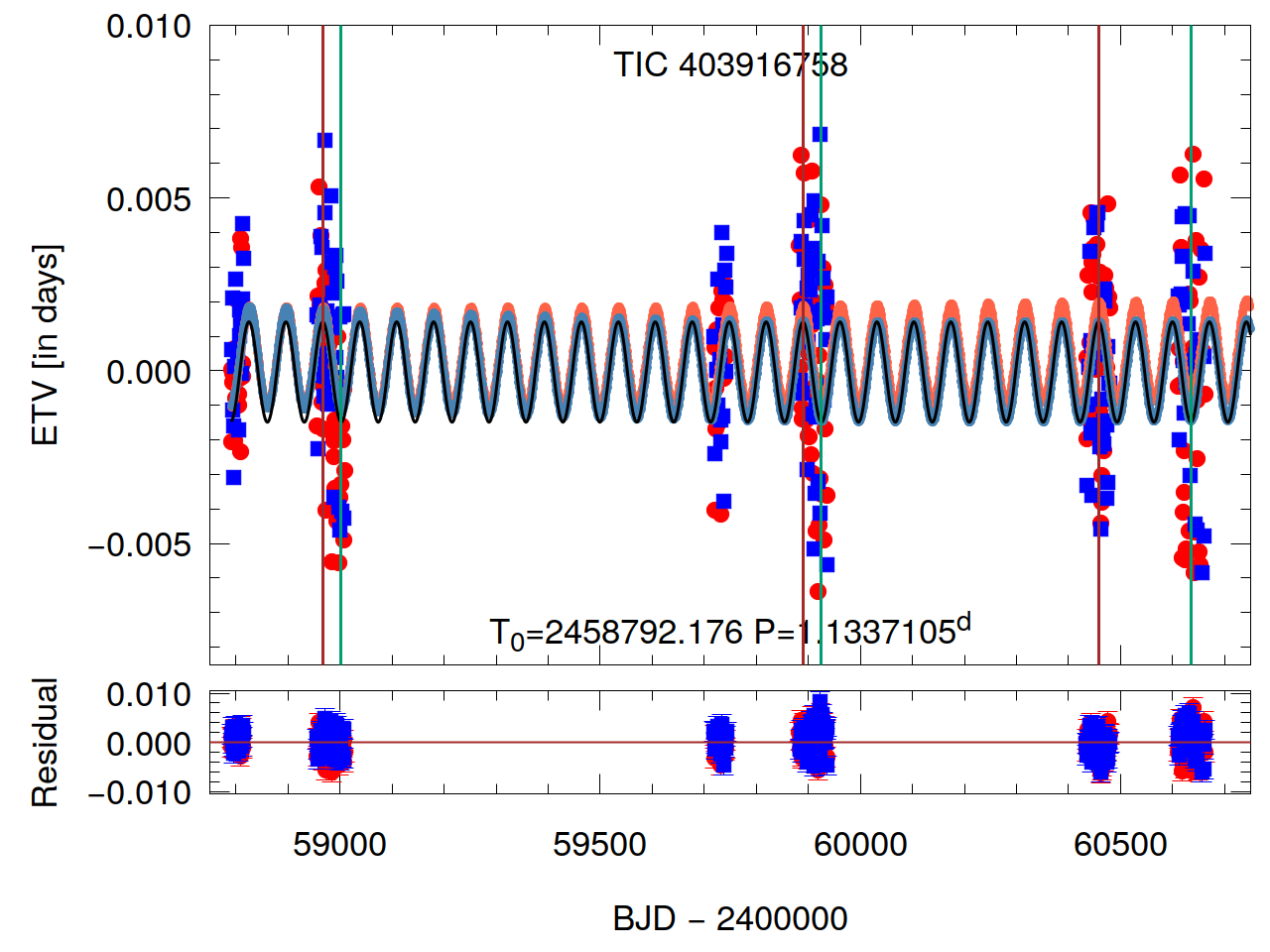}\includegraphics[width=0.32\textwidth]{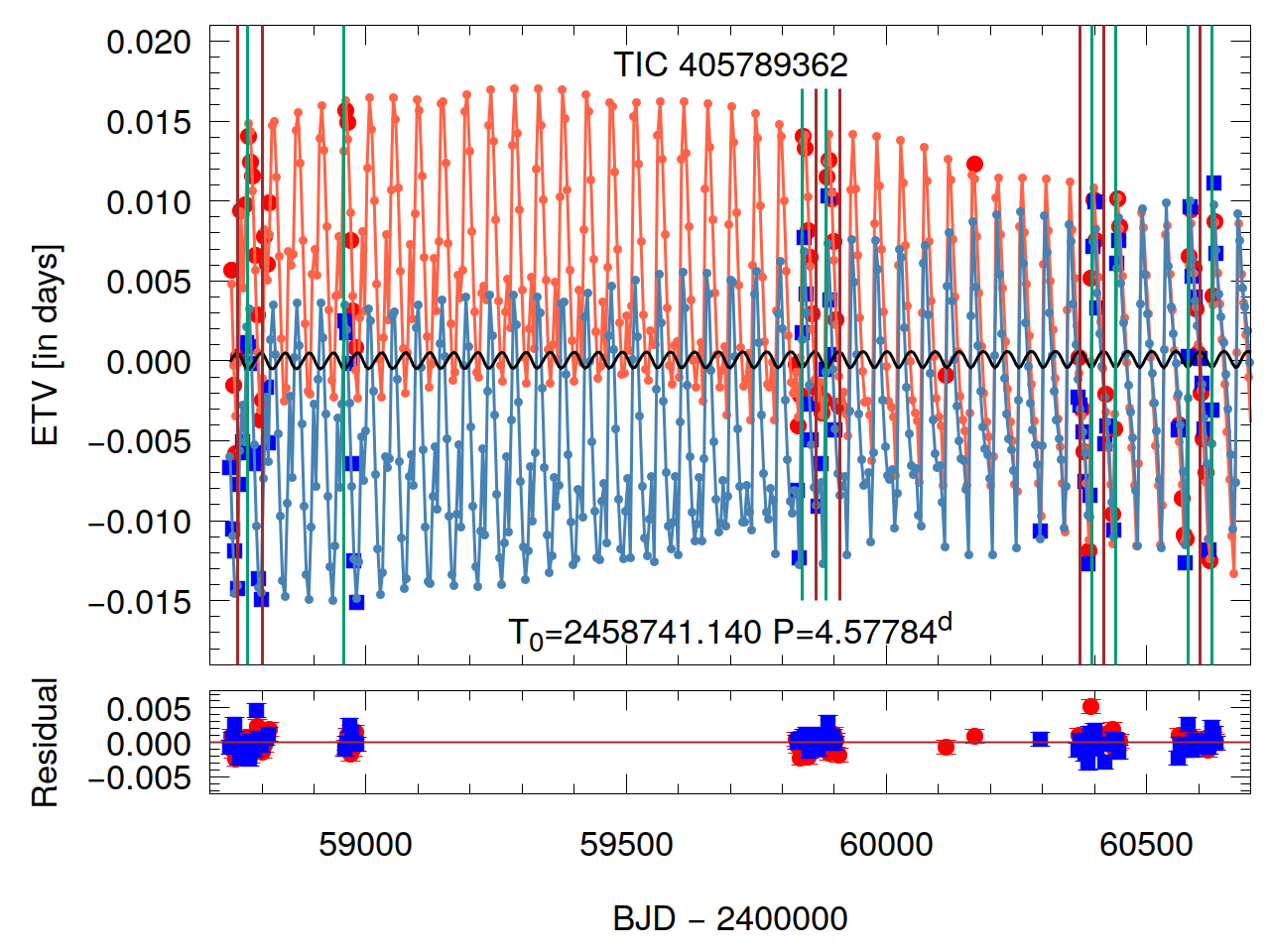}\includegraphics[width=0.32\textwidth]{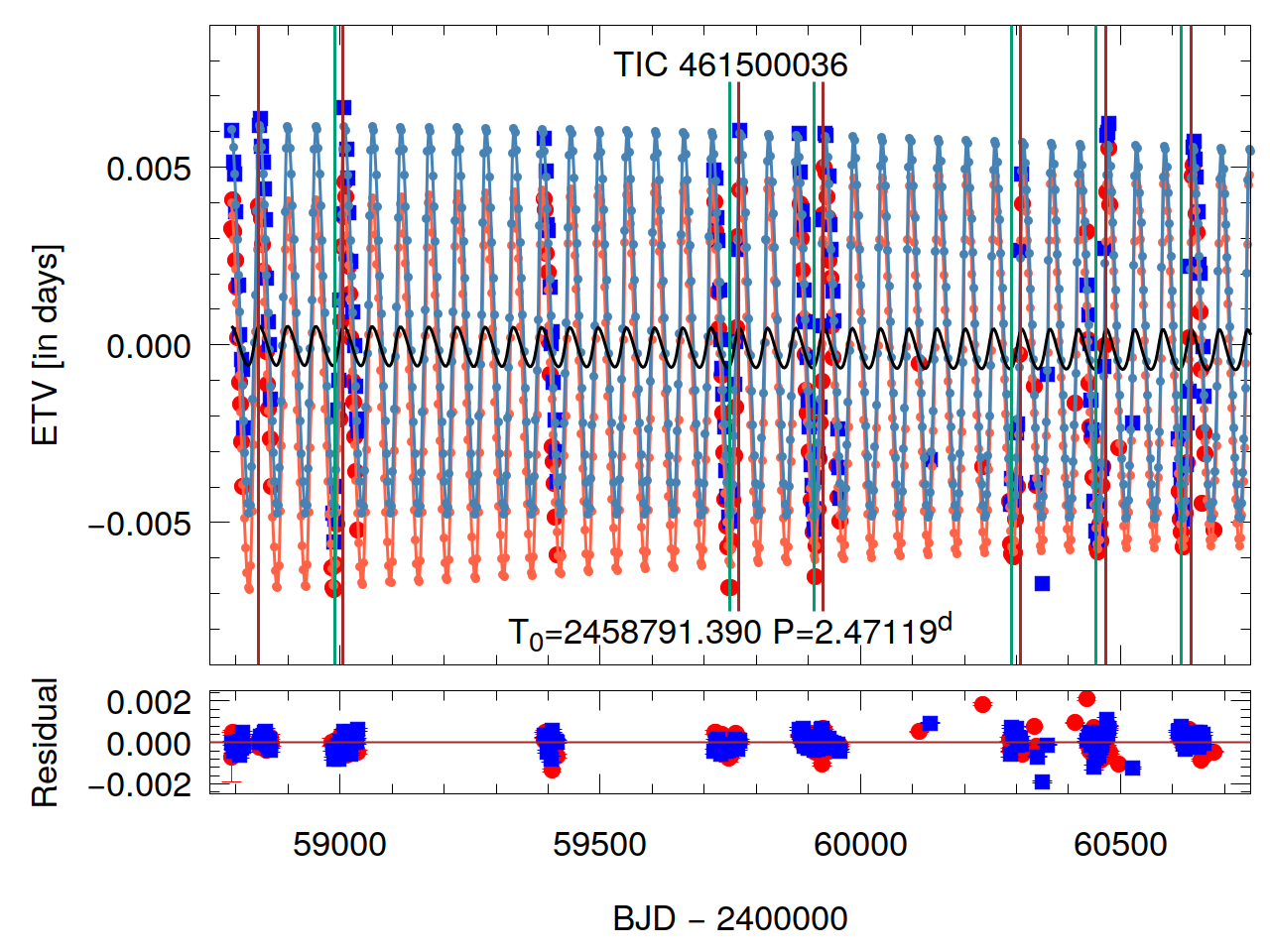}
          \caption{Primary and secondary ETV curves (red and blue circles, respectively) formed from the TESS observations with the best-fit photodynamical solution for nine targets. The horizontally centred black curve represents the pure LTTE contribution. Vertical lines mark the times of the observed outer eclipses (green -- the binary occulting the tertiary star and, brown -- vica versa).}
\label{fig:etvs}
\end{figure*}  

\FloatBarrier
\onecolumn
\section{Dynamics of TIC\,403792414}
\label{app:T403792414dyn}

In this appendix we plot some auxiliary figures about the irregular AM of TIC\,403792414. This effect is discussed in Sect.\,\ref{Sect:discussion_TIC403792414}.

\begin{figure}[ht]
\begin{center}
     \includegraphics[width=0.45\textwidth]{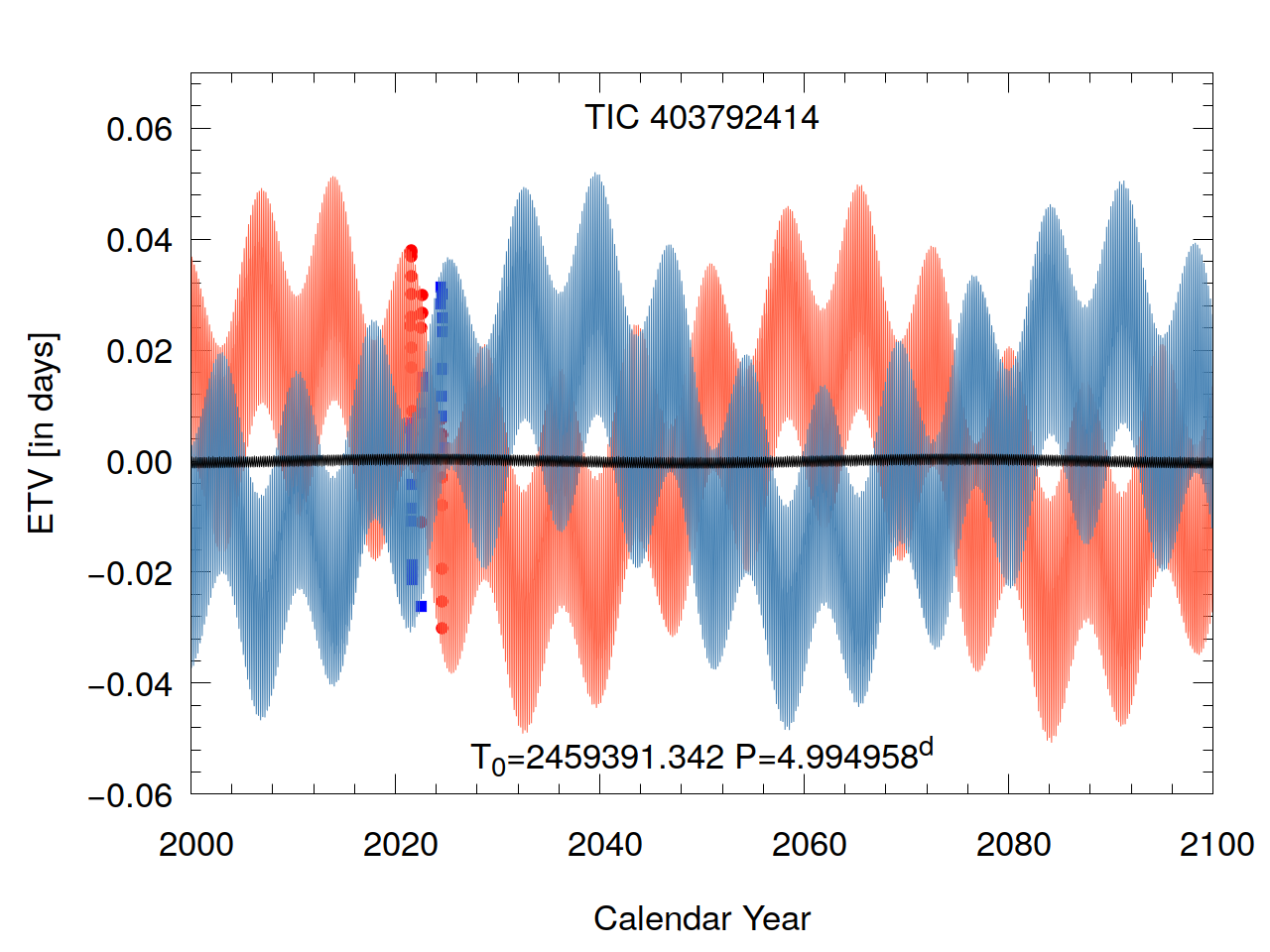}  
     \caption{Numerically generated ETV curves for TIC 403792414 spanning the current century. The red and blue curves represent the numerically calculated ETV curves. The ETV points derived from the TESS eclipse observations are plotted with red circles and blue squares. The black curve around the zero ETV level indicates the pure LTTE contribution. The hugely uneven nature of the apsidal motion is clearly visible. See the text for further details.}
\label{fig:403792414ETV-100yr}
\end{center}
\end{figure}  

\begin{figure*}[ht]
\begin{center}
     \includegraphics[width=0.45\textwidth]{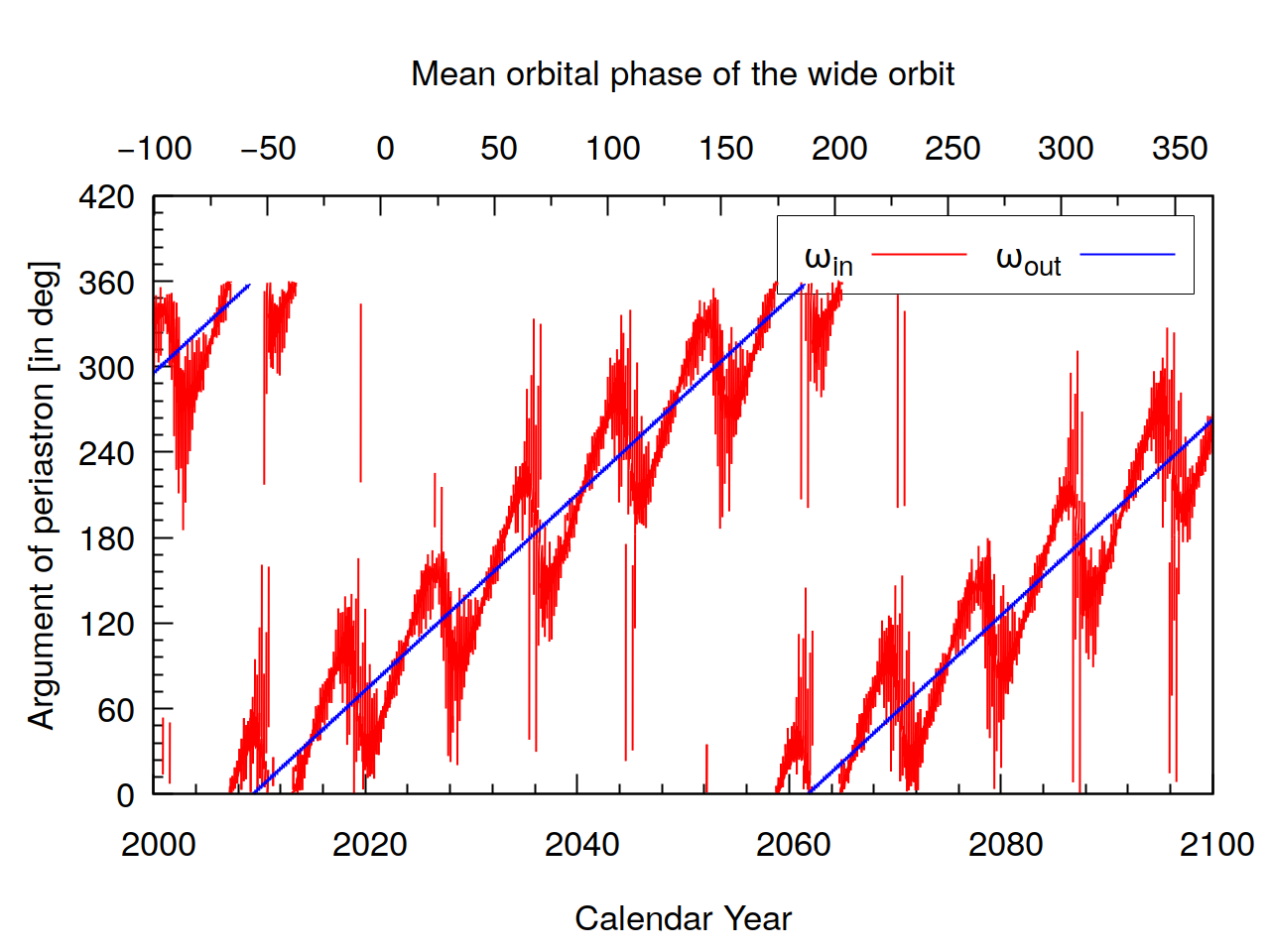}\includegraphics[width=0.45\textwidth]{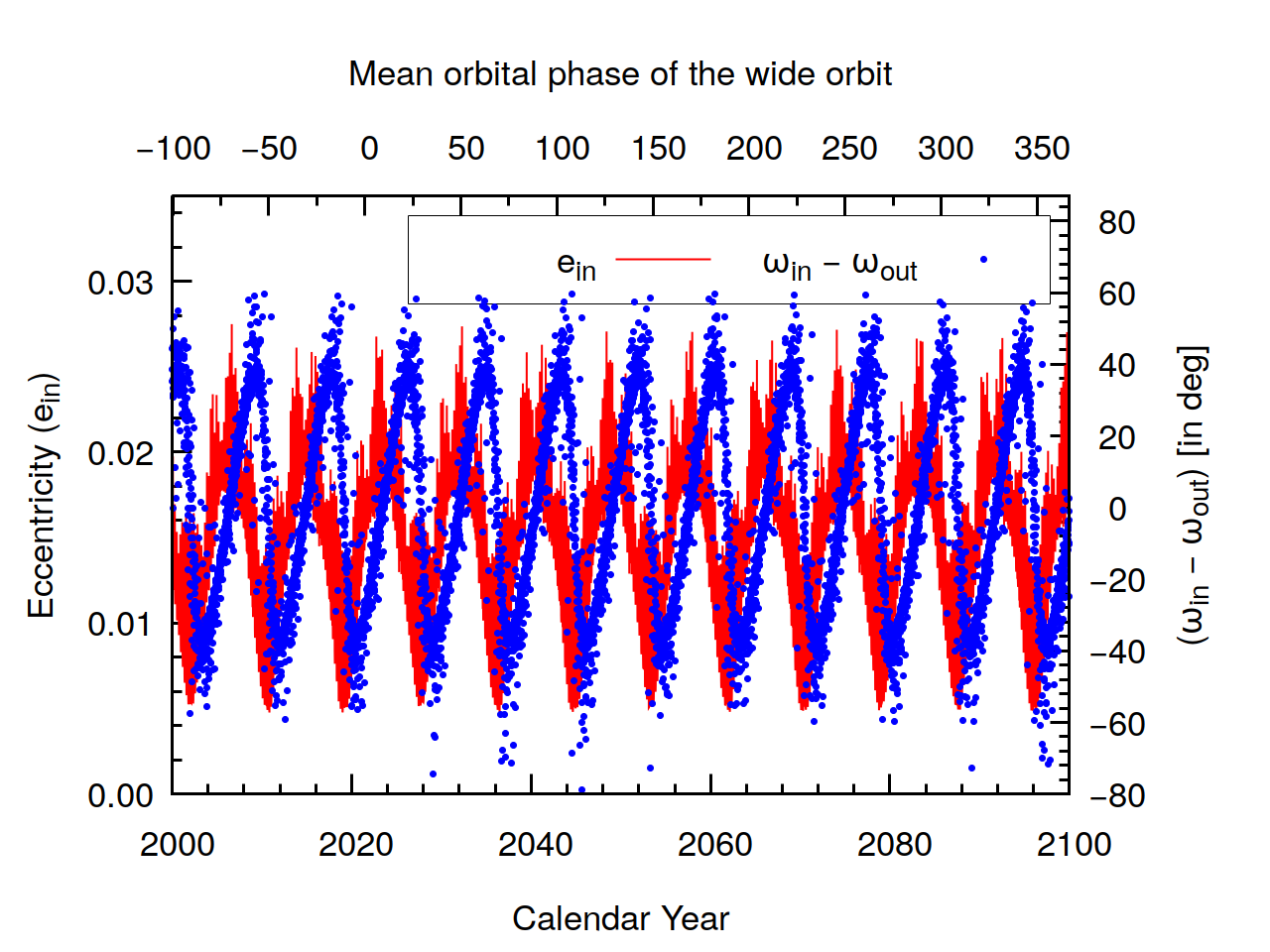}    
     \caption{The evolution of some of the orbital elements of TIC 403792414 over the current century. {\it Left panel:} The variations of the observable arguments of periastrons of the inner and the outer orbits ($\omega_\mathrm{in}$ -- red; $\omega_\mathrm{out}$ -- blue, respectively). As one can see nicely, the major axis of the inner ellipse librates around the direction of the rotating major axis of the outer orbit. {\it Right panel:} The cyclic variations of the inner eccentricity (red) and the difference of the inner and outer arguments of pericenters (blue). See the text for further details.}
\label{fig:403792414numint100yr}
\end{center}
\end{figure*}  

\FloatBarrier
\onecolumn
\section{Eclipse times of the inner EBs of the ten triples}
\label{app:ToMs}

In this appendix we tabulate the times of the individual primary and secondary eclipses of the inner EBs of the triples considered in this study.  These naturally include mostly eclipses from \textit{TESS}, plus a few that were observed from the ground (Tables~\ref{Tab:TIC_198581208_ToM}--\ref{Tab:TIC_461500036_ToM}). In the final, A\&A version, we present only the list of times of minima for TIC\,198581208. The other nine tables, together with this sample table, are available at the CDS, in machine readable form. Here, in the arXiv version, however, all the ten tables are published.

\begin{table*}[h!]
\caption{Times of minima of TIC\,198581208}
 \label{Tab:TIC_198581208_ToM}
\scalebox{0.88}{
\begin{tabular}{@{}lrllrllrllrl}
\hline
BJD & Cycle  & std. dev. & BJD & Cycle  & std. dev. & BJD & Cycle  & std. dev. & BJD & Cycle  & std. dev. \\ 
$-2\,400\,000$ & no. &   \multicolumn{1}{c}{$(d)$} & $-2\,400\,000$ & no. &   \multicolumn{1}{c}{$(d)$} & $-2\,400\,000$ & no. &   \multicolumn{1}{c}{$(d)$} & $-2\,400\,000$ & no. &   \multicolumn{1}{c}{$(d)$} \\ 
\hline
58956.18207 &    -0.5 & 0.00296 & 59012.26646 &    19.0 & 0.00026 & 59728.60025 &   268.0 & 0.00017 & 60457.89061 &   521.5 & 0.00137  \\ 
58957.61645 &     0.0 & 0.00027 & 59016.58859 &    20.5 & 0.00230 & 59730.04626 &   268.5 & 0.00135 & 60462.20733 &   523.0 & 0.00015  \\ 
58959.05198 &     0.5 & 0.00211 & 59018.02947 &    21.0 & 0.00023 & 59732.91804 &   269.5 & 0.00112 & 60463.65632 &   523.5 & 0.00060  \\ 
58960.49317 &     1.0 & 0.00031 & 59019.47592 &    21.5 & 0.00108 & 59734.35654 &   270.0 & 0.00034 & 60465.08735 &   524.0 & 0.00012  \\ 
58961.94039 &     1.5 & 0.09870 & 59020.90715 &    22.0 & 0.00028 & 59740.11718 &   272.0 & 0.00043 & 60466.53950 &   524.5 & 0.00089  \\ 
58963.37010 &     2.0 & 0.00048 & 59023.78392 &    23.0 & 0.00031 & 59741.56735 &   272.5 & 0.00106 & 60467.96765 &   525.0 & 0.00010  \\
\hline
\end{tabular}}
\textit{Notes}. Integer and half-integer cycle numbers (here and the following tables) denote primary and secondary eclipses, respectively.
\end{table*}

\begin{table*}[h!]
\caption{Times of minima of TIC\,265274458}
 \label{Tab:TIC_265274458_ToM}
\scalebox{0.88}{\begin{tabular}{@{}lrllrllrllrl}
\hline
BJD & Cycle  & std. dev. & BJD & Cycle  & std. dev. & BJD & Cycle  & std. dev. & BJD & Cycle  & std. dev. \\ 
$-2\,400\,000$ & no. &   \multicolumn{1}{c}{$(d)$} & $-2\,400\,000$ & no. &   \multicolumn{1}{c}{$(d)$} & $-2\,400\,000$ & no. &   \multicolumn{1}{c}{$(d)$} & $-2\,400\,000$ & no. &   \multicolumn{1}{c}{$(d)$} \\ 
\hline
58766.35145 &     0.0 & 0.00015 & 58979.19474 &    71.0 & 0.00007 & 59896.52796 &   377.0 & 0.00006 & 60463.11771 &   566.0 & 0.00008  \\ 
58769.34910 &     1.0 & 0.00118 & 58985.19242 &    73.0 & 0.00010 & 59899.52642 &   378.0 & 0.00008 & 60466.11444 &   567.0 & 0.00007  \\ 
58772.34703 &     2.0 & 0.00042 & 58988.19042 &    74.0 & 0.00007 & 59905.52382 &   380.0 & 0.00005 & 60469.11194 &   568.0 & 0.00011  \\ 
58775.34469 &     3.0 & 0.00164 & 58991.18941 &    75.0 & 0.00008 & 59908.52218 &   381.0 & 0.00007 & 60472.11061 &   569.0 & 0.00006  \\ 
58778.34285 &     4.0 & 0.00052 & 58994.18680 &    76.0 & 0.00006 & 59911.52121 &   382.0 & 0.00012 & 60475.10866 &   570.0 & 0.00009  \\ 
58787.33588 &     7.0 & 0.00030 & 58997.18479 &    77.0 & 0.00007 & 59914.51848 &   383.0 & 0.00007 & 60478.10704 &   571.0 & 0.00009  \\
\hline
\end{tabular}}
\end{table*}

\begin{table*}[h!]
\caption{Times of minima of TIC\,283846096}
 \label{Tab:TIC_283846096_ToM}
\scalebox{0.88}{\begin{tabular}{@{}lrllrllrllrl}
\hline
BJD & Cycle  & std. dev. & BJD & Cycle  & std. dev. & BJD & Cycle  & std. dev. & BJD & Cycle  & std. dev. \\ 
$-2\,400\,000$ & no. &   \multicolumn{1}{c}{$(d)$} & $-2\,400\,000$ & no. &   \multicolumn{1}{c}{$(d)$} & $-2\,400\,000$ & no. &   \multicolumn{1}{c}{$(d)$} & $-2\,400\,000$ & no. &   \multicolumn{1}{c}{$(d)$} \\ 
\hline
58712.43950 &     0.0 & 0.00224 & 59424.24473 &   122.5 & 0.00626 & 59801.41518 &   187.5 & 0.00226 & 60536.39001 &   314.0 & 0.00061  \\ 
58715.43348 &     0.5 & 0.05156 & 59426.74948 &   123.0 & 0.00123 & 59804.61672 &   188.0 & 0.00116 & 60539.08722 &   314.5 & 0.00155  \\ 
58718.26642 &     1.0 & 0.00390 & 59430.04019 &   123.5 & 0.00295 & 59807.21333 &   188.5 & 0.00471 & 60542.17732 &   315.0 & 0.00066  \\ 
58727.11261 &     2.5 & 0.00694 & 59432.57505 &   124.0 & 0.00116 & 59812.98631 &   189.5 & 0.00161 & 60544.90293 &   315.5 & 0.00144  \\ 
58729.93548 &     3.0 & 0.00296 & 59435.83245 &   124.5 & 0.10680 & 59816.27810 &   190.0 & 0.00137 & 60550.69154 &   316.5 & 0.00164  \\ 
58732.92584 &     3.5 & 0.02686 & 59438.41518 &   125.0 & 0.00137 & 59818.80318 &   190.5 & 0.00361 & 60553.84582 &   317.0 & 0.00061  \\ 
58735.72082 &     4.0 & 0.16640 & 59444.23570 &   126.0 & 0.00163 & 59822.05533 &   191.0 & 0.00194 & 60556.46774 &   317.5 & 0.00154  \\ 
59420.97900 &   122.0 & 0.00231 & 59798.78826 &   187.0 & 0.00099  \\ 
\hline
\end{tabular}}
\end{table*}

\begin{table*}[h!]
\caption{Times of minima of TIC\,337993842}
 \label{Tab:TIC_337993842_ToM}
\scalebox{0.88}{\begin{tabular}{@{}lrllrllrllrl}
\hline
BJD & Cycle  & std. dev. & BJD & Cycle  & std. dev. & BJD & Cycle  & std. dev. & BJD & Cycle  & std. dev. \\ 
$-2\,400\,000$ & no. &   \multicolumn{1}{c}{$(d)$} & $-2\,400\,000$ & no. &   \multicolumn{1}{c}{$(d)$} & $-2\,400\,000$ & no. &   \multicolumn{1}{c}{$(d)$} & $-2\,400\,000$ & no. &   \multicolumn{1}{c}{$(d)$} \\ 
\hline
58739.35602 &    -0.5 & 0.00041 & 58784.00912 &    16.5 & 0.00034 & 59870.35209 &   430.0 & 0.00025 & 60595.44314 &   706.0 & 0.00035  \\ 
58740.67607 &     0.0 & 0.00036 & 58785.32858 &    17.0 & 0.00034 & 59871.66113 &   430.5 & 0.00028 & 60596.75439 &   706.5 & 0.00036  \\ 
58741.98179 &     0.5 & 0.00038 & 58786.63613 &    17.5 & 0.00039 & 59872.97875 &   431.0 & 0.00025 & 60598.06902 &   707.0 & 0.00031  \\ 
58743.30286 &     1.0 & 0.00039 & 58787.95562 &    18.0 & 0.00038 & 59874.28862 &   431.5 & 0.00033 & 60599.38213 &   707.5 & 0.00031  \\ 
58744.60963 &     1.5 & 0.00045 & 58789.26379 &    18.5 & 0.00053 & 59875.60508 &   432.0 & 0.00026 & 60600.69647 &   708.0 & 0.00032  \\ 
58745.92898 &     2.0 & 0.00043 & 58956.09981 &    82.0 & 0.00058 & 59876.91544 &   432.5 & 0.00030 & 60602.01027 &   708.5 & 0.00034  \\
\hline
\end{tabular}}
\end{table*}

\begin{table*}
\caption{Times of minima of TIC\,351404069}
 \label{Tab:TIC_351404069_ToM}
\scalebox{0.88}{\begin{tabular}{@{}lrllrllrllrl}
\hline
BJD & Cycle  & std. dev. & BJD & Cycle  & std. dev. & BJD & Cycle  & std. dev. & BJD & Cycle  & std. dev. \\ 
$-2\,400\,000$ & no. &   \multicolumn{1}{c}{$(d)$} & $-2\,400\,000$ & no. &   \multicolumn{1}{c}{$(d)$} & $-2\,400\,000$ & no. &   \multicolumn{1}{c}{$(d)$} & $-2\,400\,000$ & no. &   \multicolumn{1}{c}{$(d)$} \\ 
\hline
58713.90919 &    -0.5 & 0.00051 & 59818.99070 &   123.5 & 0.00051 & 60349.16412 &   183.0 & 0.00039 & 60393.69644 &   188.0 & 0.00035  \\ 
58718.45552 &     0.0 & 0.00045 & 59823.27403 &   124.0 & 0.00041 & 60353.62223 &   183.5 & 0.00069 & 60536.35144 &   204.0 & 0.00026  \\ 
58722.81104 &     0.5 & 0.00095 & 59827.94418 &   124.5 & 0.00038 & 60358.05710 &   184.0 & 0.00043 & 60540.71828 &   204.5 & 0.00035  \\ 
58727.35608 &     1.0 & 0.00051 & 59832.17404 &   125.0 & 0.00040 & 60362.51292 &   184.5 & 0.00042 & 60545.24445 &   205.0 & 0.00034  \\ 
58731.70573 &     1.5 & 0.00078 & 59836.84659 &   125.5 & 0.00037 & 60366.95099 &   185.0 & 0.00038 & 60549.61684 &   205.5 & 0.00033  \\ 
58736.25469 &     2.0 & 0.00057 & 59841.11189 &   126.0 & 0.00040 & 60371.41166 &   185.5 & 0.00050 & 60554.13773 &   206.0 & 0.00029  \\ 
58740.59707 &     2.5 & 0.00049 & 59845.77213 &   126.5 & 0.00045 & 60375.84931 &   186.0 & 0.00039 & 60558.51710 &   206.5 & 0.00036  \\ 
58745.15324 &     3.0 & 0.00050 & 59850.06540 &   127.0 & 0.00036 & 60380.30850 &   186.5 & 0.00045 & 60563.03258 &   207.0 & 0.00033  \\ 
58754.05127 &     4.0 & 0.00044 & 60340.26817 &   182.0 & 0.00029 & 60384.75992 &   187.0 & 0.00039 & 60567.41565 &   207.5 & 0.00039  \\ 
58762.95196 &     5.0 & 0.00069 & 60344.71478 &   182.5 & 0.00051 & 60389.21318 &   187.5 & 0.00037 & 60580.85536 &   209.0 & 0.00036  \\ 
59801.18277 &   121.5 & 0.00042  \\ 
\hline
\end{tabular}}
\end{table*}

\begin{table*}[h!]
\caption{Times of minima of TIC\,378270875}
 \label{Tab:TIC_378270875_ToM}
\scalebox{0.88}{\begin{tabular}{@{}lrllrllrllrl}
\hline
BJD & Cycle  & std. dev. & BJD & Cycle  & std. dev. & BJD & Cycle  & std. dev. & BJD & Cycle  & std. dev. \\ 
$-2\,400\,000$ & no. &   \multicolumn{1}{c}{$(d)$} & $-2\,400\,000$ & no. &   \multicolumn{1}{c}{$(d)$} & $-2\,400\,000$ & no. &   \multicolumn{1}{c}{$(d)$} & $-2\,400\,000$ & no. &   \multicolumn{1}{c}{$(d)$} \\ 
\hline
58684.28515 &     0.0 & 0.00040 & 59438.03168 &   292.5 & 0.00026 & 60341.24879 &   643.0 & 0.00023 & 60517.76692 &   711.5 & 0.00024  \\ 
58685.57149 &     0.5 & 0.00067 & 59439.32149 &   293.0 & 0.00019 & 60342.53697 &   643.5 & 0.00028 & 60519.05678 &   712.0 & 0.00018  \\ 
58686.86054 &     1.0 & 0.00035 & 59440.60712 &   293.5 & 0.00029 & 60343.82678 &   644.0 & 0.00021 & 60520.34404 &   712.5 & 0.00017  \\ 
58688.14767 &     1.5 & 0.00057 & 59441.89717 &   294.0 & 0.00019 & 60345.11412 &   644.5 & 0.00025 & 60521.63392 &   713.0 & 0.00015  \\ 
58689.43640 &     2.0 & 0.00051 & 59443.18297 &   294.5 & 0.00027 & 60346.40308 &   645.0 & 0.00052 & 60522.92041 &   713.5 & 0.00015  \\ 
58690.72320 &     2.5 & 0.00062 & 59444.47278 &   295.0 & 0.00018 & 60347.69054 &   645.5 & 0.00030 & 60524.20997 &   714.0 & 0.00013  \\
\hline
\end{tabular}}
\end{table*}

\begin{table*}
\caption{Times of minima of TIC\,403792414}
 \label{Tab:TIC_403792414_ToM}
\scalebox{0.88}{\begin{tabular}{@{}lrllrllrllrl}
\hline
BJD & Cycle  & std. dev. & BJD & Cycle  & std. dev. & BJD & Cycle  & std. dev. & BJD & Cycle  & std. dev. \\ 
$-2\,400\,000$ & no. &   \multicolumn{1}{c}{$(d)$} & $-2\,400\,000$ & no. &   \multicolumn{1}{c}{$(d)$} & $-2\,400\,000$ & no. &   \multicolumn{1}{c}{$(d)$} & $-2\,400\,000$ & no. &   \multicolumn{1}{c}{$(d)$} \\ 
\hline
59391.36642 &     0.0 & 0.00084 & 59423.80076 &     6.5 & 0.00176 & 59778.45988 &    77.5 & 0.00391 & 60500.22737 &   222.0 & 0.00095  \\ 
59393.84551 &     0.5 & 0.00464 & 59426.32710 &     7.0 & 0.00063 & 59780.97284 &    78.0 & 0.00124 & 60502.74607 &   222.5 & 0.00158  \\ 
59396.37449 &     1.0 & 0.00130 & 59428.79336 &     7.5 & 0.00154 & 59785.97024 &    79.0 & 0.00059 & 60505.21473 &   223.0 & 0.00039  \\ 
59398.84138 &     1.5 & 0.00417 & 59431.31860 &     8.0 & 0.00041 & 59788.45458 &    79.5 & 0.00363 & 60507.73852 &   223.5 & 0.00093  \\ 
59401.36991 &     2.0 & 0.00096 & 59433.77914 &     8.5 & 0.01752 & 59790.96860 &    80.0 & 0.00121 & 60510.20456 &   224.0 & 0.00022  \\ 
59403.83489 &     2.5 & 0.00193 & 59436.30569 &     9.0 & 0.00079 & 59793.45113 &    80.5 & 0.00722 & 60512.72659 &   224.5 & 0.00083  \\
\hline
\end{tabular}}
\end{table*}

\begin{table*}
\caption{Times of minima of TIC\,403916758}
 \label{Tab:TIC_403916758_ToM}
\scalebox{0.88}{\begin{tabular}{@{}lrllrllrllrl}
\hline
BJD & Cycle  & std. dev. & BJD & Cycle  & std. dev. & BJD & Cycle  & std. dev. & BJD & Cycle  & std. dev. \\ 
$-2\,400\,000$ & no. &   \multicolumn{1}{c}{$(d)$} & $-2\,400\,000$ & no. &   \multicolumn{1}{c}{$(d)$} & $-2\,400\,000$ & no. &   \multicolumn{1}{c}{$(d)$} & $-2\,400\,000$ & no. &   \multicolumn{1}{c}{$(d)$} \\ 
\hline
58791.04022 &    -1.0 & 0.00303 & 58991.71010 &   176.0 & 0.00139 & 59900.39543 &   977.5 & 0.00098 & 60461.58223 &  1472.5 & 0.00190  \\ 
58791.60976 &    -0.5 & 0.00451 & 58992.28190 &   176.5 & 0.00259 & 59900.95764 &   978.0 & 0.00095 & 60462.14926 &  1473.0 & 0.00213  \\ 
58792.17603 &     0.0 & 0.00251 & 58992.84495 &   177.0 & 0.00224 & 59901.52962 &   978.5 & 0.00099 & 60463.28434 &  1474.0 & 0.00093  \\ 
58792.74172 &     0.5 & 0.00710 & 58993.41207 &   177.5 & 0.00293 & 59902.09476 &   979.0 & 0.00106 & 60463.85343 &  1474.5 & 0.00184  \\ 
58793.30940 &     1.0 & 0.00148 & 58993.97480 &   178.0 & 0.00141 & 59902.66247 &   979.5 & 0.00113 & 60464.41731 &  1475.0 & 0.00206  \\ 
58793.87868 &     1.5 & 0.00208 & 58994.54858 &   178.5 & 0.00577 & 59903.22643 &   980.0 & 0.00105 & 60466.11977 &  1476.5 & 0.00128  \\
\hline
\end{tabular}}
\end{table*}

\addtocounter{table}{-1}

\begin{table*}
\caption{Times of minima of TIC\,403916758 (continued)}
% \label{Tab:TIC_403916758_ToM}
\scalebox{0.88}{\begin{tabular}{@{}lrllrllrllrl}
\hline
BJD & Cycle  & std. dev. & BJD & Cycle  & std. dev. & BJD & Cycle  & std. dev. & BJD & Cycle  & std. dev. \\ 
$-2\,400\,000$ & no. &   \multicolumn{1}{c}{$(d)$} & $-2\,400\,000$ & no. &   \multicolumn{1}{c}{$(d)$} & $-2\,400\,000$ & no. &   \multicolumn{1}{c}{$(d)$} & $-2\,400\,000$ & no. &   \multicolumn{1}{c}{$(d)$} \\ 
\hline
58811.45291 &    17.0 & 0.00140 & 59722.39655 &   820.5 & 0.00174 & 59919.09275 &   994.0 & 0.12269 & 60611.80608 &  1605.0 & 0.00098  \\ 
58812.02046 &    17.5 & 0.00164 & 59722.96294 &   821.0 & 0.00229 & 59919.66919 &   994.5 & 0.06085 & 60612.37421 &  1605.5 & 0.00154  \\ 
58812.58289 &    18.0 & 0.00122 & 59723.53152 &   821.5 & 0.00125 & 59920.23332 &   995.0 & 0.05982 & 60612.94109 &  1606.0 & 0.00119  \\ 
58813.15199 &    18.5 & 0.00315 & 59724.09955 &   822.0 & 0.00134 & 59920.79877 &   995.5 & 0.07230 & 60613.50401 &  1606.5 & 0.00222  \\ 
58813.71698 &    19.0 & 0.00111 & 59724.66526 &   822.5 & 0.00194 & 59921.36212 &   996.0 & 0.03137 & 60614.07488 &  1607.0 & 0.00232  \\ 
58814.28690 &    19.5 & 0.00854 & 59725.23054 &   823.0 & 0.00143 & 59921.94030 &   996.5 & 0.30148 & 60614.64179 &  1607.5 & 0.00129  \\
\hline
\end{tabular}}
\end{table*}

\begin{table*}
\caption{Times of minima of TIC\,405789362}
 \label{Tab:TIC_405789362_ToM}
\scalebox{0.88}{\begin{tabular}{@{}lrllrllrllrl}
\hline
BJD & Cycle  & std. dev. & BJD & Cycle  & std. dev. & BJD & Cycle  & std. dev. & BJD & Cycle  & std. dev. \\ 
$-2\,400\,000$ & no. &   \multicolumn{1}{c}{$(d)$} & $-2\,400\,000$ & no. &   \multicolumn{1}{c}{$(d)$} & $-2\,400\,000$ & no. &   \multicolumn{1}{c}{$(d)$} & $-2\,400\,000$ & no. &   \multicolumn{1}{c}{$(d)$} \\ 
\hline
58738.84443 &    -0.5 & 0.00041 & 58963.16388 &    48.5 & 0.00030 & 59892.45421 &   251.5 & 0.00020 & 60444.08244 &   372.0 & 0.00014  \\ 
58741.14567 &     0.0 & 0.00019 & 58965.46590 &    49.0 & 0.00014 & 59894.74946 &   252.0 & 0.00020 & 60446.36875 &   372.5 & 0.00013  \\ 
58743.41839 &     0.5 & 0.00034 & 58967.73967 &    49.5 & 0.00029 & 59897.02859 &   252.5 & 0.00026 & 60448.65848 &   373.0 & 0.00024  \\ 
58745.71622 &     1.0 & 0.00025 & 58970.03628 &    50.0 & 0.00014 & 59899.32453 &   253.0 & 0.00017 & 60560.80125 &   397.5 & 0.00027  \\ 
58747.99476 &     1.5 & 0.00043 & 58972.31123 &    50.5 & 0.00026 & 59901.60165 &   253.5 & 0.00019 & 60563.09047 &   398.0 & 0.00015  \\ 
58750.28976 &     2.0 & 0.00059 & 58974.60963 &    51.0 & 0.00022 & 59903.89744 &   254.0 & 0.00013 & 60565.37913 &   398.5 & 0.00018  \\
\hline
\end{tabular}}
\textit{Notes}. $^a$: These eclipses observed with the GAO80 instrument.
\end{table*}

\begin{table*}
\caption{Times of minima of TIC\,461500036}
 \label{Tab:TIC_461500036_ToM}
\scalebox{0.88}{\begin{tabular}{@{}lrllrllrllrl}
\hline
BJD & Cycle  & std. dev. & BJD & Cycle  & std. dev. & BJD & Cycle  & std. dev. & BJD & Cycle  & std. dev. \\ 
$-2\,400\,000$ & no. &   \multicolumn{1}{c}{$(d)$} & $-2\,400\,000$ & no. &   \multicolumn{1}{c}{$(d)$} & $-2\,400\,000$ & no. &   \multicolumn{1}{c}{$(d)$} & $-2\,400\,000$ & no. &   \multicolumn{1}{c}{$(d)$} \\ 
\hline
58791.39329 &     0.0 & 0.00119 & 59032.32894 &    97.5 & 0.00015 & 59894.77700 &   446.5 & 0.00006 & 60412.48899$^a$ &   656.0 & 0.00002  \\ 
58792.63164 &     0.5 & 0.00009 & 59033.56140 &    98.0 & 0.00009 & 59896.01066 &   447.0 & 0.00008 & 60434.73454 &   665.0 & 0.00008  \\ 
58793.86528 &     1.0 & 0.00013 & 59034.79978 &    98.5 & 0.00007 & 59897.24718 &   447.5 & 0.00008 & 60435.96862 &   665.5 & 0.00007  \\ 
58795.10193 &     1.5 & 0.00011 & 59391.89327 &   243.0 & 0.00009 & 59898.48121 &   448.0 & 0.00006 & 60437.20269 &   666.0 & 0.00009  \\ 
58796.33557 &     2.0 & 0.00011 & 59393.13059 &   243.5 & 0.00010 & 59899.71715 &   448.5 & 0.00007 & 60438.43897 &   666.5 & 0.00007  \\ 
58797.57278 &     2.5 & 0.00008 & 59394.36417 &   244.0 & 0.00012 & 59900.95129 &   449.0 & 0.00007 & 60439.67264 &   667.0 & 0.00007  \\
\hline
\end{tabular}}
\textit{Notes}. Eclipses observed with the $^a$: GAO80 and $^b$: BAO80 instruments.
\end{table*}

\addtocounter{table}{-1}

\begin{table*}
\caption{Times of minima of TIC\,461500036 (continued)}
% \label{Tab:TIC_461500036_ToM}
\scalebox{0.88}{\begin{tabular}{@{}lrllrllrllrl}
\hline
BJD & Cycle  & std. dev. & BJD & Cycle  & std. dev. & BJD & Cycle  & std. dev. & BJD & Cycle  & std. dev. \\ 
$-2\,400\,000$ & no. &   \multicolumn{1}{c}{$(d)$} & $-2\,400\,000$ & no. &   \multicolumn{1}{c}{$(d)$} & $-2\,400\,000$ & no. &   \multicolumn{1}{c}{$(d)$} & $-2\,400\,000$ & no. &   \multicolumn{1}{c}{$(d)$} \\ 
\hline
58811.15678 &     8.0 & 0.00011 & 59406.71345 &   249.0 & 0.00008 & 59915.77580 &   455.0 & 0.00007 & 60454.49516 &   673.0 & 0.00007  \\ 
58812.39444 &     8.5 & 0.00007 & 59407.95227 &   249.5 & 0.00008 & 59917.01332 &   455.5 & 0.00005 & 60455.73104 &   673.5 & 0.00007  \\ 
58813.62674 &     9.0 & 0.00014 & 59409.18421 &   250.0 & 0.00008 & 59918.24803 &   456.0 & 0.00005 & 60456.96625 &   674.0 & 0.00007  \\ 
58814.86397 &     9.5 & 0.00029 & 59410.42204 &   250.5 & 0.00010 & 59919.48518 &   456.5 & 0.00006 & 60458.20215 &   674.5 & 0.00007  \\ 
58843.28889 &    21.0 & 0.00012 & 59411.65479 &   251.0 & 0.00009 & 59920.72062 &   457.0 & 0.00007 & 60459.43821 &   675.0 & 0.00010  \\ 
58844.52677 &    21.5 & 0.00014 & 59412.89216 &   251.5 & 0.00011 & 59921.95768 &   457.5 & 0.00007 & 60460.67433 &   675.5 & 0.00009  \\
\hline
\end{tabular}}
\textit{Notes}. Eclipses observed with the $^a$: GAO80 and $^b$: BAO80 instruments.
\end{table*}

\end{appendix}

\begin{thebibliography}{}

 \bibitem [\protect\citeauthoryear{Bailer-Jones et al.}{2021}]  {bailer-jonesetal21} Bailer-Jones, C. A. L., Rybizki, J., Fouesneau, M., Demleitner., M., Andrae, R., 2021, \aj, 161, 147

   \bibitem[Borkovits(2022)]{borkovits22} Borkovits, T. 2022, Galaxies, 10, 9

   \bibitem[Borkovits \& Mitnyan (2023)]{borkovitsmitnyan23} Borkovits, T., \& Mitnyan, T. 2023, Universe, 9, 485

   \bibitem[Borkovits et al. (2003)]{borkovitsetal03} Borkovits, T., \'Erdi, B., Forg\'acs-Dajka, E., Kov\'acs, T. 2003, \aap, 398, 1091
   
   \bibitem[Borkovits et al. (2013)]{borkovitsetal13} Borkovits, T., Derekas, A., Kiss, L.L., et al. 2013, \mnras, 428, 1656

  \bibitem[Borkovits et al.(2015)]{borkovitsetal15} Borkovits T., Rappaport S., Hajdu T., Sztakovics J., 2015, \mnras , 448, 946
   
  \bibitem [\protect\citeauthoryear{Borkovits et al.}{2016}]  {borkovitsetal16} Borkovits, T., Hajdu, T., Sztakovics, J., Rappaport, S., Levine, A., B\'\i r\'o, I. B., Klagyivik, P., 2016, \mnras, 455, 4136

\bibitem [\protect\citeauthoryear{Borkovits et al.}{2018}]  {borkovitsetal18} Borkovits, T., Albrecht, S., Rappaport, S., et al. 2018, \mnras, 478, 513

 \bibitem [\protect\citeauthoryear{Borkovits et al.}{2019a}]  {borkovitsetal19a} Borkovits, T., Rappaport, S., Kaye, T., et al. 2019a, \mnras, 483, 1934

 \bibitem [\protect\citeauthoryear{Borkovits et al.}{2019b}]  {borkovitsetal19b} Borkovits, T.,  Sperauskas, J., Tokovinin, A., Latham, D. W., Cs\'anyi, I., Hajdu, T., Moln\'ar, L., 2019b, \mnras, 487, 4631

 \bibitem [\protect\citeauthoryear{Borkovits et al.}{2020a}]  {borkovitsetal20a} Borkovits, T., Rappaport, S., Hajdu, T., et al. 2020a, \mnras, 493, 5005
      
 \bibitem[Borkovits et al.(2020b)]{borkovitsetal20b}Borkovits, T., Rappaport, S., Tan, T.G., et al.  2020b, \mnras, 496, 4624

  \bibitem [\protect\citeauthoryear{Borkovits et al.}{2021}]  {borkovitsetal21} Borkovits, T., Rappaport, S., Maxted, P. F. L., et al. 2021, \mnras, 503, 3759

   \bibitem[Borkovits et al.(2022a)]{borkovitsetal22a} Borkovits, T., Mitnyan, T., Rappaport, S., et al. 2022a, \mnras, 510, 1352
   
   \bibitem[Borkovits et al.(2022b)]{borkovitsetal22b} Borkovits, T., Rappaport, S. A., Toonen, S., Moe, M., Mitnyan, T., Cs\'anyi, I., 2022b, \mnras, 515, 3773
   
   \bibitem[Borkovits et al.(2025)]{borkovitsetal25} Borkovits, T., Rappaport, S. A., Mitnyan, T., et al., 2025, \aap, 695, A209
   
   \bibitem[Borucki et al.(2010)]{borucki10} Borucki W. J. et al., 2010, Science, 327, 977
   
   \bibitem[\protect\citeauthoryear{Bressan et al.}{2012}]{PARSEC} Bressan, A., Marigo, P., Girardi, L. et al. 2012, \mnras, 427, 127
   
   \bibitem[Carter et al.(2011)]{carter11} Carter, J.A., Fabrycky, D.C., Ragozzine, D., et al. 2011, Sci, 331, 562
   
   \bibitem[\protect\citeauthoryear{Correia et al.}{2011}]{correiaetal11} Correia, A.C.M, Laskar, J., Farago, F., Bou\'e, G., 2011, CeMDA, 111, 105
      
   \bibitem [\protect\citeauthoryear{Cutri et al.}{2013}]{WISE} Cutri, R.M., Wright, E.L., Conrow, T., et al.~2013, wise.rept, 1C.
   
   \bibitem[Czavalinga et al.(2023)]{czavalingaetal23} Czavalinga, D.~R., Borkovits, T.,  Mitnyan, T., Rappaport, S.~A.\& P\'al, A. 2023, \mnras, 526, 2830

  \bibitem[\protect\citeauthoryear{Derekas et al.}{2011}]{derekasetal11} Derekas, A., Kiss, L. L., Borkovits, T., et al. 2011, Sci, 332, 216

  \bibitem[\protect\citeauthoryear{Ebbighausen \& Struve}{1956}]{ebbighausenstruve956} Ebbighausen, E. G., Struve, O. 1956, \apj, 124, 507
  
  \bibitem [\protect\citeauthoryear{Ford}{2005}] {ford05} Ford, E. B., 2005, \aj, 129, 1706

   \bibitem [\protect\citeauthoryear{Gaia collaboration}{2021}]  {GaiaEDR3} Gaia Collaboration, Brown, A. G. A., Vallenari, A., Prusti, T. et al. 2021, \aap, 649, A1

  \bibitem[Gaulme et al.(2022)]{gaulme22} Gaulme, P., Borkovits, T., Appourchaux, T., et al. 2022, \aap, 668, A173

  \bibitem[\protect\citeauthoryear{Grishin \& Perets}{2022}]{grishinperets22} Grishin, E., Perets, H.B., 2022, \mnras, 512, 4993
   
   \bibitem [\protect\citeauthoryear{Henden et al.}{2015}] {APASS} Henden, A. A., Levine, S., Terrell, D., Welch, D. 2015, American Astronomical Society, AAS Meeting \#225, id.336.16
   
   \bibitem [\protect\citeauthoryear{Kisseleva-Eggleton \& Eggleton}{2010}] {kisseleva-eggletoneggleton10} Kisseleva-Eggleton, L., Eggleton, P. 2010, in {\it International Conference on Binaries: in celebration of Ron Webbink's 65th Birthday}, eds. Kalogera, V., van der Sluys, M., American Institute of Physics Conference Series, 1314, 128
         
   \bibitem[Kostov et al.(2021)]{kostovetal21} Kostov et al., 2021, \apj, 917, 93

   \bibitem[Kostov et al.(2022)]{kostovetal22} Kostov, V.B., Powell, B.P., Rappaport, S.A, et al.,  2022, \apjs, 259, 66
      
   \bibitem[Kostov et al.(2024)]{kostovetal24} Kostov, V.B., Rappaport, S.A., Borkovits, T., et al., 2024, \apj, 974, 25
         
   \bibitem[Kristiansen et al.(2022)]{kristiansen22} Kristiansen, M.H., Rappaport, S., Vanderburg, A., et al. 2022, \pasp, 134, 074401
 
   \bibitem[Lightkurve Collaboration (2018)]{2018ascl.soft12013L} Lightkurve Collaboration, Cardoso, Jos{\'e} Vin{\'\i}cius de Miranda, Hedges, C., Gully-Santiago, M., et al. 2018, Astrophysics Source Code Library, record ascl:1812.013, arXiv: 1812.013
   
   \bibitem[Lindegren et al. (2021)]{lindegren21} Lindegren, L., Bastian, U., Biermann, M., et al. 2021, A\&A, 649, 4L
   
   \bibitem[Mardling \& Aarseth (2001)]{mardlingaarseth01} Mardling R. A., \& Aarseth S. J., 2001, \mnras, 321, 398
   
   \bibitem [\protect\citeauthoryear{Mitnyan et al.}{2020}] {mitnyanetal20} Mitnyan, T., Borkovits, T., Rappaport, S., P\'al, A., Maxted, P. F. L., 2020, \mnras, 498, 6034
   
   \bibitem[Naoz (2016)]{naoz16} Naoz, S. 2016, ARA\&A, 54, 441
   
   \bibitem[Ochsenbein et al.(2000)]{ochsenbein00} Ochsenbein F., Bauer P., Marcout J., 2000, \aaps, 143, 23
   
   \bibitem [\protect\citeauthoryear{Offner~et~al.}{2023}]{offneretal23} Offner, S. S. R., Moe, M., Kratter, K. M., Sadavoy, S. I., Jensen, E. L. N., Tobin, J. J., 2023 ASP Conf. Ser. 534, 275
   
   \bibitem[Orosz (2023)]{orosz2023} Orosz, J.A. 2023, Universe, 9, 12, 505
         
   \bibitem [\protect\citeauthoryear{Paegert et al.}{2021}] {TIC8} Paegert, M. et al, 2021, arXiv:2108.04778
   
   \bibitem[P\'al (2012)] {pal12} P\'al, A., 2012, \mnras, 421, 1825

   \bibitem[Rappaport et al.(2013)]{rappaportetal13} Rappaport S., Deck K., Levine A., Borkovits T., Carter J., El Mellah I., Sanchis-Ojeda R., \& Kalomeni B., 2013, \apj , 768, 33
      
   \bibitem[Rappaport et al.(2022)]{rappaportetal22} Rappaport, S., Borkovits, T., Gagliano, R., et al. 2022, \mnras, 513, 4341
   
   \bibitem[Rappaport et al.(2023)]{rappaportetal23} Rappaport, S., Borkovits, T., Gagliano, R., et al. 2023, \mnras, 521, 558
   
   \bibitem[Rappaport et al.(2024)]{rappaportetal24} Rappaport, S., Borkovits, T., Mitnyan, T., et al. 2024, \aap, 686, A27
      
   \bibitem[Ricker et al.(2015)]{ricker15} Ricker, G.R., Winn, J.N., Vanderspek, R., et al. 2015, JATIS, 1, 014003
   
   \bibitem[Roemer (1677)]{roemer1677} Roemer, O. 1677, Philosophical Transactions (1665-1678), Volume 12, 893
   
   \bibitem [\protect\citeauthoryear{Saglia et al.}{2025}]{sagliaetal25} Saglia et al., 2025, \aap, 699, A151 
      
   \bibitem[S\"oderhjelm (1975)]{soderhjelm75} S\"oderhjelm, S. 1975, A\&A, 42, 229.
   
   \bibitem[Skrutskie et al.(2006)]{2MASS} Skrutskie, M.F., Cutri, R.M., Stiening, R., et al. 2006, \aj, 131, 1163
      
   \bibitem [\protect\citeauthoryear{Tokovinin}{2014}]{tokovinin14} Tokovinin, A., 2014, \aj, 147, 86

   \bibitem [\protect\citeauthoryear{Tokovinin}{2021}]{tokovinin21} Tokovinin, A., 2021, Universe, 7, 352

   \bibitem[Zasche et al.(2022)]{zascheetal22} Zasche, P., Henzl, Z., \& Ma\v{s}ek, M. 2022, A\&A, 664, 96

\end{thebibliography}
\end{document}